\documentclass[11pt]{article}

\usepackage{amsmath}
\allowdisplaybreaks
\usepackage{amssymb}
\usepackage{amsthm}
\usepackage{pifont}
\usepackage{hyperref}
\usepackage{bm}
\usepackage{listings}
\usepackage{lstbayes}
\usepackage{graphicx}
\usepackage{xcolor}
\usepackage{float}
\usepackage[capitalize,nameinlink,noabbrev]{cleveref}
\usepackage{authblk}
\usepackage{setspace}
\usepackage{todonotes}
\usepackage{lipsum}
\usepackage{orcidlink}

\usepackage[total={16cm, 22cm}]{geometry}

\def\spacingset#1{\renewcommand{\baselinestretch}
{#1}\small\normalsize} \spacingset{1}

\theoremstyle{definition}
\newtheorem{defn}{Definition}
\theoremstyle{plain}
\newtheorem{prop}[defn]{Proposition}

\usepackage{natbib}
\DeclareMathOperator{\cov}{\mathrm{Cov}}
\DeclareMathOperator{\var}{\mathrm{Var}}
\DeclareMathOperator{\ch}{\mathrm{chol}}
\DeclareMathOperator{\vc}{\mathrm{vec}}
\newcommand{\gp}{\mathrm{GP}}
\newcommand{\indep}{\perp \!\!\! \perp}
\newcommand{\R}{\mathbb{R}}
\newcommand{\m}[1]{\bm{#1}}

\title{Computationally efficient multi-level Gaussian process regression for functional data observed under completely or partially regular sampling designs}

\author[]{Adam Gorm Hoffmann\,\orcidlink{0009-0008-0058-3804}\thanks{Corresponding author. adam.hoffmann@sund.ku.dk, Section of Biostatistics, Department of Public Health, University of Copenhagen. Øster Farimagsgade 5. DK-1353 Copenhagen. Denmark.}}
\author[]{Claus Thorn Ekstrøm\,\orcidlink{0000-0003-1191-373X}}
\author[]{Andreas Kryger Jensen\,\orcidlink{0000-0002-8233-9176}}
\affil[]{Section of Biostatistics, Department of Public Health, University of Copenhagen}

\date{}

\begin{document}
\maketitle

\begin{abstract}
Gaussian process regression is a frequently used statistical method for flexible yet fully probabilistic non-linear regression modeling. A common obstacle is its computational complexity which scales poorly with the number of observations. This is especially an issue when applying Gaussian process models to multiple functions simultaneously in various applications of functional data analysis. 

We consider a multi-level Gaussian process regression model where a common mean function and individual subject-specific deviations are modeled simultaneously as latent Gaussian processes. We derive exact analytic and computationally efficient expressions for the log-likelihood function and the posterior distributions in the case where the observations are sampled on either a completely or partially regular grid. This enables us to fit the model to large data sets that are currently computationally inaccessible using a standard implementation. We show through a simulation study that our analytic expressions are several orders of magnitude faster compared to a standard implementation, and we provide an implementation in the probabilistic programming language Stan.
\end{abstract}

\begin{center}
  Gaussian process regression, Linear algebra
\end{center}

\newpage

\spacingset{1.3}

\section{Introduction}

A fundamental problem in functional data analysis is to estimate an underlying mean function and individual subject-specific trajectories in continuous time based on a collection of discretely sampled observations from some underlying set of functions contaminated with noise. An often used method is based on Functional Principal Component Analysis (FPCA), see e.g., \citet{rice1991estimating} and \citet{wang2016functional}. However, as FPCA in its standard form is not a fully probabilistic model, it is not straightforward to correctly account for the uncertainty in the estimates it provides. This can be problematic if e.g., the estimated functions are used downstream in other functional data methods. A particularly opportune solution to this problem is to frame the function estimation problem in a Bayesian framework based on latent Gaussian processes.

Gaussian process regression is a popular and flexible method for non-parametric but fully probabilistic curve estimation \citep{gpml}. In the standard version, Gaussian process regression is concerned with estimating a single latent function based on noise-contaminated observations, and it can be implemented as either an empirical Bayes or a fully Bayesian method. However, in both cases, the computation time can be substantial even for moderately sized problems. This is mainly due to the requirement of the inverse of a potentially large covariance matrix which grows cubicly in the number of observations. In a fully Bayesian treatment using e.g., Hamiltonian Monte Carlo methods this problem is exacerbated due to the many evaluations required of the log-likelihood. Several approaches for improving the computation time of Gaussian process regression have been considered. These include e.g., methods based on sparseness through inducing points \citep{quinonero2005unifying} or approximations of the covariance kernel through Laplace operators \citep{riutort2023practical}. However, these approaches all involve different kinds of approximations that make the model that is actually fitted deviate from its intended probabilistic definition.

Extensions of Gaussian process regression to functional data with replications have been considered by \citet{gp-anova} and as special cases of the additive Gaussian process models for longitudinal data introduced by \citet{cheng2019additive} and \citet{lgpr}. We consider a multi-level setup similar to the functional ANOVA model of \citet{gp-anova} where we simultaneously model a set of functions and their common mean function in the framework of Gaussian process priors based on vectors of discrete observations contaminated with additive Gaussian noise. The main computational obstacle in fitting such a model is calculating certain matrix operations involving the covariance matrix of the observed data used in the expressions of the log-likelihood function and the posterior distribution of the Gaussian processes. In order to make it computationally feasible, we show how a complete or partial restriction on the sampling design can lead to enormous improvements in computation time.

Specifically, we first consider the case where the observations of all the functions are taken at the same vector of sampling points which we will subsequently think of as time. We call this the completely regular sampling design. Examples of functional data where a completely regular sampling design can occur are among others: electrocardiograms measured at a fixed frequency starting at the end of the QRS complex, blood glucose measurements using a continuous glucose monitoring device (CGM), climate temperature and precipitation data, electricity consumption, absorbance values from near-infrared spectroscopy used in food quality control, energy expenditure measured from wearable devices, and mass spectrometry data used in proteomics. We provide exact and analytic expressions for the log-likelihood and the posterior distribution of the latent functions under this design by exploiting a certain block structure in the covariance matrix along with symbolic simplifications of the expressions leading to a computational complexity where most computationally expensive operations are independent of the number of latent functions. These expressions are exact and do not involve any approximations of the model. 

To relax the complete regularity requirement we also consider a partially regular sampling design where some subset of the functions are assumed to be measured at the same vector of sampling points but the remaining functions can be observed at arbitrary sampling points. In this setting, we also derive exact and analytic expressions for the log-likelihood and the posterior distributions of the latent Gaussian processes.

We have implemented these efficient expressions in the probabilistic programming language Stan \citep{stan} which enables users to either perform fully Bayesian estimation and inference or use the individual components of the implementation in other settings by exposing the Stan functions to R \citep{r} using the R package \texttt{cmdstanr} \citep{cmdstanr}. The implementation is available on the first author's GitHub site at \url{https://github.com/adamgorm/efficient-gp}.

This paper is structured as follows: In \cref{sec:methods} we introduce our model based on hierarchically defined Gaussian processes and present analytic expressions for the log-likelihood and posterior distributions under both a completely regular and a partially regular sampling design. In addition, we describe how computational efficiency can be obtained by using an iterative block Cholesky factorization. In \cref{sec:benchmarks} we empirically compare the computation time of our approach to the naïve implementation. We conclude with a discussion in \cref{sec:discussion}.

\paragraph{Notation:}
Vectors and matrices are denoted by bold lower and upper case letters, respectively. For a function $f:\R^d \to \R$ we write $\m f(\m t) := (f(\m t_1), \cdots, f(\m t_n))$ where $\m t = (\m t_1^T, \ldots, \m t_n^T)^T\in (\R^d)^n$. Similarly, for a function $K:\R^d \times \R^d \to \R$ we write $\m K(\m t, \m t')$ for the $n\times n$ matrix with entries $\m K(\m t, \m t')_{ij} = K(\m t_i, \m t'_j)$ where $\m t, \m t' \in (\R^d)^n$. Special matrices such as the $n$-dimensional square identity matrix and the $n \times m$ matrix with the value one in every entry are denoted by $\m I_{n}$ and $\m 1_{n,m}$ respectively. We write $\indep$ to denote independence, $|\cdot|$ to denote the determinant, and $\otimes$ to denote the Kronecker product. The $\vc$ operator maps a matrix $\m A \in \mathbb{R}^{n \times m}$ to a vector obtained by stacking its columns so that $\vc(\m A)  \in \mathbb{R}^{n m}.$ We write $\vc^{-1}$ for the inverse operation, which maps the column vector $\vc(\m A)$ to the matrix $\m A$. If the dimensions $n$, $m$ are not clear from the context, we write $\vc^{-1}_{n, m}$ to make clear that it is a map $\R^{n m} \to \R^{n \times m}$. The Cholesky factorization of a positive definite symmetric matrix $\m A$ is denoted $\ch \m A = \m L \m L^T$, where $\m L$ is a unique lower triangular matrix with positive diagonal elements.

\section{Methods}\label{sec:methods}

A stochastic process $(f(\m t))_{\m t\in\R^d}$ is a Gaussian process if for any finite set of index points, $\m t_1, \ldots, \m t_J$, $(f(\m t_1), \ldots, f(\m t_J))$ follows a multivariate Gaussian distribution. The distribution of a Gaussian process is completely determined by a mean function $\mu : \R^d \to \R$ and a covariance function (kernel) $K:\R^d \times \R^d \to \R$, such that for any $\m t = (\m t_1^T, \ldots, \m t_J^T)^T \in \left(\R^d\right)^J$ the distribution of $(f(\m t_1), \ldots, f(\m t_J))$ is multivariate Gaussian with mean $(\mu(\m t_1), \ldots, \mu(\m t_J))$ and covariance given by $\cov(f(\m t_i), f(\m t_j)) = K(\m t_i, \m t_j)$.  We use the notation $f \sim \gp(\mu, K)$ and note that $\m f(\m t) \sim N(\m \mu (\m t), \m K(\m t, \m t))$ using the notation introduced above. A function $K$ is a valid covariance function, if and only if it is positive semi-definite, in the sense that $\m v^T \m K(\m t, \m t) \m v \ge 0$ for any choice of $J\in \mathbb{N}$, $\m v \in \R^J$ and $\m t \in \left(\R^d\right)^J$ \citep{gpml}. Usually, the covariance function will depend on a vector of parameters and we write $K_{\m \theta}$. Here we will focus on Gaussian processes defined on a univariate, $d = 1$, compact domain which we will think of as time.

The simplest example of Gaussian process regression is where several observations $\m y(\m t) := (y(t_1), \ldots, y(t_J))$ at time points $\m t = (t_1,\ldots, t_J)$ are assumed generated from a latent Gaussian process with Gaussian measurement noise. This model can be written as 
\begin{align*}
  (\m \theta, \sigma) &\sim G_{\m \Psi}\\
  f \mid \m \theta &\sim \gp(0, K_{\m \theta})\\
  y(t_{j}) \mid f(t_j), \sigma & \stackrel{\indep}{\sim} N(f(t_j), \sigma^2)
\end{align*}
where $G$ is some prior distribution over the covariance and measurement noise parameters with hyper-parameters $\m \Psi$, and we assume without loss of generality that the latent Gaussian process has mean zero. The distribution of the observed data is given by $\m y(\m t) \mid \m \theta, \sigma \sim N(0, \m K_{\m \theta}(\m t, \m t) + \sigma^2 \m I_J)$. Let $\m \Theta = (\m \theta, \sigma)$  be the vector of parameters. Then the log-likelihood function is
\begin{align*}
\log p(\m y(\m t) \mid \m \Theta) = -0.5\left(J\log(2\pi) + \log |\m \Sigma_{\m \Theta}| +
 \m y(\m t)^T \m \Sigma_{\m \Theta}^{-1} \m y(\m t)\right)
\end{align*}
with $\m \Sigma_{\m \Theta} = \m K_{\m \theta}(\m t, \m t) + \sigma^2 \m I_J$. Prediction of the latent function $\m f(\tilde{\m t}) := (f(\tilde{t}_1), \ldots, f(\tilde{t}_{\tilde{J}}))$ at any finite vector of time points $(\tilde{t}_1, \ldots, \tilde{t}_{\tilde{J}})$ is given by considering the posterior distribution of $f \mid \m y(\m t)$. In our case with additive Gaussian measurement error, this posterior distribution is conditionally conjugate. Conditional on the covariance parameters and the measurement error variance, the posterior takes the following analytic form as a consequence of the properties of conditioning in a multivariate normal distribution: $\m f(\tilde{\m t}) \mid \m y(\m t), \m\Theta \sim
  N\left(\m \mu_{\m f(\tilde{\m t}) \mid \m y(\m t), \m \Theta}, \m \Sigma_{\m f(\tilde{\m t}) \mid \m y(\m t), \m \Theta}\right)$
where $\m \mu_{\m f(\tilde{\m t}) \mid \m y(\m t), \m \Theta} = \m K_{\m\theta}(\tilde{\m t}, \m t) \m \Sigma^{-1}_{\m \Theta} \m y(\m t)$ and $\m \Sigma_{\m f(\tilde{\m t}) \mid \m y(\m t), \m \Theta} = \m K_\theta(\tilde{\m t}, \tilde{\m t}) - \m K_{\m\theta}(\tilde{\m t}, \m t) \m \Sigma^{-1}_{\m \Theta} \m K_{\m\theta}(\m t, \tilde{\m t})$.

When fitting the model using Markov chain Monte Carlo methods it is therefore not necessary to include $\m f(\tilde{\m t})$ as parameters due to the conditional conjugacy. In practice, prior distributions will be defined for $\m \Theta$ and each posterior draw is then plugged into the analytic expression for the conditional posterior distribution. This is typically much faster than including
$\m f(\tilde{\m t})$ as parameters directly in the model. Despite the efficiency gained by having an analytic expression for the posterior distribution, there is still a computational challenge present when the number of observations, $J$, is large. Specifically, the computational bottleneck is controlled by the need to handle the covariance matrix of the observed data, $\m \Sigma_{\m \Theta}$, and calculating its log-determinant and the product of its inverse with either a vector or a matrix. This computational challenge is especially exacerbated when we consider the problem of multi-level Gaussian process regression. We will now describe how the computational burden can be greatly reduced under either a completely regular or a partially regular sampling design.

\begin{figure}[ht]
  \centering\includegraphics[scale=1]{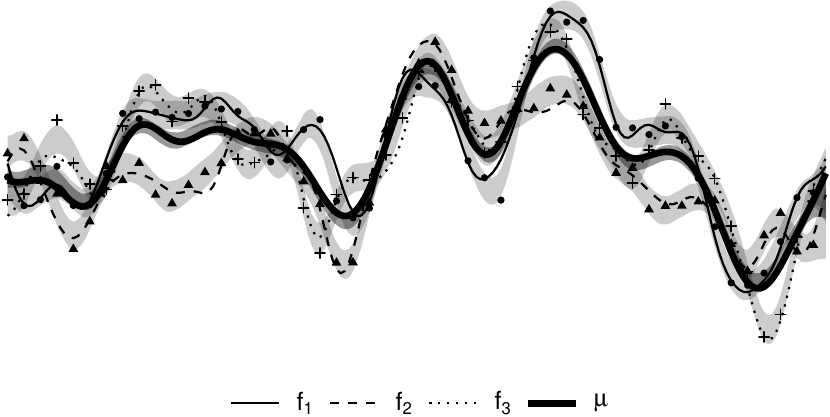}
  \caption{A random realization of a common mean function and three subject-specific functions from the multi-level GP model along with 90\% credible intervals based on 50 equidistant observations per function with Gaussian noise.}
  \label{fig:fit-example}
\end{figure}

In the rest of this manuscript, we consider a multi-level extension of the single function Gaussian process regression model just presented. Specifically, we consider $n$ latent functions $f_1, \ldots, f_n$ from which we observe noisy realizations
$\m y_i(\m t_i) = \m f_i(\m t_i) + \m \varepsilon_i \in \mathbb{R}^{J_i}$ at the
$J_i\in\mathbb{N}$ time points $\m t_i = (t_{i1}, \ldots, t_{iJ_i})$. We will consider each function as consisting of a common mean $\mu$ and an individual-specific deviation $\eta_i$, so that $f_i := \mu + \eta_i$. An example of a random draw from such a model is seen in \cref{fig:fit-example}, where the thick line shows the true common mean and the three finer lines show realizations of three functions with each their individual-specific deviations around the mean. The points are the observed data which in this example are the function values at 50 equidistant time points with additive Gaussian noise, and the bands are 90\% credible intervals.

Some identifiability condition is required as the model contains $n+1$ functions, $\mu,\eta_1,\ldots,\eta_n$, to describe
the $n$ functions $f_1,\ldots,f_n$. Similar to \citet{gp-anova} and \citet{lgpr}, we will make the model identifiable by requiring that $\sum_{i = 1}^n \eta_i = 0$. This means that, for each point of interest $t_0 \in \R$, we only need to fit $\eta_1(t_0)$, \ldots, $\eta_{n-1}(t_0)$ and then calculate $\eta_n(t_0) = -\sum_{i = 1}^{n-1} \eta_i(t_0)$. To impose this identifiability constraint, we will model $\m \eta$ as a multi-output Gaussian process. We write a multi-output Gaussian process as $(\eta_1, \ldots, \eta_n) \mid \m \theta_\eta \sim \gp_n(0, \m \Xi, K^\eta_{\m\theta_\eta})$ to denote that $\m \eta(\m t) \mid \m\theta_\eta$ is multivariate normal with mean $\m 0$ and covariances $\cov(\m \eta_i(\m t_i), \m \eta_j(\m t_j) \mid \m \theta_\eta) = \xi_{ij} \m K^\eta_{\m \theta_\eta}(\m t_i, \m t_j)$ \citep{multiGP}. The matrix $\m \Xi$ determines the covariance structure
between the different $\eta$ functions, while the kernel $K^\eta_{\m\theta_\eta}$
determines the individual covariance structure for each function and in
particular $\eta_i \mid \m\theta_\eta \sim \gp(0, \xi_{ii}K^\eta_{\m\theta_{\eta}})$. Note, that
$\m \Xi$ also influences the magnitude of the covariance of $\eta_i$, but this
can be circumvented by e.g., setting $\xi_{ii} = 1$ for all $i$. Using this definition of a multi-output Gaussian process we consider the following multi-level model for a collection of observations obtained from $n$ latent Gaussian processes
\begin{equation}
\begin{aligned}
  (\m\theta_\mu, \m\theta_\eta, \sigma) &\sim G_{\m \Psi}\\
  \mu \mid \m\theta_\mu &\sim \gp(0, K^\mu_{\m\theta_\mu})\\
  (\eta_1, \ldots, \eta_n) \mid \m\theta_\eta &\sim \gp_n(0, \m \Xi, K^\eta_{\m\theta_\eta})\\
  f_i(t_{ij}) &= \mu(t_{ij}) + \eta_i(t_{ij})\\
  y_i(t_{ij}) \mid f_i(t_{ij}), \sigma &\stackrel{\indep}{\sim} N(f_i(t_{ij}), \sigma^2)\\
\end{aligned}
\label{mod:eta}  
\end{equation}
where $K^\mu_{\m\theta_\mu}$ and $K^\eta_{\m\theta_\eta}$ are covariance functions with parameters $\m\theta_\mu$ and $\m\theta_\eta$. The model definition is completed by specifying $\m \Xi$ to get the desired property that $\sum_{i = 1}^n \eta_i = 0$. \cref{prop:xi} defines $\m \Xi$ and a proof is given in the supplementary material.

\begin{prop}      
  \label{prop:xi}  
  With $\m \Xi = (\xi_{ij}) \in \R^{n\times n}$ given by $\xi_{ii} = 1$ and
  $\xi_{ij} = -\frac{1}{n - 1}$ for $i\ne j$, \cref{mod:eta} implies that
  $P\left(\sum_{i = 1}^n \eta_i(t_0) = 0\right) = 1$ for any time point
  $t_0 \in \R$.
\end{prop}

We now turn to detail the analytic expressions for the two necessary components for implementing the multi-level model in \cref{mod:eta}; the likelihood of the observed data and the posterior distributions of $\mu$ and $\m \eta$. This enables us to fit this model using exact analytic expressions for data sets several orders of magnitude larger than what is currently possible using a naïve implementation.

\subsection{Completely Regular Sampling Design}
We consider the multi-level model defined in \cref{mod:eta} in the regular sampling case where $\m y_1, \ldots, \m y_n$ are all observed at the same time points $\m t := \m t_1 = \cdots = \m t_n \in \R^J$ and show how this leads to simplifications in the likelihood and the posterior distributions due to a Kronecker product structure in the covariance matrix of the observed data. The key insight to how computational efficiency can be obtained in the completely regular sampling design is to note that the covariance matrix of the observed data $\m y(\m t) := (\m y_1(\m t)^T, \ldots, \m y_n(\m t)^T)^T$  conditional on the parameters $\m \Theta = (\m \theta_\eta, \m \theta_\mu, \sigma)$ under the model in \cref{mod:eta} takes the following block form
\begin{align}
\m \Sigma_{\m \Theta} = \begin{bmatrix}\var[\m y_i \mid \m \Theta] & \cov[\m y_i, \m y_j \mid \m \Theta] &  \cov[\m y_i, \m y_j \mid \m \Theta] & \cdots & \cov[\m y_i, \m y_j \mid \m\Theta]\\ \cov[\m y_i, \m y_j \mid \m \Theta] & \var[\m y_i, \mid \m \Theta] & \cov[\m y_i, \m y_j \mid \m \Theta] & \cdots & \cov[\m y_i, \m y_j \mid \m \Theta]\\ \vdots & \vdots & \vdots & \ddots & \vdots \\ \cov[\m y_i, \m y_j \mid \m \Theta] & \cov[\m y_i, \m y_j \mid \m \Theta] & \cov[\m y_i, \m y_j \mid \m \Theta] & \cdots & \var[\m y_i \mid \m\Theta]\end{bmatrix}\label{eq:covRegularBlockForm}
\end{align}
where each diagonal block is equal to each other and similarly for each off-diagonal block. Specifically, all the blocks are $J \times J$ matrices of the form
\begin{align*}
  \var[\m y_i \mid \m\Theta] &= K^\mu_{\m \theta_\mu}(\m t, \m t) + K^\eta_{\m\theta_\eta}(\m t, \m t) + \sigma_\varepsilon^2 \m I_{J}\\
  \cov[\m y_i, \m y_j \mid \m\Theta] &= K^\mu_{\m\theta_\mu}(\m t, \m t) - \frac{1}{n-1}K^\eta_{\m \theta_\eta}(\m t, \m t)
\end{align*} 
as a consequence of the model specification in \cref{mod:eta} and the result of \cref{prop:xi}. A direct but crucial consequence of the regular sampling design leading to this block structure is that it can be written as the sum of two Kronecker products as
\begin{equation*}
\label{eq:covarianceKronecker}
\m \Sigma_{\m \Theta} = \m I_n \otimes \left(\var[\m y_i \mid \m \Theta] - \cov[\m y_i, \m y_j \mid \m \Theta]\right) + \m 1_{n,n} \otimes \cov[\m y_i, \m y_j \mid \m \Theta].
\end{equation*}
Using the identity of \citet[11.12]{seber2008matrix}, $\m \Sigma^{-1}_{\m \Theta}$ has the same structure as $\m \Sigma_{\m \Theta}$ being a sum of two Kronecker products. This facilitates a simplification of the terms $\log |\m \Sigma_{\m \Theta}|$ and $\m \Sigma^{-1}_{\m \Theta} \m y(\m t)$ required in the analytic expression for the likelihood function and the posterior distributions of $\mu$ and $\m \eta$ and hence a massive increase in computational efficiency. \cref{prop:loglikreg} states the simplifications of the log-likelihood function in the completely regular sampling design, and a proof is given in the supplementary material.

\begin{prop}
  \label{prop:loglikreg}
  Under the completely regular sampling design, the distribution of the observed data is $\m y(\m t) := (\m y_1(\m t)^T, \ldots, \m y_n(\m t)^T)^T \mid \m \Theta \sim N(\m 0_{nJ}, \m \Sigma_{\m \Theta})$ with $\m \Sigma_{\m \Theta}$ given in \cref{eq:covarianceKronecker}. Therefore the log-likelihood is 
  \begin{equation*}
    \log p(\m y(\m t) \mid \m \Theta) = -0.5\left(nJ\log(2\pi) + \log |\m \Sigma_{\m \Theta}| +
      \m y(\m t)^T \m \Sigma_{\m \Theta}^{-1} \m y(\m t)\right)
  \end{equation*}
  where the log-determinant simplifies as
  \begin{equation*}
    \log|\m \Sigma_{\m \Theta}| = (n-1)\log|\m \Sigma_0| + \log|\m \Sigma_1|
  \end{equation*}
  and the matrix-vector product simplifies as
  \begin{equation*}
    \label{eq:Siyreg}
    \m \Sigma_{\m \Theta}^{-1}\m y(\m t) = \vc\left(\m \Sigma_0^{-1}\vc^{-1}_{J,n}(\m y(\m t)) +
      \frac{1}{n}\left(\m \Sigma_1^{-1} -
        \m \Sigma_0^{-1}\right)\vc^{-1}_{J,n}(\m y(\m t))\m 1_{n,n} \right),
  \end{equation*}
  with
  \begin{align*}
    \m \Sigma_0 := \frac{n}{n-1}\m K_{\m \theta_\eta}^\eta(\m t, \m t) + \sigma^2 \m I_J, \quad \m \Sigma_1 := n\m K^\mu_{\m \theta_\mu}(\m t, \m t) + \sigma^2\m  I_J.
  \end{align*}
\end{prop}

The computational gain in efficiency from applying the results of \cref{prop:loglikreg} can be understood by comparing the most computationally expensive operations: i) calculating the log determinant of the covariance matrix of the observed data, $\log |\m \Sigma_{\m \Theta}| $, and ii) calculating the product of its inverse and the vector of the observed data, $\m \Sigma_{\m \Theta}^{-1} \m y(\m t)$. Using the naïve structure of the covariance matrix in \cref{eq:covRegularBlockForm}, each calculation would be of order $O(n^3 J^3)$. Applying the results from \cref{prop:loglikreg}, the computational complexity is of order $O(J^3)$ for each of the matrices $\m \Sigma_0$ and $\m \Sigma_1$ as they are both of size $J \times J$. Therefore, the most computationally expensive parts of calculating the likelihood can be made asymptotically independent of the number of functions under the completely regular design.

A further increase in both computational efficiency and numerical stability in the calculation of the likelihood function can be obtained by noting that both the log-determinant of a matrix and the product of its inverse with a vector can be calculated using a single dominating operation of order $O(J^3)$ through a Cholesky factorization. Letting $\m M$ be a positive definite symmetric $n \times n$ matrix with Cholesky factorization $\m M = \m L \m L ^T$ where $\m L$ is a unique lower triangular matrix with positive diagonal elements, then the left-division, $\m L \backslash \m b$, and right division, $\m b^T / \m L$, can
be calculated in quadratic time using forward (respectively back) substitution  \citep[Section 3.1.1 and
3.1.2]{golub2013matrix}. The left division $\m M \backslash \m b$ can be expressed as $\m M \backslash \m b = (\m L\m L^T)^{-1} \m b
    = \m L^{-T}\m L ^{-1} \m b
    = [(\m L^{-1} \m b)^T\m L^{-1}]^T
    = [(\m L \backslash \m b)^T / \m L]^T$
in order to only use these two procedures and transposition. Similarly, the log-determinant can be calculated from the Cholesky decomposition as
  $\log |\m M|
  =
  \log
  \left(
    |\m L||\m L^T|
  \right)
  =
  \log
  \left(
    |\m L|^2
  \right)
  =
  \log
  \left[
    \left(
      \prod_{i = 1}^n L_{ii}
    \right)^2
  \right]
  =
  2\sum_{i = 1}^n \log L_{ii}$.

Having obtained a computationally efficient version of the log-likelihood we show how a similar approach can be used to formulate a fast implementation of the posterior distributions of the latent Gaussian processes. Under the completely regular sampling design, the posterior distributions of the mean function, $\mu$, the subject-specific deviations, $\m \eta$, and their combination into the subject-specific functions, $\m f$, are presented in \cref{prop:postmureg}. The proof is given in the supplementary material.

\begin{prop}
  \label{prop:postmureg}
   Let $\tilde{\m t} := (\tilde{t}_1, \ldots, \tilde{t}_{J_p})^T \in \R^{J_p}$ be $J_p$ evaluation points for the posterior distributions. The posterior of the mean function, $\m \mu(\tilde{\m t}) := (\mu(\tilde{t}_1), \ldots, \mu(\tilde{t}_{J_p}))^T$, is then
    \begin{equation*}
    \m \mu(\tilde{\m t}) \mid \m y(\m t), \m \Theta \sim N\left(\m \mu_{\m \mu(\tilde{\m t}) \mid \m y(\m t), \m \Theta}, \m \Sigma_{\m \mu(\tilde{\m t}) \mid \m y(\m t), \m \Theta}\right)
  \end{equation*}
  where
  \begin{align*}
  \m \mu_{\m \mu(\tilde{\m t}) \mid \m y(\m t), \m \Theta} &=
    \m K^\mu_{\m \theta_\mu}(\tilde{\m t}, \m t) \m \Sigma_1^{-1}
    \vc^{-1}_{J,n}(\m y(\m t))\m 1_n\\
    \m \Sigma_{\m \mu(\tilde{\m t}) \mid \m y(\m t), \m \Theta} &= \m K^\mu_{\m \theta_\mu}(\tilde{\m t}, \tilde{\m t}) - n\m K^\mu_{\m \theta_\mu}(\tilde{\m t},
    \m t)\m \Sigma_1^{-1}\m K^\mu_{\m \theta_\mu}(\m t, \tilde{\m t}).
  \end{align*}
  and $\m \Sigma_0$ and $\m \Sigma_1$ are defined as in \cref{prop:loglikreg}.
  The joint posterior of the subject-specific deviations, $\m \eta(\tilde{\m t}) := (\m \eta_1(\tilde{\m t})^T, \ldots, \m
  \eta_n(\tilde{\m t})^T)^T$, is given by
  \begin{equation*}
    \m \eta(\tilde{\m t}) \mid \m y(\m t), \m \Theta
    \sim N\left(\m \mu_{\m \eta(\tilde{\m t}) \mid \m y(\m t), \m \Theta}, \m \Sigma_{\m \eta(\tilde{\m t}) \mid \m y(\m t), \m \Theta}\right)
  \end{equation*}  
  where
  \begin{align*}
    \m \mu_{\m \eta(\tilde{\m t}) \mid \m y(\m t), \m \Theta} &=
    \frac{1}{n-1}\vc
    \left(
      n\m K^\eta_{\m \theta_\eta}(\tilde{\m t},\m t)\m \Sigma_0^{-1}\vc^{-1}_{J,n}(\m y(\m t))
      -
      \m K^\eta_{\m \theta_\eta}(\tilde{\m t},\m t)\m \Sigma_0^{-1}\vc^{-1}_{J,n}(\m y(\m t))\m 1_{n,n}
    \right)\\
      \m \Sigma_{\m \eta(\tilde{\m t}) \mid \m y(\m t), \m \Theta} &=
                   \frac{n}{n-1} \m I_n \otimes
                   \left(
                   \m K^\eta_{\m \theta_\eta}(\tilde{\m t}, \tilde{\m t}) - \frac{n}{n-1}
                   \m K^\eta_{\m \theta_\eta}(\tilde{\m t}, \m t) \m \Sigma_0^{-1}
                   \m K^\eta_{\m \theta_\eta}(\m t, \tilde{\m t})
                   \right)\\
                 &\quad +
                   \frac{1}{n-1}
                   \m 1_{n,n} \otimes
                   \left(
                   \frac{n}{n-1}
                   \m K^\eta_{\m \theta_\eta}(\tilde{\m t}, \m t)
                   \m \Sigma_0^{-1}
                   \m K^\eta_{\m \theta_\eta}(\m t, \tilde{\m t})
                   -
                   \m K^\eta_{\m \theta_\eta}(\tilde{\m t}, \tilde{\m t})
                   \right).    
  \end{align*}
  Because of posterior independence between $\m \mu(\tilde{\m t})$ and $\bm \eta(\tilde{\m t})$, samples from the posterior distribution of $\m f(\tilde{\m t})$ can be generated by combining draws from the individual posterior distributions according to the definition $\m f_i(\tilde{\m t}) := \m\mu(\tilde{\m t}) + \m\eta_i(\tilde{\m t})$.
\end{prop}

It can be seen from \cref{prop:postmureg} that the computational gain when calculating the moments of the posterior distributions arises from only having to work with the inverse of $\m \Sigma_0$ and $\m \Sigma_1$ instead of the full marginal covariance matrix similar to the simplifications when calculating the log-likelihood. There is, however, still a bottleneck when sampling from $\m \eta(\tilde{\m t}) \mid \m y(\m t), \m \Theta$ as it is of size $n J_p \times n J_p$ and a naïve approach would be of order $O(n^3 J_p^3)$. We consider the problem of simulating the first $1,\ldots, n-1$ functions of  $\m \eta(\tilde{\m t})$ which we denote by $\m \eta'(\tilde{\m t})$. This is sufficient due to the zero-sum constraint of $\m \eta(t_0)$ at every $t_0$. A simulation of $\m\eta_n(\tilde{\m t})$ is then given by $\m\eta_n(\tilde{\m t}) = -\sum_{i=1}^{n-1} \m\eta_i(\tilde{\m t})$.  Let $\m \mu_{\m \eta'(\tilde{\m t}) \mid \m y(\m t), \m \Theta}$ and $\m \Sigma_{\m \eta'(\tilde{\m t}) \mid \m y(\m t), \m \Theta}$ be the posterior mean and covariance function of $\m \eta'(\tilde{\m t})$ given by $\m \mu_{\m \eta(\tilde{\m t}) \mid \m y(\m t), \m \Theta}$ and $\m \Sigma_{\m \eta(\tilde{\m t}) \mid \m y(\m t), \m \Theta}$ in \cref{prop:postmureg} but with the necessary indices removed. To draw $\m \eta'(\tilde{\m t})$ from its posterior distribution we will calculate the Cholesky factor $\m L_{\m \eta'(\tilde{\m t}) \mid \m y(\m t), \m \Theta}$ of $\m \Sigma_{\m \eta'(\tilde{\m t}) \mid \m y(\m t), \m \Theta}$, which enables us to transform a standard Gaussian draw $\m z \sim N(\m 0, \m I_{(n-1)J_p})$ to a draw $\m \eta'(\tilde{\m t}) =  \m \mu_{\m \eta'(\tilde{\m t}) \mid \m y(\m t), \m \Theta} + \m L_{\m \eta'(\tilde{\m t}) \mid \m y(\m t), \m \Theta} \m z$. The computation of the Cholesky factorization can be performed efficiently using an iterative block Cholesky factorization algorithm similar to \citet[Section 4.2.9]{golub2013matrix}, which consists of iteratively applying the observation that for any matrices $\m A, \m B, \m C$ (leading to positive definiteness), we have
\begin{equation}
  \label{eq:blockchol}
  \ch
  \begin{pmatrix}
    \m A & \m C^T\\
    \m C & \m B
  \end{pmatrix}
  =
  \begin{pmatrix}
    \ch \m A & \m 0\\
    \m C (\ch \m A)^{-T} & \ch \m S
  \end{pmatrix},
\end{equation}
where $\m S = \m B - \m C \m A ^{-1} \m C ^T$ is the Schur complement of $\m
A$. From the Kronecker product expression for $\m \Sigma_{\m \eta'(\tilde{\m t}) \mid \m y(\m t), \m \Theta}$ in \cref{prop:postmureg} it is seen that all of its diagonal blocks and all of its off-diagonal blocks are equal, so when iteratively applying \cref{eq:blockchol} we can greatly limit the amount of calculations in the matrix division $\m C (\ch \m A)^{-T}$ by updating results from earlier steps, which is the reason that the iterative block Cholesky factorization algorithm will be faster than the standard Cholesky factorization algorithm in our setting---see the supplementary material for details.

\subsection{Partially Regular Sampling Design}

Under the completely regular sampling design, we assumed that all $n$ curves were observed at the same $J$ time points. We will now deal with the partially regular case, where we relax this assumption by saying that we have $n_a$ responses $y^a_1, \ldots, y^a_{n_a}$
that are, as in the regular setting, measured at the same $J_a$ time points
$\m t^a = (t^a_1, \ldots, t^a_{J_a}) \in \R^{J_a}$, but where we now further
allow $n_b$ responses $y^b_1, \ldots, y^b_{n_b}$ where $y^b_i$ is measured at
individual $J^b_i$ time points
$\m t^b_i = (t^b_{i, 1}, \ldots, t^b_{i, J^b_i}) \in \R^{J^b_i}$. We write
$\m y^a(\m t^a) = (\m y^a_1(\m t^a)^T, \ldots, \m y^a_{n_a}(\m t^a)^T)^T$, $\m t^b = ((\m t^b_1)^T, \ldots, (\m t^b_{n_b})^T)^T \in \R^{\sum_{i = 1}^n J_i}$,
$\m y^b(\m t^b) = (\m y^b_1(\m t^b_1)^T, \ldots, \m y^b_{n_b}(\m
t^b_{n_b})^T)^T$, and $\m y(\m t) = (\m y^a(\m t^a)^T, \m y^b(\m t^b)^T)^T$. We
let $N = n_aJ_a + \sum_{i = 1}^{n_b} J^b_i$ be the total number of observations
and $n = n_a + n_b$ the total number of functions. Note that the special case with $n_a = 1$ corresponds to a completely irregular sampling design (the baseline case with no assumptions of a regular sampling grid for any of the functions).

By partitioning the observed data into completely regular and partially regular sampled functions, the distribution of the observed data, $\m y(\m t)$, can be written in a block form as
\begin{equation}
\label{eq:irreg-sigma-blocks}
\begin{pmatrix}
      \m y^a(\m t^a) \\
      \m y^b(\m t^b)
    \end{pmatrix}
    \mid \m \Theta
    \sim
    N
    \left(
      \m 0_{N},
      \m \Sigma_{\m \Theta}
    \right), \quad \m \Sigma_{\m \Theta} := \begin{bmatrix}
      \m A & \m C^T\\
      \m C & \m B
    \end{bmatrix}
\end{equation} 
where the blocks are given by
\begin{align}\label{eq:irregblocks}
\begin{aligned}
    \m A &:=
           \m I_{n_a} \otimes \left(\frac{n}{n-1}\m K^\eta_{\m \theta_\eta}(\m t^a, \m t^a) + \sigma^2\m I_{J_a}\right) +
           \m 1_{n_a,n_a} \otimes \left(\m K^\mu_{\m \theta_\mu}(\m t^a, \m t^a) -
           \frac{1}{n-1}\m K^\eta_{\m \theta_\eta}(\m t^a, \m t^a)\right)\\
    \m B_{i,j} &:=   \begin{cases}
    \m K^\mu_{\m \theta_\mu}(\m t^b_i, \m t^b_i) + \m K^\eta_{\m \theta_\eta}(\m t^b_i, \m t^b_i) + \sigma^2 \m I_{J^b_i}, & i = j \\
    \m K^\mu_{\m \theta_\mu}(\m t^b_i, \m t^b_j) - \frac{1}{n-1} \m K^\eta_{\m \theta_\eta}(\m t^b_i, \m t^b_j), & i \ne j
  \end{cases}\\
  \m C^T &:= \m 1_{n_a} \otimes \left(\m K^\mu_{\m \theta_\mu}(\m t^a, \m t^b) - \frac{1}{n-1}\m K^\eta_{\m \theta_\eta}(\m t^a, \m t^b)\right)  
\end{aligned}
\end{align}
according to the model definition in \cref{mod:eta}. Due to the completely regular sampling grid for $\m y^a$, the computations involving $\m A$ and $\m C$ can be simplified similar to the completely regular sampling design above, but since $\m B$ only has to do with $\m y^b$, upon which no regularity assumptions are imposed, it cannot be simplified. The computational gains from the simplifications will therefore depend on the proportion of functions that are sampled completely regularly.

Using this block representation and simplifications for $\m A$ and $\m C$ similar to in the completely regular sampling design, \cref{prop:irreg-loglik} states the simplified form of the log-likelihood in the partially regular sampling design. The proof is given in the supplementary material.

\begin{prop}\label{prop:irreg-loglik}
Under the partially regular sampling design, the distribution of the observed data is $\m y(\m t) \mid \m \Theta \sim N(\m 0_N, \m \Sigma_{\m \Theta})$ and therefore the log-likelihood is
  \begin{equation*}
    \log p(\m y(\m t) \mid \m \Theta) = -0.5
    \left(
      N\log(2\pi) +
      \log |\m \Sigma_{\m \Theta}| + \m y(\m t)^T \m \Sigma_{\m \Theta}^{-1}\m y(\m t)
    \right)
  \end{equation*}
where the log-determinant simplifies as
\begin{equation*}
    \log|\m \Sigma_{\m \Theta}| = (n_a - 1)\log|\m A_0| + \log|\m A_1| + \log\left|\m S\right|,
\end{equation*}
with
\begin{alignat*}{2}
    & \m A_0 := \frac{n}{n-1}\m K^\eta_{\m \theta_\eta}(\m t^a, \m t^a) + \sigma^2\m I_{J_a},  \quad && \m A_1 := \sigma^2\m I_{J_a} + n_a \m K^\mu_{\m \theta_\mu}(\m t^a, \m t^a) + \frac{n_b}{n - 1}\m K^\eta_{\m \theta_\eta}(\m t^a, \m t^a)\\
    & \m S := \m B - n_a(\m C^b)^T \m A_1^{-1} \m C^b, && \m C^b := \m K^\mu_{\m \theta_\mu}(\m t^a, \m t^b) - \frac{1}{n-1}\m K^\eta_{\m \theta_\eta}(\m t^a, \m t^b)
\end{alignat*}
and $\m B$ is a block matrix with elements $\m B_{i,j}$ as defined in \cref{eq:irregblocks}.
Using the notation
\begin{equation*}
    \m P_1 := \begin{pmatrix}
      \m A^{-1} + \m A^{-1}\m C^T\m S^{-1}\m C\m A^{-1} & - \m A^{-1}\m C^T\m S^{-1}      
    \end{pmatrix},\quad
    \m P_2 := \begin{pmatrix}
    -\m S^{-1}\m C\m A^{-1} & \m S^{-1}
    \end{pmatrix},
\end{equation*}
both of which are $1\times 2$ block matrices, the matrix-vector product can be written as
  \begin{equation*}
    \m \Sigma_{\m \Theta}^{-1}\m y(\m t) =
    \begin{pmatrix}
        \m P_1\\
        \m P_2
    \end{pmatrix}
    \begin{pmatrix}
        \m y^a(\m t^a)\\
        \m y^b(\m t^b)
    \end{pmatrix}
    =
    \begin{pmatrix}
      \m A^{-1}\m y^a(\m t^a) + \m A^{-1}\m C^T\m S^{-1}\m C\m A^{-1}\m y^a(\m t^a) -
      \m A^{-1}\m C^T\m S^{-1}\m y^b(\m t^b)\\
      -\m S^{-1}\m C\m A^{-1}\m y^a(\m t^a) + \m S^{-1}\m y^b(\m t^b)
    \end{pmatrix}
  \end{equation*}
with the following simplifications
  \begin{align*}
    \m A^{-1}\m y^a(\m t^a) &= \vc
                              \left(
                              \m A_0^{-1}\vc^{-1}_{J_a, n_a}(\m y^a(\m t^a))
                              +
                              \frac{1}{n_a}(\m A_1^{-1} - \m A_0^{-1})\vc^{-1}_{J_a,
                              n_a}(\m y^a(\m t^a)) \m 1_{n_a,n_a}
                              \right)\\
    \m C^T &= \m 1_{n_a} \otimes \m C^b, \quad \m A^{-1} \m C^T = \m 1_{n_a} \otimes \m A_1^{-1} \m C^b, \quad
    \m C \m A^{-1} \m C^T = n_a (\m C^b)^T \m A_1^{-1} \m C^b.
  \end{align*}
\end{prop}

Similar to the completely regular case, this leads to a computational gain since we no longer need to work with the inverse and log-determinants of the $n_aJ_a + \sum_{i=1}^{n_b} J_i^b$ dimensional matrix $\m \Sigma_{\m \Theta}$, but only of the two $J_a$ dimensional matrices $\m A_0$ and $\m A_1$ and the $\sum_{i=1}^{n_b} J_i^b$ dimensional matrix $\m S$. Computing the Cholesky factorization of $\m \Sigma_{\m \Theta}$ is of order $O((n_aJ_a + \sum_{i=1}^{n_b} J_i^b)^3)$, while computing it for $\m A_0$, $\m A_1$ and $\m S$ is of order $O(J_a^3 + (\sum_{i=1}^{n_b} J_i^b)^3)$ all together, thus asymptotically removing the dependence of the runtime on the number of completely regularly sampled functions $n_a$.

Contrary to the posterior expressions in the case of the completely regular sampling design given in \cref{prop:postmureg}, $\m \mu(\tilde{\m t})$ and $\m \eta(\tilde{\m t})$ are no longer independent in the posterior distribution. Therefore, their cross-covariance needs to be accounted for when drawing samples from $\m f(\tilde{\m t})$. \cref{prop:postirreg} and its proof in the supplementary material give the joint posterior distribution of $\m \mu(\tilde{\m t})$ and $\m \eta(\tilde{\m t})$.

\begin{prop}\label{prop:postirreg}
Let $\tilde{\m t} := (\tilde{t}_1, \ldots, \tilde{t}_{J_p})^T \in \R^{J_p}$ be the $J_p$ time points of interest. Then the joint posterior of the mean function, $\m \mu(\tilde{\m t}) := (\mu(\tilde{t}_1), \ldots, \mu(\tilde{t}_{J_p}))^T$, and the subject-specific deviations, $\m \eta(\tilde{\m t}) := (\m \eta_1(\tilde{\m t})^T, \ldots, \m
  \eta_n(\tilde{\m t})^T)^T$, is
\begin{equation*}
  \begin{pmatrix}
    \m \eta(\tilde{\m t})\\
    \m \mu(\tilde{\m t})
  \end{pmatrix}
  \mid \m y(\m t), \m \Theta
  \sim
  N\left(
  \begin{pmatrix}
    \m \mu_{\m \eta(\tilde{\m t}) \mid \m y(\m t), \m \Theta}\\
    \m \mu_{\m \mu(\tilde{\m t}) \mid \m y(\m t), \m \Theta}
  \end{pmatrix},
  \begin{pmatrix}
      \m \Sigma_{\m \eta(\tilde{\m t}) \mid \m y(\m t), \m \Theta} & \m \Sigma_{\m \eta(\tilde{\m t}), \m \mu(\tilde{\m t}) \mid \m y(\m t), \m \Theta}\\
      \m \Sigma^T_{\m \eta(\tilde{\m t}), \m \mu(\tilde{\m t}) \mid \m y(\m t), \m \Theta} & \m \Sigma_{\m \mu(\tilde{\m t}) \mid \m y(\m t), \m \Theta},
  \end{pmatrix}
  \right)
\end{equation*}
where the posterior moments for $\m \mu(\tilde{\m t})$ are given by
\begin{align*}
    \m \mu_{\m \mu(\tilde{\m t}) \mid \m y(\m t), \m \Theta} &= \m K^\mu_{\m \theta_\mu}(\tilde{\m t}, \m t^a) \vc^{-1}_{J_a, n_a} (
    \m P_1 \m y(\m t)) \m 1_{n_a} + \m K^\mu_{\m \theta_\mu}(\tilde{\m t}, \m t^b)\m
               P_2 \m y(\m t)
    \\
    \m \Sigma_{\m \mu(\tilde{\m t}) \mid \m y(\m t), \m \Theta} &=
                    \m K^\mu_{\m \theta_\mu}(\tilde{\m t}, \tilde{\m t}) - \m C^{\mu, a} \m A^{-1} (\m C^{\mu, a})^T - \m C^{\mu, a}\m A^{-1} \m C^T \m S^{-1} \m C \m A^{-1}(\m
               C^{\mu, a})^T\\
             &\quad+ \m C^{\mu, a}\m A^{-1}\m C^T\m S^{-1}(\m C^{\mu, b})^T + \m C^{\mu, b} \m S^{-1}\m C\m A^{-1}(\m C^{\mu, a})^T - \m C^{\mu, b}\m S^{-1} (\m C^{\mu, b})^T
  \end{align*}
  with the simplifications
  \begin{align*}
    \m C^{\mu, a} \m A^{-1} (\m C^{\mu, a})^T &= n_a \m K^\mu_{\m \theta_\mu}(\tilde{\m t}, \m t^a) \m A_1^{-1} \m
                                            K^\mu_{\m \theta_\mu}(\m t^a, \tilde{\m t})\\
    \m C \m A^{-1}(\m C^{\mu, a})^T &= n_a (\m C^b)^T \m A_1^{-1} \m K^\mu_{\m \theta_\mu}(\m t^a,
                               \tilde{\m t})
  \end{align*}
and where the cross-covariance is given by
\begin{equation*}
    \m \Sigma^T_{\m \eta(\tilde{\m t}), \m \mu(\tilde{\m t}) \mid \m y(\m t), \m \Theta} = (- \m C^{\mu, a}\m P_1 - \m C^{\mu, b}\m P_2)\m C^{y, \eta}
\end{equation*}
with
\begin{equation*}
      \m C^{\mu, a}\m P_1 =
    \begin{pmatrix}
        \m H_1 + \m H_2 & \m H_3
      \end{pmatrix},\quad
    \m C^{\mu, b} \m P_2 =
    \begin{pmatrix}
      - \m C^{\mu, b} \m S^{-1} \m C \m A^{-1} & \m C^{\mu, b} \m S^{-1}
    \end{pmatrix}
\end{equation*}
where
\begin{equation*}
    \m H_1
    =
      \m 1_{n_a}^T \otimes \m K^\mu_{\m \theta_\mu}(\tilde{\m t}, \m t^a)\m A_1 ^{-1},\quad
          \m H_2
    =
      - \m H_3 \m C \m A^{-1},\quad
    \m H_3
    =
      - n_a \m K^\mu_{\m \theta_\mu}(\tilde{\m t}, \m t^a) \m A^{-1}_1 \m C^b \m S^{-1}.
\end{equation*}
Similar, but longer, simplifications have been derived for $\m \mu_{\m \eta(\tilde{\m t}) \mid \m y(\m t), \m \Theta}$ and $\m \Sigma_{\m \eta(\tilde{\m t}) \mid \m y(\m t), \m \Theta}$ and can be found in the supplementary material.

\end{prop}

The structure derived in \cref{prop:postirreg} allows efficient computation of the mean and covariance of the joint posterior distribution of the mean function, $\m \mu(\tilde{\m t})$, and the subject-specific deviations, $\m \eta(\tilde{\m t})$. However, similar to the completely regular sampling design, we need the Cholesky factorization of the posterior covariance matrix in order to sample from the posterior distribution. This can again be computed efficiently by using an iterative block Cholesky factorization algorithm based on \cref{eq:blockchol}, similar to that described in the completely regular sampling design above but with a few extra steps---see the supplementary material for details.

\section{Simulation Results}\label{sec:benchmarks}

We now show benchmark results of our implementations on simulated data and compare them against a naïve baseline implementation not using any algebraic optimizations. We will measure execution time both for a single evaluation of the log-likelihood function, for a single posterior simulation of
$\m f(\tilde{\m t})$, and for a full fit where we run Hamiltonian Monte Carlo (HMC) for 1000 warm-up iterations and 1000 sampling iterations using Stan \citep{stan}.  For the log-likelihood calculations we compare a baseline implementation (baseline) that does not make use of any optimizations against our implementation (efficient) that uses the optimizations derived in this paper. For the posterior simulations we compare the baseline to an implementation of the optimizations derived in this paper either with (efficient) or without (intermediary efficient) the iterative block Cholesky decomposition. For the full HMC we compare different combinations of log-likelihood and posterior implementations. We consider either a baseline log-likelihood implementation with no posterior draws of $\m f(\tilde{\m t})$ (baseline + none), the efficient implementation of the log-likelihood and the intermediary efficient implementation of the posterior draws (efficient + intermediary efficient), and both the efficient log-likelihood and posterior simulation (efficient + efficient). In the baseline + none case, posterior simulation of $\m f(\tilde{\m t})$ is omitted due to infeasible run times. For all quantities we vary either the number of functions ($n$), the number of observations per function ($J$), or the number of prediction points for the posterior ($J_p$) in the completely regular sampling design. In the partially regular sampling design, we also vary the number of irregularly sampled functions ($n_b$). In the case of the partially regular sampling design we do not consider and compare to the intermediary efficient implementation as it will already be seen to be superior in the completely regular sampling setting. The baseline + none configuration is also not compared to in the full HMC setting, as the irregularity does not have an effect on its performance. The simulations were performed on a server with an AMD EPYC\textsuperscript{TM} 7702P 64-core CPU and 512 GiB of RAM, and each result is the average of the run times across a number of independent replicates.

\subsection{Completely Regular Sampling Design}
\begin{figure}[ht]
  \centering
  \includegraphics[scale=0.65]{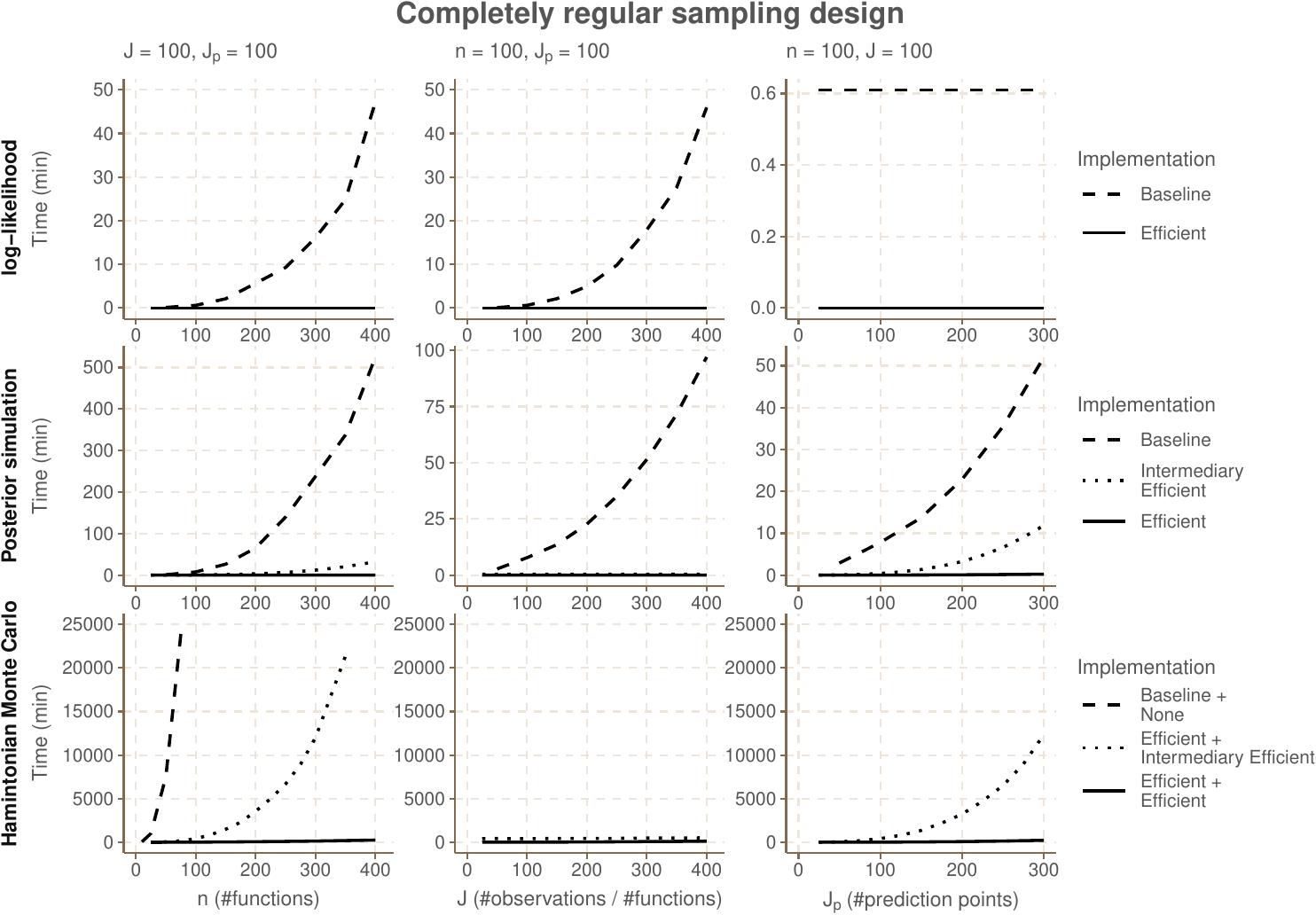}
  \vspace{-0.5cm}
  \caption{Benchmarks of different implementations of the log-likelihood, posterior draws, and full HMC in the completely regular sampling design for varying number of functions, observations per function, and prediction points. Note that the $y$-axes are not the same.}
  \label{fig:bench-reg}
\end{figure}

The results for the completely regular sampling design are seen in \cref{fig:bench-reg}. The rows correspond to the different quantities compared, and the columns are for different varying parameters. When varying the number of functions, we fix the observations per function and the number of prediction points at 100. When varying the number of observations per function we fix the number of functions and number of prediction points at 100. When varying the number of prediction points, we fix the number of functions and number of observations per function at 100.

From the figure we see that our optimized log-likelihood implementation is about $1{,}000$-$100{,}000$ times faster than the baseline implementation and that this performance gain becomes larger for increasing number of functions and observations per function. Our optimized posterior draw implementation is about $100$-$1{,}000$ times faster than the baseline implementation and the performance gain also becomes larger for increasing number of functions and observations per function but decreases as the number of prediction points increases. For the baseline HMC we only ran a few settings due to it being very slow. For instance, running it with $n = 75$, $J=100$, and $J_p = 100$ takes 350 hours, which is 3500 times slower than our optimized implementation that takes 6 minutes at the same setting. Therefore it is only included in the lower left panel of the figure. Comparing the efficient + intermediary efficient against the efficient + efficient implementation for the full HMC clearly shows the added performance increase from using the iterative block Cholesky decomposition.

\subsection{Partially Regular Sampling Design}
\begin{figure}[ht]
  \centering
  \includegraphics[width=\textwidth]{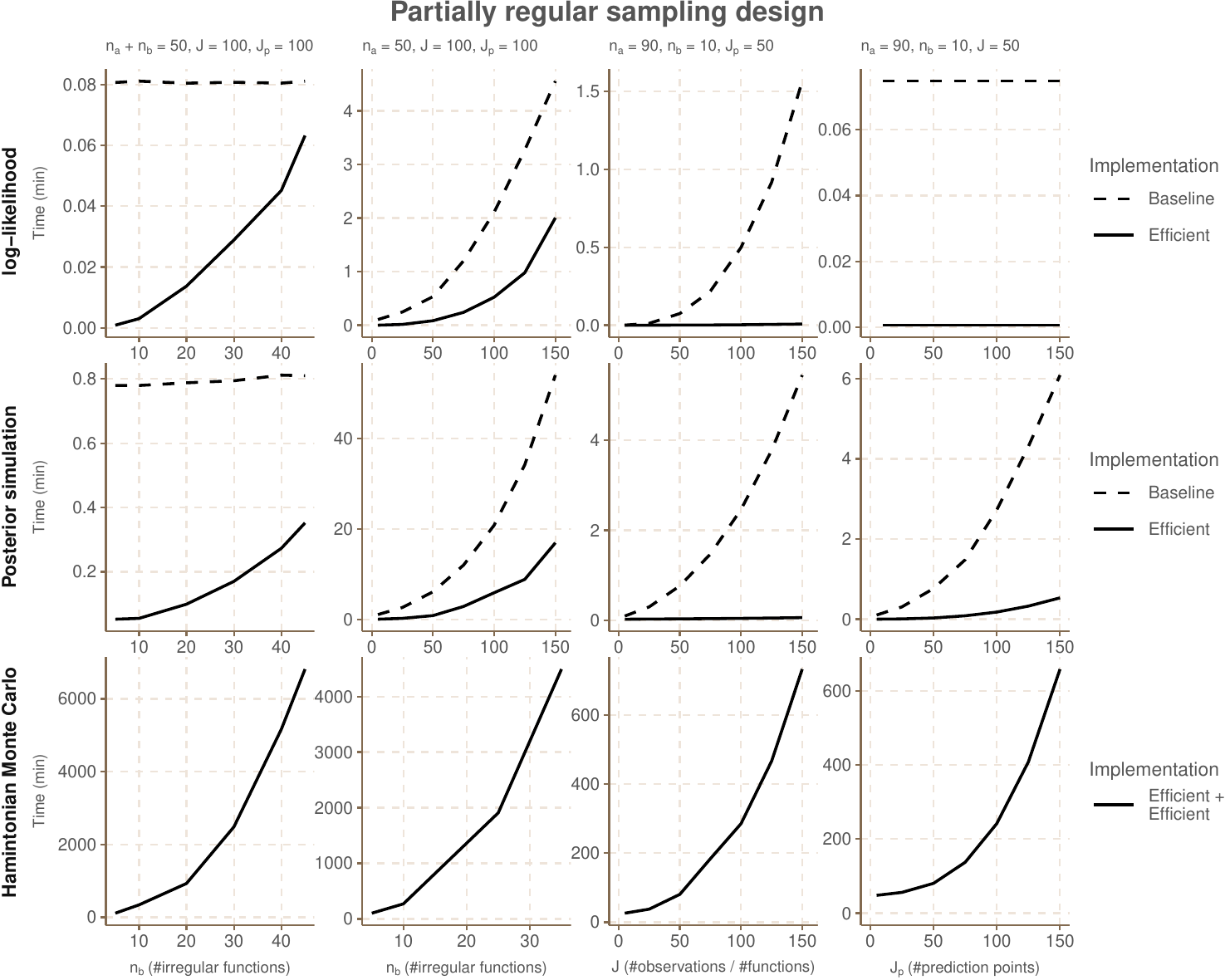}
  \vspace{-0.5cm}
  \caption{
  Benchmarks of our different implementations of the log-likelihood, posterior draws, and full HMC in the partially regular sampling design for varying number of irregularly sampled functions, number of observations per function, and number of prediction points. Note that the $y$-axes are not the same.}
  \label{fig:bench-irreg}
\end{figure}

\cref{fig:bench-irreg} shows the benchmark results in the partially regular sampling design. As before the rows show the different quantities compared, and here the columns are for varying number of irregularly observed functions, number of observations per function, and the number of prediction points. In the first column we vary the number of irregularly observed functions ($n_b$) while fixing the total number of functions ($n_a + n_b$) at 50 and fixing both the number of observations and prediction points at 100. In the second column we also vary the number of irregularly observed functions but keep the number of regularly sampled functions ($n_a$) at 50 and both the number of observations per function and prediction points equal to 100. In the third column we vary the number of observations per function while fixing the number of regularly sampled functions at 90, the number of irregularly sampled functions at 10, and fix the number of prediction points at 50. In the fourth column we vary the number of prediction points with the same number of regularly and irregularly sampled functions as in the previous column and fix the number of observations per function to 50.
In all cases we use the same number of observations per function for all regularly and irregularly sampled functions, i.e., $J_a = J^b_1 = \cdots = J^b_{n_b} = J$, but our implementation is not restricted to that.

From the results it is clear that the improvements in execution time depend heavily on the proportion of irregularly sampled functions, and the extreme case with only irregularly observed functions is equivalent to the general baseline case. With 90 regularly and 10 irregularly observed functions, the optimized log-likelihood implementation is about 100 times faster than the baseline, and the performance gain increases with the number of observations per function. The implementation of the optimized posterior is also about 10-100 faster when keeping the number of observations per function fixed and varying the number of prediction points, but this performance gain decreases with increasing number of prediction points.

\section{Discussion}\label{sec:discussion}

In this paper we have derived and implemented simplified expressions for the log-likelihood and the conditional posterior distribution for a hierarchical Gaussian process regression model using a Kronecker product structure that arises in the covariance matrix under a completely or partially regular sampling design. These simplifications greatly reduce the amount of computations required. Benchmarks of our implementation compared to a baseline implementation show that computation time is reduced by several orders of magnitude which permits applying the model to data of sizes often encountered in real-world examples.

The simplifications given in this paper are, however, only applicable when some of the functions have been measured on the same sampling grid, while the baseline implementation does not require this assumption. Furthermore, our approach relies on the conditional conjugacy for the measurement error terms which allows us to drastically speed up the computations by finding the conditional posterior of the latent functions analytically. This conjugacy requires a Gaussian likelihood.  Even without a Gaussian likelihood, however, we conjecture that a partially regular sampling design will give rise to a similar Kronecker product structure that can be exploited for an efficient implementation of the hierarchical Gaussian process likelihood. A direct extension of our approach is to consider student-$t$ processes for which closed form expressions for the marginal
likelihood and predictive distribution also exist \citep{shah2014student}. The student-$t$ process is the most general elliptical process for which a density exists in analytical form.

Another extension of interest would be to add an additional layer to the hierarchy. This would for instance allow modeling within-subject replications, where an experiment is repeated multiple times for each subject. Such data arises in the context of wearable devices where e.g., daily activity measurements is collected at a high sampling frequency across multiple days for several subjects. Each subject would thus have one latent function per repetition, all of which would lie around a common subject-specific mean function. All of the subject-specific mean functions would in turn lie around a population-wide common mean function. Under the assumption of a (partially) regular sampling design, a similar but more general Kronecker product structure would arise leading to similar simplifications as presented in this paper.

\section*{Disclosure Statement}

The authors report there are no competing interests to declare.

\bibliography{refs.bib}

\appendix

\section*{Appendix}

\renewcommand{\theequation}{S\arabic{equation}}

\subsection*{Proof of \cref{prop:xi}}
  Consider a specific, single time point $t_0 \in \R$. We will prove that
  $P(\sum_{i = 1}^n \eta_i(t_0) = 0) = 1$. From the model specification in \cref{mod:eta}, we have that $\m \eta(t_0) = (\eta_1(t_0), \ldots, \eta_n(t_0))^T \mid \m\theta_\eta \sim N(0, K^\eta_{\m\theta_\eta}(t_0, t_0) \m
  \Xi)$
  since $K^\eta_{\m\theta_\eta}(t_0, t_0)$ is just a single real number.
  We now find a matrix square root of $\m \Xi$. To this end, define the matrix
  $\m \Omega = (\omega_{ij}) \in \R^{n\times n}$ by
  $\omega_{ii} = \sqrt{\frac{n-1}{n}}$ and
  $\omega_{ij} = -\frac{1}{\sqrt{n(n-1)}}$ for $i\ne j$. The matrix $\m \Omega$
  is symmetric and we see that
  \begin{equation*}
    \begin{split}
      \m \Omega^2_{ii} = \m \Omega_{i\cdot }\m \Omega_{\cdot i} = \sum_{j = 1}^n
      \omega_{ij} \omega_{ji} &= \omega_{ii}^2 + \sum_{\substack{1\le j\le n\\j \ne i}}
      \omega_{ij}\omega_{ji}\\ &= \frac{n-1}{n} + (n-1)\frac{1}{n(n-1)} = 1 = \xi_{ii}
    \end{split}
  \end{equation*}
  and for $i\ne i'$ we get
  \begin{equation*}
    \begin{split}
      \m \Omega^2_{ii'}
      =
        \m \Omega_{i\cdot }\m \Omega_{\cdot i'}
      =
        \sum_{j = 1}^n \omega_{ij} \omega_{ji'}
      &=
        \omega_{ii}\omega_{ii'} + \omega_{ii'}\omega_{i'i'} + \sum_{\substack{1 \le
        j \le n\\j \not \in \{i, i'\}}} \omega_{ij} \omega_{ji'}\\
      &=
        2\left(-\frac{1}{n}\right) + (n-2)\frac{1}{n(n-1)}
      =
        -\frac{1}{n-1}
      =
        \xi_{ii'}
    \end{split}
  \end{equation*}
  so $\m \Omega^2 = \m \Xi$. This also implies that $\m\Xi$ is positive
  semi-definite, since
  $\m v^T \m \Xi \m v = (\m\Omega \m v)^T\m\Omega \m v \ge 0$ for any
  $\m v \in \R^n$.
  We can now use that $\m \eta(t_0) \mid \m\theta_\eta$ has the same degenerate
  Gaussian distribution as $\sqrt{K^\eta_{\m\theta_\eta}(t_0, t_0)}\m \Omega \m z$
  for $\m z = (z_1, \ldots, z_n) \sim N(\m 0_n, \m I_n)$. Since
  \begin{equation*}
    (\m 1_n^T\m \Omega)_j = \sum_{i = 1}^n \omega_{ij} = \omega_{ii} + \sum_{\substack{1\le j\le n\\j \ne i}}
    \omega_{ij} = \sqrt{\frac{n-1}{n}} - (n-1)\frac{1}{\sqrt{n(n-1)}} = 0,
  \end{equation*}
  we have that $\m 1_n^T\m \Omega = \m 0_n^T$ and thus $\m 1_n^T
  \sqrt{K^\eta_{\m\theta_\eta}(t_0, t_0)} \m \Omega \m z =
  \sqrt{K^\eta_{\m\theta_\eta}(t_0, t_0)}(\m
  1_n^T \m \Omega)\m z = 0$, so
  \begin{equation*}
    P\left(\sum_{i = 1}^n\eta_i(t_0) = \m 1_n^T\m \eta(t_0) = 0 \mid \m\theta_\eta\right) = P\left(\m 1_n^T
      \sqrt{K^\eta_{\m \theta_\eta}(t_0, t_0)} \m \Omega \m z = 0\right) = 1.
  \end{equation*}
  By integrating out $\m\theta_\eta$, we now get that
  \begin{equation*}
    P
    \left(
      \sum_{i = 1}^n \eta_i(t_0) = 0
    \right) =
    \int
    P
    \left(
      \sum_{i = 1}^n \eta_i(t_0) = 0 \mid \m\theta_\eta
    \right)p(\m\theta_\eta)
    d\m\theta_\eta =
    \int p(\m\theta_\eta) d\m\theta_\eta = 1,
  \end{equation*}
  as desired, and the proof is completed.

\subsection*{Proof of \cref{prop:loglikreg}}
  Since $\m f(\m t) \mid \m\theta_\mu, \m\theta_\eta$ is zero-mean Gaussian according to the model specification in \cref{mod:eta} we can write $\m y(\m t) = \m f(\m t) + \m \varepsilon$
  with $\m \varepsilon \mid \sigma \sim N(\m 0_{nJ}, \sigma^2 \m I_J)$ and $\m f(\m t) \indep \m \varepsilon \mid \m\Theta$. We directly have that $\m y(\m t) = (\m y_1(\m t)^T, \ldots, \m y_n(\m t)^T)^T \mid \m \Theta \sim N(\m 0_{nJ}, \m \Sigma_{\m \Theta})$ with some covariance matrix $\m \Sigma_{\m \Theta}$. We now derive the expression for $\m \Sigma_{\m \Theta}$. For any $i\in \{1, \ldots, n\}$ we find that
\begin{equation*}
  \label{eq:Areg}
  \begin{split}
    \m G
    &:= \cov(\m y_i(\m t), \m y_i(\m t) \mid \m \Theta)\\
    &=
      \cov(\m \mu(\m t) +
      \m \eta_i(\m t) +
      \m \varepsilon_i,
      \m \mu(\m t) +
      \m \eta_i(\m t) +
      \m \varepsilon_i(\m t) \mid
      \m \Theta)\\
    &= \cov(\m \mu(\m t),
      \m \mu(\m t) \mid \m\Theta) + \cov(\m \eta_i(\m t),
      \m \eta_i(\m t) \mid \m \Theta) +
      \cov(\m \varepsilon_i,
      \m \varepsilon_i
      \mid \m \Theta)\\
    &= \m K^\mu_{\m \theta_\mu}(\m t, \m t) + \m K^\eta_{\m \theta_\eta}(\m t, \m t) + \sigma^2\m I_J
  \end{split}
\end{equation*}
and for any $i,j\in\{1,\ldots,n\}$ with $i\ne j$ we similarly find that
\begin{equation*}
  \label{eq:Breg}
  \begin{split}
    \m H &:= \cov(\m y_i(\m t), \m y_j(\m t)\mid \m \Theta)\\
                               &= \cov(\m \mu(\m t), \m \mu(\m t) \mid \m \Theta) + \cov(\m \eta_i(\m t), \m
                                 \eta_j(\m t)\mid \m\Theta)\\
                               &= \m K^\mu_{\m \theta_\mu}(\m t, \m t) - \frac{1}{n - 1}\m
                                 K^\eta_{\m \theta_\eta}(\m t, \m t)
  \end{split}
\end{equation*}
so therefore  $\m \Sigma_{\m \Theta} := \m I_n \otimes (\m G - \m H) + \m 1_{n,n} \otimes \m H$. From \citet[11.12]{seber2008matrix} it follows that
\begin{equation*}
  \begin{split}
    \m \Sigma_{\m \Theta}^{-1} &= \m I_n^{-1} \otimes (\m G - \m H)^{-1} \\
                   &\quad-\m I_n^{-1} \m 1_n\m 1_n^T \m I_n^{-1}
                   \otimes \frac{1}{\m 1_n^T \m I_n^{-1} \m 1_n} \left((\m G - \m H)^{-1} - (\m G - \m H + \m 1_n^T
                  \m I_n^{-1}\m 1_n \m H)^{-1}\right)\\
                &= \m I_n \otimes (\m G - \m H)^{-1} - \frac{1}{n}\m 1_{n,n} \otimes
                  \left(
                  (\m G - \m H)^{-1} -
                  (\m G + (n-1)\m H)^{-1}
                  \right)\\
                &= \m I_n \otimes (\m G - \m H)^{-1} + \frac{1}{n}\m 1_{n,n} \otimes
                  \left(
                  (\m G + (n-1)\m H)^{-1} -
                  (\m G - \m H)^{-1}
                  \right)\\
                   &= \m I_n \otimes \m \Sigma_0^{-1} +
                     \frac{1}{n}\m 1_{n,n} \otimes
                     \left(
                     \m \Sigma_1^{-1} - \m \Sigma_0^{-1}
                     \right),
  \end{split}
\end{equation*}
where
\begin{align*}
  \m \Sigma_0 &:= \m G - \m H\\
  &= \m K^\mu_{\m \theta_\mu}(\m t, \m t) + \m K^\eta_{\m \theta_\eta}(\m t, \m t) + \sigma^2 \m I_J - \m K^\mu_{\m \theta_\mu}(\m t, \m t) +
  \frac{1}{n-1} \m K^\eta_{\m \theta_\eta}(\m t, \m t)\\ &=\frac{n}{n-1}\m K^\eta_{\m \theta_\eta}(\m t, \m t) + \sigma^2 \m I_J
\end{align*}
and
\begin{align*}
  \m \Sigma_1 &:= \m G + (n-1)\m H\\
  &= \m K^\mu_{\m \theta_\mu}(\m t, \m t) + \m K^\eta_{\m \theta_\eta}(\m t, \m t) + \sigma^2 \m I_J + (n-1) \left(\m K^\mu_{\m \theta_\mu}(\m t, \m t)
             - \frac{1}{n-1} \m K^\eta_{\m \theta_\eta}(\m t, \m t)\right)\\
           &= n\m K^\mu_{\m \theta_\mu}(\m t, \m t) + \sigma^2\m  I_J.
\end{align*}
From there it follows that
\begin{equation*}
  \begin{split}
    \m \Sigma_{\m \Theta}^{-1}\m y(\m t) &= \left(\m I_n \otimes \m \Sigma_0^{-1} +
                               \frac{1}{n} \m 1_{n,n} \otimes
                               \left(\m \Sigma_1^{-1} -
                               \m \Sigma_0^{-1}\right)\right)\m y(\m t)\\
                 &= \left(\m I_n \otimes \m \Sigma_0^{-1}\right)\m y(\m t) +
                   \left(\frac{1}{n} \m 1_{n,n} \otimes
                   \left(\m \Sigma_1^{-1} - \m \Sigma_0^{-1}\right)\right)\m
                   y(\m t).
  \end{split}
\end{equation*}
and since $\vc(\vc^{-1}_{J, n}(\m y(\m t))) = \m y(\m t)$, the result from \citet[11.16(b)]{seber2008matrix} gives us that
\begin{equation*}
  \begin{split}
    \m \Sigma_{\m \Theta}^{-1}\m y(\m t) &=
                               \vc\left(\m \Sigma_0^{-1}\vc^{-1}_{J,n}(\m y(\m
                               t))\right) +
                               \vc\left(\frac{1}{n}\left(\m \Sigma_1^{-1} -
                               \m \Sigma_0^{-1}\right)\vc^{-1}_{J,n}(\m y(\m t))\m
                               1_{n,n}\right)\\
                             &= \vc\left(\m \Sigma_0^{-1}\vc^{-1}_{J,n}(\m y(\m t)) +
                               \frac{1}{n}\left(\m \Sigma_1^{-1} -
                               \m \Sigma_0^{-1}\right)\vc^{-1}_{J,n}(\m y(\m t))\m 1_{n,n} \right).
  \end{split}
\end{equation*}
From \citet[11.12]{seber2008matrix} it also follows that
\begin{equation*}
  \begin{split}
    |\m \Sigma_{\m \Theta}| &= |\m I_n|^J|\m G - \m H|^{n-1}|\m G - \m H + \m 1_n^T \m
                  I_n^{-1}\m 1_n \m H|\\
             &= |\m G-\m H|^{n-1}|\m G - \m H + n\m H|\\
                &= |\m G-\m H|^{n-1}|\m G + (n-1)\m H|\\
                &= |\m \Sigma_0|^{n-1}|\m \Sigma_1|
  \end{split}
\end{equation*}
so therefore
\begin{equation*}
  \log |\m \Sigma_{\m \Theta}| =
  (n-1)\log|\m \Sigma_0| + \log|\m \Sigma_1|
\end{equation*}
which completes the proof.

\subsection*{Proof of \cref{prop:postmureg}}
The proof of \cref{prop:postmureg} has three parts. First we prove the expressions for the posterior distribution of $\m \mu(\tilde{\m t})$, then we prove the expressions for the posterior of $\m \eta(\tilde{\m t})$, and finally we prove the independence of $\m \mu(\tilde{\m t})$ and $\m \eta(\tilde{\m t})$ in the posterior distribution.

\subsubsection*{Posterior distribution of $\m \mu(\tilde{\m t})$}
As a direct consequence of the model specification in \cref{mod:eta} we have that
  \begin{equation*}
    \begin{pmatrix}
      \m y(\m t)\\
      \m \mu(\tilde{\m t})
    \end{pmatrix}\mid
    \m \Theta
    \sim
    N
    \left(
      \m 0_{nJ + J_p},
      \begin{pmatrix}
        \m \Sigma_{\m \Theta} & (\m C^{\mu, y})^T\\
        \m C^{\mu, y} & \m K^\mu_{\m \theta_\mu}(\tilde{\m t}, \tilde{\m t})
      \end{pmatrix}
    \right).
  \end{equation*}
and because $\cov(\m \mu(\tilde{\m t}), \m y_i(\m t)\mid \m\Theta) = \m K^\mu_{\m \theta_\mu}(\tilde{\m t}, \m t)$ then
\begin{equation*}
    \m C^{\mu, y} := \cov(\m \mu(\tilde{\m t}), \m y(\m t)\mid \m\Theta)
    = (\m K^\mu_{\m \theta_\mu}(\tilde{\m t}, \m t), \ldots, \m K^\mu_{\m \theta_\mu}(\tilde{\m t}, \m t)) = \m
    1_n^T \otimes \m K^\mu_{\m \theta_\mu}(\tilde{\m t}, \m t).
\end{equation*}
Using the properties of conditioning in the multivariate normal distribution, it follows that
  \begin{equation*}
    \m \mu(\tilde{\m t}) \mid \m y(\m t), \m \Theta \sim N\left(\m
      C^{\mu, y}\m \Sigma_{\m \Theta}^{-1}\m y(\m t), \m K^\mu_{\m \theta_\mu}(\tilde{\m t}, \tilde{\m t}) - \m C^{\mu, y} \m
      \Sigma_{\m \Theta}^{-1} (\m C^{\mu, y})^T\right).
  \end{equation*}
To prove the simplified expressions for the posterior mean and covariance we start by noting that $\m C^{\mu, y} \m \Sigma_{\m \Theta}^{-1} \m y(\m t) = \left(\m 1_n^T \otimes \m K^\mu_{\m \theta_\mu}(\tilde{\m t}, \m t)\right) \m \Sigma_{\m \Theta}^{-1} \m y(\m t)$ so by using the result of \citet[11.16(b)]{seber2008matrix} we have that
  \begin{equation*}
    \begin{split}
      \m C^{\mu, y} \m \Sigma_{\m \Theta}^{-1} \m y(\m t) &= \vc
                                                \left(
                                                \m K^\mu_{\m \theta_\mu}(\tilde{\m t}, \m t)\vc^{-1}_{J, n}
                                                \left(
                                                \m \Sigma_{\m \Theta}^{-1}\m y(\m t)
                                                \right)\m 1_n
                                                \right)\\
                                              &=
                                                \m K^\mu_{\m \theta_\mu}(\tilde{\m t}, \m t)\vc^{-1}_{J,n}
                                                \left(
                                                \m \Sigma_{\m \Theta}^{-1}\m y(\m t)
                                                \right)\m 1_n,
    \end{split}
  \end{equation*}
  where $\vc$ disappears, since it is already a column vector. Plugging in the simplification of $\m \Sigma_{\m \Theta}^{-1} \m y(\m t)$ given in \cref{prop:loglikreg} and using the fact that $\m 1_{n,n}\m 1_n = \m 1_n(\m 1_n^T\m 1_n) = \m 1_nn$ gives
  \begin{equation*}
    \begin{split}
      \m C^{\mu, y} \m \Sigma_{\m \Theta}^{-1} \m y(\m t) &=
                                                \m K^\mu_{\m \theta_\mu}(\tilde{\m t}, \m t)
                                                \left(
                                                \m \Sigma_0^{-1}\vc^{-1}_{J,n}(\m y(\m t)) + \frac{1}{n}\left(\m \Sigma_1^{-1} -
                                                \m \Sigma_0^{-1}\right)\vc^{-1}_{J,n}(\m y(\m t))\m 1_{n,n}
                                                \right)
                                                \m 1_n\\
                                              &=
                                                \m K^\mu_{\m \theta_\mu}(\tilde{\m t}, \m t)\left(
                                                \m \Sigma_0^{-1} \vc^{-1}_{J,n}(\m y(\m t))\m 1_n +
                                                \left(
                                                \m \Sigma_1^{-1} -
                                                \m \Sigma_0^{-1}
                                                \right)\vc^{-1}_{J,n}(\m y(\m t))\m 1_n\right)\\
                                              &=
                                                \m K^\mu_{\m \theta_\mu}(\tilde{\m t}, \m t) \m \Sigma_1^{-1}
                                                \vc^{-1}_{J,n}(\m y(\m t))\m 1_n,
    \end{split}
  \end{equation*}
  which proves the simplified expression for the posterior mean of $\m \mu(\tilde{\m t})$. Looking at the simplification of the posterior covariance of $\m \mu(\tilde{\m t})$, we plug in the expression for $\m \Sigma_{\m \Theta}^{-1}$ from the proof of \cref{prop:loglikreg} to find that
  \begin{equation*}
    \begin{split}
      \m K^\mu_{\m \theta_\mu} t&(\tilde{\m t}, \tilde{\m t}) - \m C^{\mu, y}\m \Sigma_{\m \Theta}^{-1}(\m C^{\mu, y})^T\\
              &= \m K^\mu_{\m \theta_\mu}(\tilde{\m t}, \tilde{\m t}) -
                \left(
                \m 1_n^T \otimes \m K^\mu_{\m \theta_\mu}(\tilde{\m t}, \m t)
                \right)
                \m \Sigma_{\m \Theta}^{-1}
                \left(
                \m 1_n \otimes \m K^\mu_{\m \theta_\mu}(\m t, \tilde{\m t})
                \right)\\
              &=
                \m K^\mu_{\m \theta_\mu}(\tilde{\m t}, \tilde{\m t}) -
                \left(
                \m 1_n^T \otimes \m K^\mu_{\m \theta_\mu}(\tilde{\m t}, \m t)
                \right)
                \left(
                \m I_n \otimes \m \Sigma_0^{-1}
                +
                \frac{1}{n} \m 1_{n,n}
                \otimes
                \left(
                \m \Sigma_1^{-1} -
                \m \Sigma_0^{-1}
                \right)
                \right)
                \left(
                \m 1_n \otimes \m K^\mu_{\m \theta_\mu}(\m t, \tilde{\m t})
                \right)
    \end{split}
  \end{equation*}
  The result from \citet[11.11]{seber2008matrix} and the fact that
  $\m 1_n^T\m 1_{n,n}\m 1_n = (\m 1_n^T\m 1_n)(\m 1_n^T\m 1_n) = n^2$ gives
  \begin{equation*}
    \begin{split}
      \m K^\mu&(\tilde{\m t}, \tilde{\m t}) - \m C^{\mu, y}\m \Sigma_{\m \Theta}^{-1}(\m C^{\mu, y})^T\\
              &=
                \m K^\mu_{\m \theta_\mu}(\tilde{\m t}, \tilde{\m t}) -
                \m 1_n^T\m I_n\m 1_n \otimes
                \m K^\mu_{\m \theta_\mu}(\tilde{\m t},
                \m t)\m \Sigma_0^{-1}\m K^\mu_{\m \theta_\mu}(\m t,\tilde{\m t})\\
              &\quad-
                \frac{1}{n} \m 1_n^T\m 1_{n,n}\m 1_n \otimes
                \m K^\mu_{\m \theta_\mu}(\tilde{\m t}, \m t)
                \left(
                \m \Sigma_1^{-1} -
                \m \Sigma_0^{-1}
                \right)\m K^\mu_{\m \theta_\mu}(\m t, \tilde{\m t})\\
              &=
                \m K^\mu_{\m \theta_\mu}(\tilde{\m t}, \tilde{\m t}) -
                n\m K^\mu_{\m \theta_\mu}(\tilde{\m t},
                \m t)\m \Sigma_0^{-1}\m K^\mu_{\m \theta_\mu}(\m t,\tilde{\m t})
                -
                n\m K^\mu_{\m \theta_\mu}(\tilde{\m t}, \m t)
                \left(
                \m \Sigma_1^{-1} -
                \m \Sigma_0^{-1}
                \right)\m K^\mu_{\m \theta_\mu}(\m t, \tilde{\m t})\\
              &= \m K^\mu_{\m \theta_\mu}(\tilde{\m t}, \tilde{\m t}) - n\m K^\mu_{\m \theta_\mu}(\tilde{\m t},
                \m t)\m \Sigma_1^{-1}\m K^\mu_{\m \theta_\mu}(\m t, \tilde{\m t}),
    \end{split}
  \end{equation*}
which proves the simplified expression for the posterior covariance of $\m \mu(\tilde{\m t})$.

\subsubsection*{Posterior distribution of $\m \eta(\tilde{\m t})$}
Turning to prove the expressions for the posterior distribution of $\m \eta(\tilde{\m t})$ it directly follows from the model specification in \cref{mod:eta} that
  \begin{equation*}
    \begin{pmatrix}
      \m y(\m t)\\
      \m \eta(\tilde{\m t})
    \end{pmatrix}
    \sim
    N
    \left(
      \m 0_{nJ + nJ_p},
      \begin{pmatrix}
        \m \Sigma_{\m \Theta} & (\m C^{\eta, y})^T \\
        \m C^{\eta, y} & \m C^\eta
      \end{pmatrix}
    \right).
  \end{equation*}
Since $\cov(\m \eta_i(\tilde{\m t}), \m y_i(\m t) \mid \m \Theta) = \cov(\m \eta_i(\tilde{\m t}), \m \eta_i(\m t) \mid \m \Theta) = \m K^\eta_{\m \theta_\eta}(\tilde{\m t}, \m t)$ and $\cov(\m \eta_i(\tilde{\m t}), \m y_j(\m t) \mid \m \Theta) = \cov(\m \eta_i(\tilde{\m t}), \m \eta_j(\m t) \mid \m \Theta) = -\frac{1}{n-1}\m K^\eta_{\m \theta_\eta}(\tilde{\m t}, \m t)$, we have that
  \begin{equation*}
    \label{eq:Cetay}
    \begin{split}
      \m C^{\eta, y} &= \m I_n \otimes
                       \left(
                       \m K^\eta_{\m \theta_\eta}(\tilde{\m t}, \m t) +
                       \frac{1}{n-1}\m K^\eta_{\m \theta_\eta}(\tilde{\m t}, \m t)
                       \right) +
                       \m 1_{n,n} \otimes \left(-\frac{1}{n-1}\m K^\eta_{\m \theta_\eta}(\tilde{\m t},
                       \m t)\right)\\
                     &=
                       \frac{n}{n-1} \m I_n \otimes \m K^\eta_{\m \theta_\eta}(\tilde{\m t}, \m t) -
                       \frac{1}{n-1}\m 1_{n,n} \otimes \m K^\eta_{\m \theta_\eta}(\tilde{\m t}, \m t).
    \end{split}
  \end{equation*}  
Similarly, because $\cov(\m \eta_i(\tilde{\m t}), \m \eta_i(\tilde{\m t}) \mid \m\Theta) = \m K^{\eta}(\tilde{\m t}, \tilde{\m t})$
  and $\cov(\m \eta_i(\tilde{\m t}), \m \eta_j(\tilde{\m t}) \mid \m\Theta) = -\frac{1}{n-1}\m K^{\eta}(\tilde{\m t}, \tilde{\m t})$ we have that
  \begin{equation*}
    \begin{split}
      \m C^\eta &=
               \m I_n \otimes
               \left(
               \m K^\eta_{\m \theta_\eta}(\tilde{\m t}, \tilde{\m t}) + \frac{1}{n-1} \m
                  K^\eta_{\m \theta_\eta}(\tilde{\m t}, \tilde{\m t})
               \right)
               +
               \m 1_{n,n} \otimes
               \left(
               -\frac{1}{n-1}\m K^\eta_{\m \theta_\eta}(\tilde{\m t}, \tilde{\m t})
               \right)\\
             &= \frac{n}{n-1} \m I_n \otimes \m K^\eta_{\m \theta_\eta}(\tilde{\m t}, \tilde{\m t}) -
               \frac{1}{n-1} \m 1_{n,n}
               \otimes \m K^\eta_{\m \theta_\eta}(\tilde{\m t}, \tilde{\m t}).
    \end{split}
  \end{equation*}
Using the properties of conditioning in the multivariate normal distribution, it follows that
  \begin{equation*}
    \m \eta(\tilde{\m t}) \mid \m y(\m t) \sim N\left(\m C^{\eta, y}\m
      \Sigma_{\m \Theta}^{-1}\m y(\m t),
      \m C^{\eta} - \m C^{\eta, y}\m \Sigma_{\m \Theta}^{-1}(\m C^{\eta, y})^T\right).
  \end{equation*}  
To prove the simplified expressions for the posterior mean and covariance of $\m \eta(\tilde{\m t})$ we start by noting that
  \begin{equation*}
    \begin{split}
      \m C^{\eta, y}\m \Sigma_{\m \Theta}^{-1}\m y(\m t) = \left(\frac{n}{n-1} \m I_n \otimes \m K^\eta_{\m \theta_\eta}(\tilde{\m t}, \m t) -
                   \frac{1}{n-1}\m 1_{n,n} \otimes
                   \m K^\eta_{\m \theta_\eta}(\tilde{\m t}, \m t)\right)
                   \m \Sigma_{\m \Theta}^{-1}\m y(\m t)
    \end{split}
  \end{equation*}
and by using the result from \citet[11.16(b)]{seber2008matrix} it follows that
  \begin{equation*}
    \begin{split}
      \m C^{\eta, y}&\m \Sigma_{\m \Theta}^{-1}\m y(\m t)\\
                 &=
                   \vc
                   \left(
                   \m K^\eta_{\m \theta_\eta}(\tilde{\m t},\m t)
                   \vc^{-1}_{J,n}
                   \left(
                   \m \Sigma_{\m \Theta}^{-1}\m y(\m t)
                   \right)
                   \frac{n}{n-1}\m I_n^T
                   \right) -
                   \vc
                   \left(
                   \m K^\eta_{\m \theta_\eta}(\tilde{\m t}, \m t)
                   \vc^{-1}_{J,n}
                   \left(
                   \m \Sigma_{\m \Theta}^{-1}\m y(\m t)
                   \right)
                   \frac{1}{n-1}\m 1_{n,n}^T
                   \right)\\
                 &=
                   \frac{1}{n-1}\vc
                   \left(
                   \m K^\eta_{\m \theta_\eta}(\tilde{\m t},\m t)
                   \vc^{-1}_{J,n}
                   \left(
                   \m \Sigma_{\m \Theta}^{-1}\m y(\m t)
                   \right)
                   (n\m I_n - \m 1_{n,n})
                   \right).
    \end{split}
  \end{equation*}
  Plugging in the expression for $\m \Sigma_{\m \Theta}^{-1} \m y(\m t)$ given in \cref{prop:loglikreg} and using that $\m 1_{n,n}\m 1_{n,n} = \m 1_n(\m 1_n^T\m 1_n)\m 1_n^T = n\m 1_n\m 1_n^T = n\m 1_{n,n}$, we get
  \begin{equation*}
    \begin{split}
      \m C^{\eta, y}&\m \Sigma_{\m \Theta}^{-1}\m y(\m t)\\
                 &=
                   \frac{1}{n-1}
                   \vc
                   \Bigg(
                   \m K^\eta_{\m \theta_\eta}(\tilde{\m t}, \m t)
                   \left(
                   \m \Sigma_0^{-1}\vc^{-1}_{J,n}(\m y(\m t)) + \frac{1}{n}\left(\m \Sigma_1^{-1} -
                   \m \Sigma_0^{-1}\right)\vc^{-1}_{J,n}(\m y(\m t))\m 1_{n,n}
                   \right)\\
                   &\qquad\quad\qquad\qquad(n\m I_n - \m 1_{n,n})
                   \Bigg)\\
                 &= \frac{1}{n-1}
                   \vc
                   \biggl[
                   n\m K^\eta_{\m \theta_\eta}(\tilde{\m t},
                   \m t)\m \Sigma_0^{-1}\vc^{-1}_{J,n}(\m y(\m t))
                   +
                   \m K^\eta_{\m \theta_\eta}(\tilde{\m t},
                   \m t)
                   \left(
                   \m \Sigma_1^{-1} - \m \Sigma_0^{-1}
                   \right)\vc^{-1}_{J,n}(\m y(\m t))\m 1_{n,n}\\
                 &\qquad\qquad\qquad-
                   \m K^\eta_{\m \theta_\eta}(\tilde{\m t},
                   \m t)\m \Sigma_0^{-1}\vc^{-1}_{J,n}(\m y(\m t))\m 1_{n,n}\\
                   &\qquad\qquad\qquad-
                   \m K^\eta_{\m \theta_\eta}(\tilde{\m t},
                   \m t)
                   \frac{1}{n}
                   \left(
                   \m \Sigma_1^{-1} - \m \Sigma_0^{-1}
                   \right)\vc^{-1}_{J,n}(\m y(\m t))\m 1_{n,n}\m 1_{n,n}
                   \biggr]\\
                 &=
                   \frac{1}{n-1}\vc\biggl[
                   n\m K^\eta_{\m \theta_\eta}(\tilde{\m t},
                   \m t)\m \Sigma_0^{-1}\vc^{-1}_{J,n}(\m y(\m t))
                   +
                   \m K^\eta_{\m \theta_\eta}(\tilde{\m t},
                   \m t)\m \Sigma_1^{-1}\vc^{-1}_{J,n}(\m y(\m t))\m 1_{n,n}\\
                 &\qquad\qquad\qquad
                   -
                   \m K^\eta_{\m \theta_\eta}(\tilde{\m t},
                   \m t)\m \Sigma_0^{-1}\vc^{-1}_{J,n}(\m y(\m t))\m 1_{n,n}
                   -
                   \m K^\eta_{\m \theta_\eta}(\tilde{\m t},
                   \m t)\m \Sigma_0^{-1}\vc^{-1}_{J,n}(\m y(\m t))\m 1_{n,n}\\
                 &\qquad\qquad\qquad
                   -
                   \m K^\eta_{\m \theta_\eta}(\tilde{\m t},
                   \m t)\m \Sigma_1^{-1}\vc^{-1}_{J,n}(\m y(\m t))\m 1_{n,n}
                   +
                   \m K^\eta_{\m \theta_\eta}(\tilde{\m t},
                   \m t)\m \Sigma_0^{-1}\vc^{-1}_{J,n}(\m y(\m t))\m 1_{n,n}
                   \biggr]\\
                 &= \frac{1}{n-1}\vc
                   \left(
                   n\m K^\eta_{\m \theta_\eta}(\tilde{\m t},\m t)\m \Sigma_0^{-1}\vc^{-1}_{J,n}(\m y(\m t))
                   -
                   \m K^\eta_{\m \theta_\eta}(\tilde{\m t},\m t)\m \Sigma_0^{-1}\vc^{-1}_{J,n}(\m y(\m t))\m 1_{n,n}
                   \right),
    \end{split}
  \end{equation*}
  which proves the expression for the posterior mean of $\m \eta(\tilde{\m t})$.
  To prove the simplification of the posterior covariance of $\m \eta(\tilde{\m t})$ we start by plugging in previously derived expressions to get that
  \begin{align*}
      &\m C^{\eta, y} \m \Sigma_{\m \Theta}^{-1} (\m C^{\eta, y})^T\\
      &=
        \left(
        \frac{n}{n-1} \m I_n \otimes
        \m K^\eta_{\m \theta_\eta}(\tilde{\m t}, \m t) - \frac{1}{n-1}
        \m 1_{n,n} \otimes \m K^\eta_{\m \theta_\eta}(\tilde{\m t},
        \m t)
        \right)\left(
        \m I_n \otimes \m \Sigma_0^{-1} +
        \m 1_{n,n}\otimes \frac{1}{n}
        \left(
        \m \Sigma_1^{-1} - \m \Sigma_0^{-1}
        \right)
        \right)\\
      &\quad
        \left(
        \frac{n}{n-1} \m I_n \otimes
        \m K^\eta_{\m \theta_\eta}(\m t, \tilde{\m t}) - \frac{1}{n-1}
        \m 1_{n,n} \otimes \m K^\eta_{\m \theta_\eta}(\m t,
        \tilde{\m t})
        \right)\\
      &=
        \left(
        \frac{n}{n-1}
        \right)^2 \m I_n \otimes
        \m K^\eta_{\m \theta_\eta}(\tilde{\m t}, \m t)
        \m \Sigma_0^{-1}
        \m K^\eta_{\m \theta_\eta}(\m t, \tilde{\m t}) + \frac{n}{(n-1)^2} \m 1_{n,n}
        \otimes \m K^\eta_{\m \theta_\eta}(\tilde{\m t},\m t)
        \left(
        \m \Sigma_1^{-1} - \m \Sigma_0^{-1}
        \right)
        \m K^\eta_{\m \theta_\eta}(\m t, \tilde{\m t})\\
      &\quad - \frac{n}{(n-1)^2} \m 1_{n,n} \otimes
        \m K^\eta_{\m \theta_\eta}(\tilde{\m t}, \m t)
        \m \Sigma_0^{-1}\m K^\eta_{\m \theta_\eta}(\m t, \tilde{\m t}) -
        \frac{n}{(n-1)^2} \m 1_{n,n}
        \otimes \m K^\eta_{\m \theta_\eta}(\tilde{\m t},\m t)
        \left(
        \m \Sigma_1^{-1} - \m \Sigma_0^{-1}
        \right)
        \m K^\eta_{\m \theta_\eta}(\m t, \tilde{\m t})\\
      &\quad -
        \frac{n}{(n-1)^2} \m 1_{n,n} \otimes
        \m K^\eta_{\m \theta_\eta}(\tilde{\m t}, \m t)
        \m \Sigma_0^{-1}\m K^\eta_{\m \theta_\eta}(\m t, \tilde{\m t}) -
        \frac{n}{(n-1)^2} \m 1_{n,n} \otimes
        \m K^\eta_{\m \theta_\eta}(\tilde{\m t}, \m t)
        \left(
        \m \Sigma_1^{-1} - \m \Sigma_0^{-1}
        \right)\m K^\eta_{\m \theta_\eta}(\m t, \tilde{\m t})\\
      &\quad +
        \frac{n}{(n-1)^2} \m 1_{n,n} \otimes
        \m K^\eta_{\m \theta_\eta}(\tilde{\m t}, \m t) \m \Sigma_0^{-1}
        \m K^\eta_{\m \theta_\eta}(\m t, \tilde{\m t}) +
        \frac{n}{(n-1)^2} \m 1_{n,n} \otimes
        \m K^\eta_{\m \theta_\eta}(\tilde{\m t}, \m t)
        \left(
        \m \Sigma_1^{-1} - \m \Sigma_0^{-1}
        \right)
        \m K^\eta_{\m \theta_\eta}(\m t, \tilde{\m t})\\
      &=
        \left(
        \frac{n}{n-1}
        \right)^2
        \m I_n \otimes \m K^\eta_{\m \theta_\eta}(\tilde{\m t}, \m t)
        \m \Sigma_0^{-1}\m K^\eta_{\m \theta_\eta}(\m t, \tilde{\m t})
        -
        \frac{n}{(n-1)^2} \m 1_{n,n} \otimes
        \m K^\eta_{\m \theta_\eta}(\tilde{\m t}, \m t) \m \Sigma_0^{-1}
        \m K^\eta_{\m \theta_\eta}(\m t, \tilde{\m t})\\
      &=
        \frac{n}{(n-1)^2}
        \left(
        n\m I_n \otimes \m K^\eta_{\m \theta_\eta}(\tilde{\m t},
        \m t)\m \Sigma_0^{-1}\m K^\eta_{\m \theta_\eta}(\m t,\tilde{\m t})
        -
        \m 1_{n,n} \otimes \m K^\eta_{\m \theta_\eta}(\tilde{\m t}, \m t)
        \m \Sigma_0^{-1} \m K^\eta_{\m \theta_\eta}(\m t, \tilde{\m t})
        \right).
  \end{align*}
  Using this result along with the expression for $\m C^\eta$ gives us that
  \begin{equation*}
    \begin{split}
      \m C^\eta - &\m C^{\eta, y} \m \Sigma_{\m \Theta}^{-1}(\m C^{\eta, y})^T\\
      &=
        \frac{1}{n-1}
        \left(
        n\m I_n \otimes \m K^\eta_{\m \theta_\eta}(\tilde{\m t}, \tilde{\m t}) - \m 1_{n,n}
        \otimes \m K^\eta_{\m \theta_\eta}(\tilde{\m t}, \tilde{\m t})  
        \right)\\
      &\quad -
        \frac{n}{(n-1)^2}
        \left(
        n\m I_n \otimes \m K^\eta_{\m \theta_\eta}(\tilde{\m t},
        \m t)\m \Sigma_0^{-1}\m K^\eta_{\m \theta_\eta}(\m t,\tilde{\m t})
        -
        \m 1_{n,n} \otimes \m K^\eta_{\m \theta_\eta}(\tilde{\m t}, \m t)
        \m \Sigma_0^{-1} \m K^\eta_{\m \theta_\eta}(\m t, \tilde{\m t})
        \right)\\
      & =
        \frac{n}{n-1}
        \m I_n \otimes
        \left(
        \m K^\eta_{\m \theta_\eta}(\tilde{\m t}, \tilde{\m t}) - \frac{n}{n-1}
        \m K^\eta_{\m \theta_\eta}(\tilde{\m t}, \m t) \m \Sigma_0^{-1}
        \m K^\eta_{\m \theta_\eta}(\m t, \tilde{\m t})
        \right)\\
      &\quad +
        \frac{1}{n-1}
        \m 1_{n,n} \otimes
        \left(
        \frac{n}{n-1}
        \m K^\eta_{\m \theta_\eta}(\tilde{\m t}, \m t)
        \m \Sigma_0^{-1}
        \m K^\eta_{\m \theta_\eta}(\m t, \tilde{\m t})
        -
        \m K^\eta_{\m \theta_\eta}(\tilde{\m t}, \tilde{\m t})
        \right),
    \end{split}
  \end{equation*}
  which proves the expression for the posterior covariance  of $\m \eta(\tilde{\m t})$.

\subsubsection*{Independence of the posterior distributions of $\m \mu(\tilde{\m t})$ and $\m \eta(\tilde{\m t})$}

According to the model specification in \cref{mod:eta} we directly have that
\begin{equation*}
  \begin{pmatrix}
    \m \eta(\tilde{\m t})\\
    \m \mu(\tilde{\m t})\\
    \m y(\m t)
  \end{pmatrix}
  \mid \m \Theta
  \sim
  N
  \left(
    \begin{pmatrix}
      \m 0\\
      \m 0\\
      \m 0
    \end{pmatrix},
    \begin{pmatrix}
      \m C^\eta & \m 0 & \m C^{\eta, y}\\
      \m 0 & \m C^\mu & \m C^{\mu, y}\\
      \m C^{y, \eta} & \m C^{y, \mu} & \m \Sigma_{\m \Theta}
    \end{pmatrix}
  \right).
\end{equation*}
and using the properties of conditioning in the multivariate normal distribution, it follows that
\begin{equation*}
  \begin{split}
    \cov((\m \eta(\tilde{\m t}), \m \mu(\tilde{\m t})) \mid \m y(\m t), \m \Theta) &=
           \begin{pmatrix}
             \m C^\eta & \m 0\\
             \m 0 & \m C^\mu
           \end{pmatrix}
           -
           \begin{pmatrix}
             \m C^{\eta, y}\\
             \m C^{\mu, y}
           \end{pmatrix}
           \m \Sigma_{\m \Theta}^{-1}
           \begin{pmatrix}
             \m C^{y, \eta} & \m C^{y, \mu}
           \end{pmatrix}\\
         &=
           \begin{pmatrix}
             \m C^\eta & \m 0\\
             \m 0 & \m C^\mu
           \end{pmatrix}
           -
           \begin{pmatrix}
             \m C^{\eta, y}\m \Sigma_{\m \Theta}^{-1}\m C^{y,\eta} &\m C^{\eta, y}\m
                                                         \Sigma_{\m \Theta}^{-1}\m C^{y,\mu}\\
             \m C^{\mu, y}\m \Sigma_{\m \Theta}^{-1}\m C^{y,\eta} &\m C^{\mu, y}\m \Sigma_{\m \Theta}^{-1}\m C^{y,\mu}
           \end{pmatrix}.
  \end{split}
\end{equation*}
In order to show that $\m \mu(\tilde{\m t})$ and $\m \eta(\tilde{\m t})$ are independent in the posterior distribution, we thus need to show that $\m C^{\eta, y}\m \Sigma_{\m \Theta}^{-1}\m C^{y,\mu} = \m 0$. Plugging in the expressions derived in the previous stages of the proof of this proposition, we find that 
\begin{equation*}
  \begin{split}
    \m C^{\eta, y}\m \Sigma_{\m \Theta}^{-1}\m C^{y,\mu}
    &=
      \left(\frac{n}{n-1}\m I_n \otimes \m K^\eta_{\m \theta_\eta}(\tilde{\m t}, \m t) - \frac{1}{n-1}\m
      1_{n.n} \otimes \m K^\eta_{\m \theta_\eta}(\tilde{\m t}, \m t)\right)\\
    &\qquad
      \left(\m I_n \otimes \m \Sigma_0^{-1} + \frac{1}{n} \m 1_{n,n}(\m \Sigma_1^{-1}
      - \m \Sigma_0^{-1})\right)\\
    &\qquad
      \left(
      \m 1_n \otimes \m K^\mu_{\m \theta_\mu}(\m t, \tilde{\m t})
      \right)\\
    &=
      \frac{n}{n-1} \m 1_n\otimes \m K^\eta_{\m \theta_\eta}(\tilde{\m t}, \m t)\m
      \Sigma_0^{-1}\m K^\mu_{\m \theta_\mu}(\m t, \tilde{\m t})\\
    &\quad + \frac{n}{n-1}\m 1_n\otimes \m K^\eta_{\m \theta_\eta}(\tilde{\m t}, \m t)(\m \Sigma_1^{-1}
      - \m \Sigma_0^{-1})\m K^\mu_{\m \theta_\mu}(\m t, \tilde{\m t})\\
    &\quad - \frac{n}{n-1}\m 1_n \otimes \m K^\eta_{\m \theta_\eta}(\tilde{\m t}, \m t)\m
      \Sigma_0^{-1}\m K^\mu_{\m \theta_\mu}(\m t, \tilde{\m t})\\
    &\quad - \frac{n}{n-1}\m 1_n \otimes \m K^\eta_{\m \theta_\eta}(\tilde{\m t}, \m t)(\m \Sigma_1^{-1}
      - \m \Sigma_0^{-1})\m K^\mu_{\m \theta_\mu}(\m t, \tilde{\m t})\\
    &= \m 0
  \end{split}
\end{equation*}
which shows that $\m \eta(\tilde{\m t}) \indep \m \mu(\tilde{\m t}) \mid \m y(\m t), \m \Theta$, and we may therefore sample from the posterior of $\m f_i(\tilde{\m t}) = \m \mu(\tilde{\m t}) + \m \eta_i(\tilde{\m t})$ by drawing $\m \mu(\tilde{\m t})$ and
$\m \eta(\tilde{\m t})$ separately. This concludes the proof of \cref{prop:postmureg}.

\subsection*{Proof of \cref{prop:irreg-loglik}}

  Similar to the proof of \cref{prop:loglikreg}, we get from the model that
  $\m y(\m t) \mid \m \Theta$ is zero-mean Gaussian with some covariance matrix $\m \Sigma_{\m \Theta}$, which can be divided into blocks as in \cref{eq:irreg-sigma-blocks}. It follows from the model definition in \cref{mod:eta} that
  \begin{equation*}
    \begin{split}
      \m A &= \m I_{n_a} \otimes \left(\m K^\mu_{\m \theta_\mu}(\m t^a, \m t^a) + \m K^\eta_{\m \theta_\eta}(\m t^a,
             \m t^a) + \sigma^2\m I_{J_a} - \left(\m K^\mu_{\m \theta_\mu}(\m t^a, \m t^a) -
          \frac{1}{n-1}\m K^\eta_{\m \theta_\eta}(\m t^a, \m t^a)\right)\right)\\
          &\quad + \m 1_{n_a,n_a} \otimes \left(\m K^\mu_{\m \theta_\mu}(\m t^a, \m t^a) -
          \frac{1}{n-1}\m K^\eta_{\m \theta_\eta}(\m t^a, \m t^a)\right)\\
        &=
          \m I_{n_a} \otimes \left(\frac{n}{n-1}\m K^\eta_{\m \theta_\eta}(\m t^a, \m t^a) + \sigma^2\m I_{J_a}\right) +
          \m 1_{n_a,n_a} \otimes \left(\m K^\mu_{\m \theta_\mu}(\m t^a, \m t^a) -
          \frac{1}{n-1}\m K^\eta_{\m \theta_\eta}(\m t^a, \m t^a)\right).
    \end{split}
  \end{equation*}
  We now find an expression for $\m C^T$. It follows from \cref{mod:eta} that
  \begin{equation*}
    \begin{split}
      \cov(\m y_i^a(\m t^a), &\m y^b_j(\m t^b_j) \mid \m \Theta)\\
      &=
        \cov(\m \mu(\m t^a) + \m \eta^a_i(\m t^a),
        \m \mu(\m t^b_j) + \m \eta_j^b(\m t^b_j) \mid
        \m \Theta)\\
      &=
        \m K^\mu_{\m \theta_\mu}(\m t^a, \m t^b_j) - \frac{1}{n - 1}\m K^\eta_{\m \theta_\eta}(\m t^a, \m t^b_j).      
    \end{split}
  \end{equation*}
  Note that this doesn't depend on $i$, so the $j$'th block-column of $\m C^T$ is
  $n_a$ block-rows of the above (one for each value $i = 1, \ldots, n_a$). This
  means that if we let
  \begin{equation*}
    \m C^b := \m K^\mu_{\m \theta_\mu}(\m t^a, \m t^b) - \frac{1}{n-1}\m K^\eta_{\m \theta_\eta}(\m t^a, \m t^b)
  \end{equation*}
  then
  \begin{equation*}
    \m C^T = \m 1_{n_a} \otimes \m C^b.
  \end{equation*}
  Similar to the proof for \cref{prop:loglikreg}, we find that $\m B$ is a block matrix with $i,i$
  block
  \begin{equation*}
    \cov(\m y^b_i(\m t^b_i), \m y^b_i(\m t^b_i) \mid \m \Theta) = \m K^\mu_{\m \theta_\mu}(\m t^b_i, \m t^b_i) + \m K^\eta_{\m \theta_\eta}(\m t^b_i, \m t^b_i) + \sigma^2 \m I_{J^b_i}
  \end{equation*}
  and $i,j$ ($i \ne j$) block
  \begin{equation*}
    \cov(\m y^b_i(\m t^b_i), \m y^b_j(\m t^b_j) \mid \m \Theta) = \m K^\mu_{\m \theta_\mu}(\m t^b_i, \m t^b_j) - \frac{1}{n-1} \m K^\eta_{\m \theta_\eta}(\m t^b_i, \m t^b_j),
  \end{equation*}
  which gives us the desired covariance matrix $\m \Sigma_{\m \Theta}$.

  We now prove the simplifications. The
  expression for $\m \Sigma_{\m \Theta}^{-1}$ follows from \citet[14.11(a)]{seber2008matrix}. Using \citet[11.12]{seber2008matrix} gives that
  \begin{equation*}
    \begin{split}
      |\m A| &= |\m I_{n_a}|^{J_a}\left|\frac{n}{n-1}\m K^\eta_{\m \theta_\eta}(\m t^a, \m t^a) +
            \sigma^2\m I_{J_a}\right|^{n_a - 1}\\
      &\quad\left|\frac{n}{n-1}\m K^\eta_{\m \theta_\eta}(\m t^a, \m t^a) +
            \sigma^2\m I_{J_a} + \m 1_{n_a}^T\m I_{n_a, n_a}^{-1}\m
        1_{n_a}\left(\m K^\mu_{\m \theta_\mu}(\m t^a, \m t^a) -
            \frac{1}{n-1}\m K^\eta_{\m \theta_\eta}(\m t^a, \m t^a)\right)\right|\\
          &= \left|\m A_0\right|^{n_a - 1}\left|\m A_1\right|
    \end{split}
  \end{equation*}
  where
  \begin{equation*}
    \begin{split}
      \m A_0 &:= \frac{n}{n-1}\m K^\eta_{\m \theta_\eta}(\m t^a, \m t^a) +
            \sigma^2\m I_{J_a}\\
      \m A_1 &:= \frac{n}{n-1}\m K^\eta_{\m \theta_\eta}(\m t^a, \m t^a) +
            \sigma^2\m I_{J_a} + \m 1_{n_a}^T\m I_{n_a, n_a}^{-1}\m
               1_{n_a}\left(\m K^\mu_{\m \theta_\mu}(\m t^a, \m t^a) -
            \frac{1}{n-1}\m K^\eta_{\m \theta_\eta}(\m t^a, \m t^a)\right)\\
          &= \sigma^2\m I_{J_a} + n_a \m K^\mu_{\m \theta_\mu}(\m t^a, \m t^a) + \frac{n -
            n_a}{n - 1}\m K^\eta_{\m \theta_\eta}(\m t^a, \m t^a)\\
          &= \sigma^2\m I_{J_a} + n_a \m K^\mu_{\m \theta_\mu}(\m t^a, \m t^a) + \frac{n_b}{n -
            1}\m K^\eta_{\m \theta_\eta}(\m t^a, \m t^a).
    \end{split}
  \end{equation*}
  \citet[14.1]{seber2008matrix} then gives
  \begin{equation*}
    \log|\m \Sigma_{\m \Theta}| = \log|\m A| + \log|\m S| = (n_a - 1)\log|\m A_0| + \log|\m A_1| +
    \log|\m S|,
  \end{equation*}
  where $\m S = \m B - \m C \m A^{-1} \m C^T$ is the Schur complement of $\m A$ in $\m \Sigma_{\m \Theta}$. \citet[11.12]{seber2008matrix} gives
  that
  \begin{equation*}
    \m A^{-1} = \m I_{n_a} \otimes \m A_0^{-1} +
                  \frac{1}{n_a} \m 1_{n_a,n_a} \otimes (\m A_1^{-1} - \m A_0^{-1}).
  \end{equation*}
  Plugging in the expressions above and using the Kronecker product rule
  \citet[11.11]{seber2008matrix} gives
  \begin{equation}
    \label{eq:AiCt}
    \begin{split}
      \m A^{-1}\m C^T &=
                  \left(
                  \m I_{n_a} \otimes \m A_0^{-1} +
                  \m 1_{n_a,n_a} \otimes \frac{1}{n_a} (\m A_1^{-1} - \m A_0^{-1})
                  \right)
                  \left(
                  \m 1_{n_a}
                  \otimes
                  \m C^b
                  \right)\\
                &= \m I_{n_a}\m 1_{n_a} \otimes \m A_0^{-1} \m C^b +
                  \frac{1}{n_a}\m 1_{n_a, n_a}\m 1_{n_a} \otimes (\m A_1^{-1} - \m A_0^{-1})
                  \m C^b\\
                &= \m 1_{n_a} \otimes \m A_0^{-1} \m C^b +
                  \m 1_{n_a} \otimes (\m A_1^{-1} - \m A_0^{-1})
                  \m C^b\\
                &= \m 1_{n_a} \otimes \m A_1^{-1}\m C^b.
    \end{split}
  \end{equation}
  Using this result and the Kronecker product rule once more then gives
  \begin{equation*}
    \begin{split}
      \m C\m A^{-1}\m C^T &=
                   \left(
                   \m 1_{n_a}^T \otimes (\m C^b)^T
                   \right)
                   \left(
                   \m 1_{n_a} \otimes \m A^{-1}_1\m C^b
                   \right)\\
                 &= \m 1_{n_a}^T\m 1_{n_a} \otimes
                   (\m C^b)^T\m A_1^{-1}\m C^b\\
                 &= n_a(\m C^b)^T \m A_1^{-1} \m C^b.
    \end{split}
  \end{equation*}
  This shows that $\m S = \m B - n_a(\m C^b)^T \m A_1^{-1} \m C^b$, which gives the desired log-determinant simplification.
  We now simplify $\m A^{-1}\m y^a(\m t^a)$ very similarly to the regular
  case. We have
  \begin{equation*}
    \begin{split}
      \m A^{-1}\m y^a(\m t^a)
      &= \left(
        \m I_{n_a} \otimes \m A_0^{-1} +
        \frac{1}{n_a} \m 1_{n_a,n_a} \otimes (\m A_1^{-1} - \m A_0^{-1})
        \right)\m y^a(\m t^a)\\
      &= (\m I_{n_a} \otimes \m A_0^{-1}) \m y^a(\m t^a) +
        \left(
        \frac{1}{n_a}\m 1_{n_a, n_a} \otimes (\m A_1^{-1} - \m A_0^{-1})
        \right)
        \m y^a(\m t^a).
    \end{split}
  \end{equation*}
  From this and \citet[11.16(b)]{seber2008matrix} follows that
  \begin{equation*}
    \begin{split}
      \m A^{-1}&\m y^a(\m t^a)\\
      &= \vc
        \left(
        \m A_0^{-1} \vc^{-1}_{J_a, n_a}(\m y^a(\m t^a)) 
        \right)
        +
        \vc
        \left(
        \frac{1}{n_a}(\m A_1^{-1} - \m A_0^{-1})\vc^{-1}_{J_a, n_a}(\m y^a(\m t^a)) \m 1_{n_a, n_a}
        \right)\\
      &= \vc
        \left(
        \m A_0^{-1} \vc^{-1}_{J_a, n_a}(\m y^a(\m t^a)) 
        +
        \frac{1}{n_a}(\m A_1^{-1} - \m A_0^{-1})\vc^{-1}_{J_a, n_a}(\m y^a(\m t^a)) \m 1_{n_a, n_a}
        \right),
    \end{split}
  \end{equation*}
  which completes the proof.

\subsection*{The Rest of \cref{prop:postirreg}: Posterior Simplifications for $\m \eta(\tilde{\m t})$}

The posterior mean for $\m \eta(\tilde{\m t})$ simplifies as
    $\m \mu_{\m \eta(\tilde{\m t}) \mid \m y(\m t), \m \Theta} = (\m m_{\eta^a}^T, \m m_{\eta^b}^T)^T$ with
  \begin{equation*}
    \begin{split}
      \m m_{\eta^a} = \frac{1}{n-1} \cdot
      \Big[&n \vc\left(\m K^\eta_{\m \theta_\eta}(\tilde{\m t}, \m t^a) \vc^{-1}_{J_a,
        n_a}\left(\m P_1 \m y(\m t)\right)\right)\\
      &-\vc
        \left(
        \m K^\eta_{\m \theta_\eta}(\tilde{\m t}, \m t^a) \vc^{-1}_{J_a, n_a}(\m P_1 \m y(\m
        t)) \m 1_{n_a, n_a}
        \right)\\
      &-\m 1_{n_a} \otimes \m K^\eta_{\m \theta_\eta}(\tilde{\m t}, \m t^b)
        \m P_2 \m y(\m t)\Big]
    \end{split}
  \end{equation*}
  and
  \begin{equation*}
      \m m_{\eta^b} = -\frac{1}{n-1}\vc
    \left(
      \m K^\eta_{\m \theta_\eta}(\tilde{\m t}, \m t^a)\vc^{-1}_{J_a, n_a}(\m P_1 \m y(\m
      t))
      \m 1_{n_a, n_b}
        \right) + \m C^{\eta^b, y^b}\m P_2\m y(\m t)
  \end{equation*}
  and
  \begin{equation*}
    \m C^{\eta^b, y^b} =
    \begin{pmatrix}
      \m K^\eta_{\m \theta_\eta}(\tilde{\m t}, \m t^b_1) & -\frac{1}{n-1}\m K^\eta_{\m \theta_\eta}(\tilde{\m t}, \m t^b_2) &
                                                                               \cdots
      & -\frac{1}{n-1}\m K^\eta_{\m \theta_\eta}(\tilde{\m t}, \m t^b_{n_b}) \\
      - \frac{1}{n-1}\m K^\eta_{\m \theta_\eta}(\tilde{\m t}, \m t^b_1) & \m K^\eta_{\m \theta_\eta}(\tilde{\m t},
                                                         \m t^b_2) & \ddots &
                                                                              \vdots\\
      \vdots & \ddots & \ddots & \vdots\\
      -\frac{1}{n-1}\m K^\eta_{\m \theta_\eta}(\tilde{\m t}, \m t^b_1) & -\frac{1}{n-1}\m
                                                        K^\eta_{\m \theta_\eta}(\tilde{\m t}, \m
                                                        t^b_2) & \cdots & \m K^\eta_{\m \theta_\eta}(\tilde{\m t}, \m t^b_{n_b})
    \end{pmatrix}
  \end{equation*}
  and the posterior covariance matrix simplifies as
  \begin{equation*}
    \m \Sigma_{\m \eta(\tilde{\m t}) \mid \m y(\m t), \m \Theta} =
    \begin{pmatrix}
      \m \Sigma_{\eta,1,1} & \m \Sigma_{\eta,1,2}\\
      \m \Sigma_{\eta,2,1} & \m \Sigma_{\eta,2,2}
    \end{pmatrix}
  \end{equation*}
  where
  \begin{equation*}
    \begin{split}
      \m \Sigma_{\eta,1,1}
      &=\m C^{\eta^a}\\
      & - \Big[\m C^{\eta^a, y^a} \m A^{-1} (\m C^{\eta^a, y^a})^T\\
      &\quad + \m C^{\eta^a, y^a}\m A^{-1} \m C^T \m S^{-1} \m C \m A^{-1} (\m
        C^{\eta^a, y^a})^T\\
      &\quad - \m C^{\eta^a, y^a}\m A^{-1}\m C^T \m S^{-1} (\m C^{\eta^a,y^b})^T\\
      &\quad - \m C^{\eta^a, y^b}\m S^{-1}\m C\m A^{-1}(\m C^{\eta^a,y^a})^T\\
      &\quad + \m C^{\eta^a, y^b}\m S^{-1}(\m C^{\eta^a,y^b})^T\Big]
    \end{split}
  \end{equation*}
  and
  \begin{equation*}
    \begin{split}
      \m \Sigma_{\eta,2,1}
      &=\m C^{\eta^b,\eta^a}\\
      &-\Big[\m C^{\eta^b, y^a} \m A^{-1} (\m C^{\eta^a, y^a})^T\\
      &\quad+ \m C^{\eta^b, y^a} \m A^{-1} \m C^T \m S^{-1} \m C \m A^{-1} (\m
        C^{\eta^a, y^a})^T\\
      &\quad- \m C^{\eta^b, y^a}\m A^{-1}\m C^T \m S^{-1} (\m C^{\eta^a,y^b})^T\\
      &\quad- \m C^{\eta^b, y^b}\m S^{-1}\m C\m A^{-1}(\m C^{\eta^a,y^a})^T\\
      &\quad+ \m C^{\eta^b, y^b}\m S^{-1}(\m C^{\eta^a,y^b})^T\Big]
    \end{split}
  \end{equation*}
  and
  \begin{equation*}
    \begin{split}
      \m \Sigma_{\eta,1,2}
      &=
        \m C^{\eta^a,\eta^b}\\
      &- \Big[\m C^{\eta^a, y^a} \m A^{-1} (\m C^{\eta^b, y^a})^T\\
      &\quad + \m C^{\eta^a, y^a}\m A^{-1} \m C^T \m S^{-1} \m C \m A^{-1} (\m
        C^{\eta^b, y^a})^T\\
      &\quad - \m C^{\eta^a, y^a}\m A^{-1}\m C^T \m S^{-1} (\m C^{\eta^b,y^b})^T\\
      &\quad - \m C^{\eta^a, y^b}\m S^{-1}\m C\m A^{-1}(\m C^{\eta^b,y^a})^T\\
      &\quad + \m C^{\eta^a, y^b}\m S^{-1}(\m C^{\eta^b,y^b})^T\Big]
    \end{split}
  \end{equation*}
  and
  \begin{equation*}
    \begin{split}
      \m \Sigma_{\eta,2,2}
      &=
        \m C^{\eta^b}\\
      &- \Big[\m C^{\eta^b, y^a} \m A^{-1} (\m C^{\eta^b, y^a})^T\\
      &\quad + \m C^{\eta^b, y^a}\m A^{-1} \m C^T \m S^{-1} \m C \m A^{-1} (\m
        C^{\eta^b, y^a})^T\\
      &\quad - \m C^{\eta^b, y^a}\m A^{-1}\m C^T \m S^{-1} (\m C^{\eta^b,y^b})^T\\
      &\quad - \m C^{\eta^b, y^b}\m S^{-1}\m C\m A^{-1}(\m C^{\eta^b,y^a})^T\\
      &\quad + \m C^{\eta^b, y^b}\m S^{-1}(\m C^{\eta^b,y^b})^T\Big]
    \end{split}
  \end{equation*}
  with
  \begin{equation*}
    \m C^\eta =
    \begin{pmatrix}
      \m C^{\eta^a} & \m C^{\eta^a, \eta^b}\\
      \m C^{\eta^b, \eta^a} & \m C^{\eta^b}
    \end{pmatrix}
    =
    \frac{n}{n-1} \m I_{n} \otimes \m K^\eta_{\m \theta_\eta}(\tilde{\m t}, \tilde{\m t}) -
    \frac{1}{n-1} \m 1_{n,n} \otimes \m K^\eta_{\m \theta_\eta}(\tilde{\m t}, \tilde{\m t})
  \end{equation*}
  where
  \begin{align*}
    \m C^{\eta^a} &= \frac{n}{n-1} \m I_{n_a} \otimes \m K^\eta_{\m \theta_\eta}(\tilde{\m t}, \tilde{\m t}) -
                  \frac{1}{n-1} \m 1_{n_a,n_a} \otimes \m K^\eta_{\m \theta_\eta}(\tilde{\m t}, \tilde{\m t})\\
    \m C^{\eta^b} &= \frac{n}{n-1} \m I_{n_b} \otimes \m K^\eta_{\m \theta_\eta}(\tilde{\m t}, \tilde{\m t}) -
                  \frac{1}{n-1} \m 1_{n_b,n_b} \otimes \m K^\eta_{\m \theta_\eta}(\tilde{\m t}, \tilde{\m t})\\
    \m C^{\eta^b, \eta^a} &= - \frac{1}{n-1} \m 1_{n_b,n_a} \otimes \m
                            K^\eta_{\m \theta_\eta}(\tilde{\m t}, \tilde{\m t})
  \end{align*}
  and with the simplifications
  \begin{align*}
      \m C^{\eta^a,y^b} \m S^{-1} (\m C^{\eta^a,y^b})^T
      &=
        \frac{1}{(n-1)^2}
        \m 1_{n_a, n_a} \otimes \m K^\eta_{\m \theta_\eta}(\tilde{\m t}, \m t^b)
        \m S^{-1} \m K^\eta_{\m \theta_\eta}(\m t^b, \tilde{\m t})\\
      \m C^{\eta^a,y^a} \m A^{-1} \m C ^T \m S^{-1} \m C\m A^{-1} (\m
      C^{\eta^a,y^a})^T &=
                          \left(
                          \frac{n-n_a}{n-1}
                          \right)^2
                          \m 1_{n_a,n_a} \otimes\\
      &\qquad \m K^\eta_{\m \theta_\eta}(\tilde{\m t}, \m t^a)
        \m A^{-1}_1 \m C^b \m S^{-1} (\m C^b)^T
        \m A^{-1}_1 \m K^\eta_{\m \theta_\eta}(\m t^a, \tilde{\m t})\\
      \m C^{\eta^a,y^a}\m A^{-1}\m C ^T \m S^{-1} (\m C^{\eta^a,y^b})^T
      &=
        -\frac{n - n_a}{(n-1)^2} \m 1_{n_a, n_a} \otimes
        \m K^\eta_{\m \theta_\eta}(\tilde{\m t}, \m t^a)\m A^{-1}_1 \m C^b \m S^{-1} \m K^\eta_{\m \theta_\eta}(\m
        t^b, \tilde{\m t})\\
      \m C^{\eta^a,y^b}\m S^{-1}\m C \m A^{-1} (\m C^{\eta^a,y^a})^T
      &=
        -\frac{n - n_a}{(n-1)^2} \m 1_{n_a, n_a} \otimes
        \m K^\eta_{\m \theta_\eta}(\tilde{\m t}, \m t^b) \m S^{-1} (\m C^b)^T\m A^{-1}_1 \m K^\eta_{\m \theta_\eta}(\m
        t^a, \tilde{\m t})\\
      \m C^{\eta^a,y^a} \m A^{-1} (\m C^{\eta^a,y^a})^T
      &=
        \left(
        \frac{n}{n-1}
        \right)^2 \m I_{n_a} \otimes \m K^\eta_{\m \theta_\eta}(\tilde{\m t}, \m t^a) \m A^{-1}_0
        \m K^\eta_{\m \theta_\eta}(\m t^a, \tilde{\m t})\\
      &\quad +
        \m 1_{n_a, n_a}\otimes\bigg[\frac{(n-n_a)^2}{n_a(n-1)^2} \m K^\eta_{\m \theta_\eta}(\tilde{\m t}, \m t^a) \m A^{-1}_1
        \m K^\eta_{\m \theta_\eta}(\m t^a, \tilde{\m t})\\
      &\qquad\qquad\qquad - \frac{n^2}{n_a(n-1)^2} \m K^\eta_{\m \theta_\eta}(\tilde{\m t}, \m t^a) \m
        A^{-1}_0 \m K^\eta_{\m \theta_\eta}(\m t^a, \tilde{\m t}) \bigg]\\
      \m C^{\eta^b,y^a} \m A^{-1} (\m C^{\eta^a,y^a})^T
      &=
        - \frac{n-n_a}{(n-1)^2} \m 1_{n_b,n_a}\otimes \m K^\eta_{\m \theta_\eta}(\tilde{\m t}, \m
        t^a)\m A^{-1}_1 \m K^\eta_{\m \theta_\eta}(\m t^a, \tilde{\m t})\\
      \m C^{\eta^b,y^a} \m A^{-1} \m C ^T \m S^{-1} \m C \m A^{-1} (\m
      C^{\eta^a,y^a})^T
      &=
        - \frac{n_a(n-n_a)}{(n-1)^2} \m 1_{n_b, n_a} \otimes\\
        &\qquad \m K^\eta_{\m \theta_\eta}(\tilde{\m t}, \m t^a) \m A^{-1}_1
        \m C^b \m S^{-1} (\m C^b)^T \m A^{-1}_1 \m K^\eta_{\m \theta_\eta}(\m t^a, \tilde{\m
        t})\\
      - \m C^{\eta^b,y^a}\m A^{-1} \m C ^T \m S^{-1} (\m C^{\eta^a,y^b})^T
      &=
        -\frac{n_a}{(n-1)^2} \m 1_{n_b, n_a} \otimes \m K^\eta_{\m \theta_\eta}(\tilde{\m t},
        \m t^a) \m A^{-1}_1 \m C^b \m S^{-1} \m K^\eta_{\m \theta_\eta}(\m t^b, \tilde{\m t}) \\
      \m S^{-1}\m C \m A^{-1} (\m C^{\eta^a,y^a})^T
      &=
        \frac{n-n_a}{n-1}\m 1_{n_a}^T \otimes \m S^{-1}(\m C^b)^T \m A^{-1}_1 \m
        K^\eta_{\m \theta_\eta}(\m t^a, \tilde{\m t})\\
      \m S^{-1} (\m C^{\eta^a,y^b})^T
      &=
        -\frac{1}{n-1}\m 1_{n_a}^T \otimes \m S^{-1} \m K^\eta_{\m \theta_\eta}(\m t^b, \tilde{\m
        t})\\
      \m C^{\eta^b,y^a} \m A^{-1} (\m C^{\eta^b,y^a})^T
      &=
        \frac{n_a}{(n-1)^2} \m 1_{n_b, n_b} \otimes \m K^\eta_{\m \theta_\eta}(\tilde{\m t}, \m t^a) \m
        A^{-1}_1 \m K^\eta_{\m \theta_\eta}(\m t^a, \tilde{\m t})\\
      \m C^{\eta^b,y^a} \m A^{-1} \m C ^T \m S^{-1} \m C \m A^{-1} (\m
      C^{\eta^b,y^a})^T
      &=
        \left(
        \frac{n_a}{n-1}
        \right)^2
        \m 1_{n_b,n_b}
        \otimes\\
      &\qquad
        \m K^\eta_{\m \theta_\eta}(\tilde{\m t}, \m t^a)\m A^{-1}_1
        \m C^b \m S^{-1} (\m C^b)^T
        \m A^{-1}_1
        \m K^\eta_{\m \theta_\eta}(\m t^a, \tilde{\m t})\\
      -\m C^{\eta^b,y^a} \m A^{-1} \m C^T \m S^{-1}
      &=
        \frac{n_a}{n-1} \m 1_{n_b} \otimes \m K^\eta_{\m \theta_\eta}(\tilde{\m t}, \m t^a)
        \m A^{-1}_1
        \m C^b
        \m S^{-1}.
  \end{align*}
  Note that $n - n_a = n_b$, allowing to further rewrite these simplifications.

\subsection*{Proof of \cref{prop:postirreg}}

Similar to the proof of \cref{prop:postmureg}, the proof of \cref{prop:postirreg} has three parts. First we prove the expressions for the posterior distribution of $\m \mu(\tilde{\m t})$, then we prove the expressions for the posterior distribution of $\m \eta(\tilde{\m t})$, and finally we prove the expression for the cross-covariance of $\m \mu(\tilde{\m t})$ and $\m \eta(\tilde{\m t})$ in the posterior distribution.

\subsubsection*{Posterior distribution of $\m \mu(\tilde{\m t})$}

  From the model specification in \cref{mod:eta} it directly follows that
  \begin{equation*}
    \begin{pmatrix}
      \m y(\m t)\\
      \m \mu(\tilde{\m t})
    \end{pmatrix}
    \sim
    N
    \left(
      \begin{pmatrix}
        \m 0_N\\
        \m 0_{J_p}
      \end{pmatrix}
      ,
      \begin{pmatrix}
        \m \Sigma_{\m \Theta} & (\m C^{\mu, y})^T\\
        \m C^{\mu, y} & \m K^\mu_{\m \theta_\mu}(\tilde{\m t}, \tilde{\m t})
      \end{pmatrix}
    \right)
  \end{equation*}
  where we will now find $\m C^{\mu, y}$ which we split into two blocks
  \begin{equation*}
    \m C^{\mu, y} =
    \begin{pmatrix}
      \m C^{\mu, a} & \m C^{\mu, b}
    \end{pmatrix}.
  \end{equation*}
  The model specification \cref{mod:eta} gives that
  \begin{equation*}
    \cov(\m \mu(\tilde{\m t}), \m y^a_i(\m t^a) \mid \m \Theta) =
    \cov(\m \mu(\tilde{\m t}), \m \mu(\m t^a)\mid \m \Theta) =
    \m K^\mu_{\m \theta_\mu}(\tilde{\m t}, \m t^a)
  \end{equation*}
  for all $i \in \{1, \ldots, n_a\}$, so
  $\m C^{\mu, a} = \m 1_{n_a}^T \otimes \m K^\mu_{\m \theta_\mu}(\tilde{\m t}, \m t^a)$. Furthermore it gives that
  \begin{equation*}
    \cov(\m \mu(\tilde{\m t}), \m y^b_i(\m t^b_i) \mid \m \Theta) =
    \cov(\m \mu(\tilde{\m t}), \m \mu(\m t^b_i) \mid \m \Theta) = 
    \m K^\mu_{\m \theta_\mu}(\tilde{\m t}, \m t^b_i),
  \end{equation*}
  so $\m C^{\mu, b} = \m K^\mu_{\m \theta_\mu}(\tilde{\m t}, \m t^b)$. By conditioning in the multivariate normal distribution, it follows that
  \begin{equation*}
    \m \mu(\tilde{\m t}) \mid \m y(\m t), \m \Theta
    \sim
    N
    \left(
      \m \mu_{\m \mu(\tilde{\m t}) \mid \m y(\m t), \m \Theta},
      \m \Sigma_{\m \mu(\tilde{\m t}) \mid \m y(\m t), \m \Theta}
    \right)
  \end{equation*}
  with
  \begin{align*}
    \m \mu_{\m \mu(\tilde{\m t}) \mid \m y(\m t), \m \Theta}
    &=
      \m C^{\mu, y}\m \Sigma_{\m \Theta}^{-1}\m y(\m t)\\
    \m \Sigma_{\m \mu(\tilde{\m t}) \mid \m y(\m t), \m \Theta}
    &=
      \m K^\mu_{\m \theta_\mu}(\tilde{\m t}, \tilde{\m t}) - \m C^{\mu, y} \m \Sigma_{\m \Theta}^{-1}(\m C^{\mu, y})^T.
  \end{align*}
  Plugging in the expressions above and the expression for $\m \Sigma_{\m \Theta}^{-1}\m y(\m t)$ given in \cref{prop:irreg-loglik} gives that
  \begin{equation*}
    \begin{split}
      \m \mu_{\m \mu(\tilde{\m t}) \mid \m y(\m t), \m \Theta}
      &=
        \m C^{\mu, y}\m \Sigma_{\m \Theta}^{-1}\m y(\m t)\\
      &= \m C^{\mu, a}\m P_1 \m y(\m t) + \m C^{\mu, b}\m P_2 \m y(\m t)\\
      &= (\m 1_{n_a}^T \otimes \m K^\mu_{\m \theta_\mu}(\tilde{\m t}, \m t^a))\m P_1 \m y(\m
        t)
        +
        \m K^\mu_{\m \theta_\mu}(\tilde{\m t}, \m t^b)\m P_2 \m y(\m t).
    \end{split}
  \end{equation*}
  Using \citet[11.16(b)]{seber2008matrix} for the first term, we get
  \begin{equation*}
    \begin{split}
      \m \mu_{\m \mu(\tilde{\m t}) \mid \m y(\m t), \m \Theta} &=
      \vc
      \left(
      \m K^\mu_{\m \theta_\mu}(\tilde{\m t}, \m t^a)
      \vc^{-1}_{J_a,n_a}(\m P_1\m y(\m t))
      \m 1_{n_a}
                 \right)
      +
        \m K^\mu_{\m \theta_\mu}(\tilde{\m t}, \m t^b)\m P_2 \m y(\m t)\\
      &=
      \m K^\mu_{\m \theta_\mu}(\tilde{\m t}, \m t^a)
      \vc^{-1}_{J_a,n_a}(\m P_1\m y(\m t))
        \m 1_{n_a}
        +
        \m K^\mu_{\m \theta_\mu}(\tilde{\m t}, \m t^b)\m P_2 \m y(\m t),
    \end{split}
  \end{equation*}  
  since the first term is already a column vector. This completes the proof of the simplified posterior mean of $\m \mu(\tilde{\m t})$.
  
  For $\m \Sigma_{\m \mu(\tilde{\m t}) \mid \m y(\m t), \m \Theta}$, plugging in the expression for $\m \Sigma_{\m \Theta}^{-1}$ and simplifying gives that
  \begin{equation*}
    \begin{split}
      \m \Sigma_{\m \mu(\tilde{\m t}) \mid \m y(\m t), \m \Theta}
      &=
        \m K^\mu_{\m \theta_\mu}(\tilde{\m t}, \tilde{\m t}) - \m C^{\mu, y} \m \Sigma_{\m \Theta}^{-1}(\m C^{\mu, y})^T\\
      &= \m K^\mu_{\m \theta_\mu}(\tilde{\m t}, \tilde{\m t}) -
        \begin{pmatrix}
          \m C^{\mu, a} & \m C^{\mu, b}
        \end{pmatrix}
        \begin{pmatrix}
          \m A^{-1} + \m A^{-1}\m C^T\m S^{-1}\m C\m A^{-1} & - \m A^{-1}\m C^T\m S^{-1}\\
          -\m S^{-1}\m C\m A^{-1} & \m S^{-1}
        \end{pmatrix}
        \begin{pmatrix}
          (\m C^{\mu, a})^T \\ (\m C^{\mu, b})^T
        \end{pmatrix}\\
      &=                     \m K^\mu_{\m \theta_\mu}(\tilde{\m t}, \tilde{\m t})\\
                    &\quad - \m C^{\mu, a} \m A^{-1} (\m C^{\mu, a})^T\\
             &\quad - \m C^{\mu, a}\m A^{-1} \m C^T \m S^{-1} \m C \m A^{-1}(\m
               C^{\mu, a})^T\\
             &\quad + \m C^{\mu, a}\m A^{-1}\m C^T\m S^{-1}(\m C^{\mu, b})^T\\
             &\quad + \m C^{\mu, b} \m S^{-1}\m C\m A^{-1}(\m C^{\mu, a})^T\\
             &\quad - \m C^{\mu, b}\m S^{-1} (\m C^{\mu, b})^T.
    \end{split}
  \end{equation*}
  Using the expression for $\m A^{-1}$ from the proof of \cref{prop:irreg-loglik} we get that
  \begin{equation*}
    \begin{split}
      \m A^{-1} &(\m C^{\mu, a})^T\\
                &=
      (\m I_{n_a} \otimes \m A_0^{-1} + \frac{1}{n_a} \m 1_{n_a, n_a} \otimes (\m
      A_1^{-1} - \m A_0^{-1}))(\m 1_{n_a} \otimes \m K^\mu_{\m \theta_\mu}(\m t^a, \tilde{\m
                  t}))\\
                &= \m I_{n_a} \m 1_{n_a} \otimes \m A_0^{-1} \m K^\mu_{\m \theta_\mu}(\m t^a,
                  \tilde{\m t}) +
      \frac{1}{n_a}\m 1_{n_a, n_a}\m 1_{n_a} \otimes (\m A_1^{-1} - \m
                  A_0^{-1})\m K^\mu_{\m \theta_\mu}(\m t^a, \tilde{\m t})\\
                &= \m 1_{n_a} \otimes \m A_0^{-1}\m K^\mu_{\m \theta_\mu}(\m t^a, \tilde{\m t})
                  + \m 1_{n_a} \otimes (\m A_1^{-1} - \m A_0^{-1}) \m K^\mu_{\m \theta_\mu}(\m t^a,
                  \tilde{\m t})\\
                &= \m 1_{n_a} \otimes \m A_1^{-1} \m K^\mu_{\m \theta_\mu}(\m t^a, \tilde{\m t})
    \end{split}
  \end{equation*}
  and thus finally
  \begin{equation*}
    \begin{split}
      \m C^{\mu, a} &\m A^{-1} (\m C^{\mu, a})^T\\
                    &= (\m 1_{n_a}^T \otimes \m K^\mu_{\m \theta_\mu}(\tilde{\m t}, \m t^a))
                      (\m 1_{n_a} \otimes \m A_1^{-1} \m K^\mu_{\m \theta_\mu}(\m t^a, \tilde{\m
                      t}))\\
                    &= \m 1_{n_a}^T\m 1_{n_a} \otimes \m K^\mu_{\m \theta_\mu}(\tilde{\m t}, \m t^a)\m
                      A_1^{-1} \m K^\mu_{\m \theta_\mu}(\m t^a, \tilde{\m t})\\
                    &= n_a \m K^\mu_{\m \theta_\mu}(\tilde{\m t}, \m t^a)\m
                      A_1^{-1} \m K^\mu_{\m \theta_\mu}(\m t^a, \tilde{\m t})
    \end{split}
  \end{equation*}
  and
  \begin{equation*}
    \begin{split}
      \m C&\m A^{-1} (\m C^{\mu, a})^T\\
                    &= (\m 1_{n_a}^T \otimes (\m C^b)^T)
                      (\m 1_{n_a} \otimes \m A_1^{-1} \m K^\mu_{\m \theta_\mu}(\m t^a, \tilde{\m
                      t}))\\
          &= \m 1_{n_a}^T\m 1_{n_a} \otimes (\m C^b)^T\m A_1^{-1} \m K^\mu_{\m \theta_\mu}(\m t^a,
            \tilde{\m t})\\
          &= n_a(\m C^b)^T\m A_1^{-1} \m K^\mu_{\m \theta_\mu}(\m t^a, \tilde{\m t}),
    \end{split}
  \end{equation*}
  which completes the proof of the simplified posterior covariance of $\m \mu(\tilde{\m t})$.

\subsubsection*{Posterior distribution of $\m \eta(\tilde{\m t})$}

  It follows from the model specification in \cref{mod:eta} that
  \begin{equation*}
    \begin{pmatrix}
      \m y(\m t)\\
      \m \eta(\tilde{\m t})
    \end{pmatrix}
    \mid \m \Theta
    \sim
    N
    \left(
      \begin{pmatrix}
        \m 0_N\\
        \m 0_{nJ_p}
      \end{pmatrix}
      ,
      \begin{pmatrix}
        \m \Sigma_{\m \Theta} & (\m C^{\eta, y})^T\\
        \m C^{\eta, y} & \m C^\eta
      \end{pmatrix}
    \right)
  \end{equation*}
  whereby conditioning in the multivariate normal distribution gives that
  \begin{equation*}
    \m \eta(\tilde{\m t}) \mid \m y(\m t), \m \Theta
    \sim
    N
    \left(
      \m \mu_{\m \eta(\tilde{\m t}) \mid \m y(\m t), \m \Theta},
      \m \Sigma_{\m \eta(\tilde{\m t}) \mid \m y(\m t), \m \Theta}
    \right)
  \end{equation*}
  with
  \begin{equation*}
    \m \mu_{\m \eta(\tilde{\m t}) \mid \m y(\m t), \m \Theta}
    =
    \m C^{\eta, y} \m \Sigma_{\m \Theta}^{-1} \m y(\m t)
  \end{equation*}
  and
  \begin{equation}
    \label{eq:Sigmaetastart}
    \m \Sigma_{\m \eta(\tilde{\m t}) \mid \m y(\m t), \m \Theta} = \m C^\eta - \m C^{\eta, y}\m \Sigma_{\m \Theta}^{-1} (\m C^{\eta, y})^T.
  \end{equation}
  We will now find expressions for
  \begin{equation*}
    \m C^{\eta, y} =
    \begin{pmatrix}
      \m C^{\eta^a, y^a} & \m C^{\eta^a, y^b}\\
      \m C^{\eta^b, y^a} & \m C^{\eta^b, y^b}
    \end{pmatrix}.
  \end{equation*}
  Exactly as in the completely regular case in the proof of \cref{prop:postmureg}, we find that
  \begin{equation*}
    \m C^{\eta^a, y^a} = \frac{n}{n-1} \m I_{n_a} \otimes \m K^\eta_{\m \theta_\eta}(\tilde{\m
      t}, \m t^a) - \frac{1}{n-1} \m 1_{n_a, n_a} \otimes \m K^\eta_{\m \theta_\eta}(\tilde{\m
      t}, \m t^a).
  \end{equation*}
  We have
  \begin{equation*}
      \cov(\m \eta^a_i(\tilde{\m t}), \m y^b_j(\m t^b_j) \mid \m \Theta)
           = \cov(\m \eta^a_i(\tilde{\m t}), \m \eta^b_j(\m
             t^b_j) \mid \m \Theta)
           = -\frac{1}{n-1} \m K^\eta_{\m \theta_\eta}(\tilde{\m t}, \m t^b_j),
  \end{equation*}
  so
  \begin{equation*}
    \m C^{\eta^a, y^b} = \m 1_{n_a} \otimes
    \left(-\frac{1}{n-1} \m K^\eta_{\m \theta_\eta}(\tilde{\m t}, \m t^b)\right).
  \end{equation*}
  Similarly, we have
  \begin{equation*}
    \cov(\m \eta^b_i(\tilde{\m t}), \m y^a_j(\m t^a) \mid \m \Theta) = - \frac{1}{n-1} \m K^\eta_{\m \theta_\eta}(\tilde{\m t}, \m t^a),
  \end{equation*}
  so
  \begin{equation*}
    \m C^{\eta^b, y^a} =
    \m 1_{n_b, n_a} \otimes
    \left(
      - \frac{1}{n-1} \m K^\eta_{\m \theta_\eta}(\tilde{\m t}, \m t^a)
    \right).
  \end{equation*}
  The expression for $\m C^{\eta^b, y^b}$ follows from the covariances
  \begin{align*}
    \cov(\m \eta^b_i(\tilde{\m t}), \m y^b_i(\m t^b_i) \mid \m \Theta)
    &=
      \cov(\m \eta^b_i(\tilde{\m t}), \m \eta^b_i(\m t^b_i) \mid \m \Theta)
    =
    \m K^\eta_{\m \theta_\eta}(\tilde{\m t}, \m t^b_i)\\
    \cov(\m \eta^b_i(\tilde{\m t}), \m y^b_j(\m t^b_j) \mid \m \Theta)
    &=
      \cov(\m \eta^b_i(\tilde{\m t}), \m \eta^b_j(\m t^b_j) \mid \m \Theta)
      =
      -\frac{1}{n-1} \m K^\eta_{\m \theta_\eta}(\tilde{\m t}, \m t^b_j).
  \end{align*}
  Collecting the above, we now have
  \begin{equation*}
    \begin{split}
      \m \mu_{\m \eta(\tilde{\m t}) \mid \m y(\m t), \m \Theta} &= \m C^{\eta, y}\m \Sigma_{\m \Theta}^{-1} \m y(\m t)\\
                &=
                  \begin{pmatrix}
                    \m C^{\eta^a, y^a} & \m C^{\eta^a, y^b}\\
                    \m C^{\eta^b, y^a} & \m C^{\eta^b, y^b}
                  \end{pmatrix}
                  \begin{pmatrix}
                    \m P_1 \m y(\m t)\\
                    \m P_2 \m y(\m t)
                  \end{pmatrix}\\
                &=
                  \begin{pmatrix}
                    \m C^{\eta^a, y^a}\m P_1 \m y(\m t) + \m C^{\eta^a,
                    y^b}\m P_2 \m y(\m t)\\
                    \m C^{\eta^b, y^a}\m P_1 \m y(\m t) + \m C^{\eta^b, y^b}\m P_2 \m y(\m t)
                  \end{pmatrix}
                  =:
                  \begin{pmatrix}
                      \m m_{\eta_a}\\
                      \m m_{\eta_b}
                  \end{pmatrix}
    \end{split}
  \end{equation*}
  which gives
  \begin{equation}
    \label{eq:mapartsimp}
    \begin{split}
     \m m_{\eta^a} = \frac{1}{n-1}\Big[&
          (n \m I_{n_a} \otimes \m K^\eta_{\m \theta_\eta}(\tilde{\m t}, \m t^a)) \m P_1
                     \m y(\m t)\\
                     &-(\m 1_{n_a, n_a} \otimes \m
                    K^\eta_{\m \theta_\eta}(\tilde{\m t}, \m t^a)) \m P_1 \m y(\m t)\\
                    &-
                    \left(\m 1_{n_a} \otimes
                         \m K^\eta_{\m \theta_\eta}(\tilde{\m t}, \m t^b)
                         \right)
                         \m P_2 \m y(\m t)\Big]
    \end{split}
  \end{equation}
  and
  \begin{equation}
    \label{eq:mbpartsimp}
      \m m_{\eta^b} = -\frac{1}{n-1}
             \left(\m 1_{n_b, n_a} \otimes \m K^\eta_{\m \theta_\eta}(\tilde{\m t}, \m t^a)
             \right)
             \m P_1 \m y(\m t)
           + \m C^{\eta^b, y^b}\m P_2 \m y(\m t),
  \end{equation}
  which we now further simplify. Using \citet[11.16(b)]{seber2008matrix}), we get
  \begin{equation*}
    \begin{split}
      (n &\m I_{n_a} \otimes \m K^\eta_{\m \theta_\eta}(\tilde{\m t}, \m t^a)) \m P_1
                   \m y(\m t)\\
                 &= n \vc\left(\m K^\eta_{\m \theta_\eta}(\tilde{\m t}, \m t^a) \vc^{-1}_{J_a,
                   n_a}\left(\m P_1 \m y(\m t)\right) \m I_{n_a}^T\right)\\
              &= n \vc\left(\m K^\eta_{\m \theta_\eta}(\tilde{\m t}, \m t^a) \vc^{-1}_{J_a,
                n_a}\left(\m P_1 \m y(\m t)\right)\right)
    \end{split}
  \end{equation*}
  and
  \begin{equation*}
    (-\m 1_{n_a, n_a} \otimes \m K^\eta_{\m \theta_\eta}(\tilde{\m t}, \m t^a)) \m P_1 \m
    y(\m t) =
    -\vc
    \left(
      \m K^\eta_{\m \theta_\eta}(\tilde{\m t}, \m t^a) \vc^{-1}_{J_a, n_a}(\m P_1 \m y(\m
      t)) \m 1_{n_a, n_a}
    \right).
  \end{equation*}
  We get
  \begin{equation*}
    \begin{split}
      \big(-\m 1_{n_a} &\otimes
                         \m K^\eta_{\m \theta_\eta}(\tilde{\m t}, \m t^b)
                         \big)
                         \m P_2 \m y(\m t)\\
                       &=
                         \left(
                         -\m 1_{n_a} \otimes \m K^\eta_{\m \theta_\eta}(\tilde{\m t}, \m t^b)
                         \right)
                         \left(
                         1 \otimes
                         \m P_2 \m y(\m t)
                         \right)\\
                       &= -\m 1_{n_a}1 \otimes \m K^\eta_{\m \theta_\eta}(\tilde{\m t}, \m t^b)
                         \m P_2 \m y(\m t)\\
                       &= -\m 1_{n_a} \otimes \m K^\eta_{\m \theta_\eta}(\tilde{\m t}, \m t^b)
                         \m P_2 \m y(\m t).
    \end{split}
  \end{equation*}
  Using \citet[11.16(b)]{seber2008matrix}), we get
  \begin{equation*}
    \left(
      \m 1_{n_b, n_a} \otimes \m K^\eta_{\m \theta_\eta}(\tilde{\m t}, \m t^a)
    \right)
    \m P_1 \m y(\m t) =
    \vc
    \left(
      \m K^\eta_{\m \theta_\eta}(\tilde{\m t}, \m t^a)\vc^{-1}_{J_a, n_a}(\m P_1 \m y(\m
      t))
      \m 1_{n_a, n_b}
    \right).
  \end{equation*}
  Plugging these simplifications into \cref{eq:mapartsimp} and \cref{eq:mbpartsimp} now
  gives the desired expressions for $\m \mu_{\m \eta(\tilde{\m t}) \mid \m y(\m t), \m \Theta}$.

  We now derive the simplifications for $\m \Sigma_{\m \eta(\tilde{\m t}) \mid \m y(\m t), \m \Theta}$. Writing
  \begin{equation*}
    \m C^\eta =
    \begin{pmatrix}
      \m C^{\eta^a} & \m C^{\eta^a, \eta^b}\\
      \m C^{\eta^b, \eta^a} & \m C^{\eta^b}
    \end{pmatrix}
  \end{equation*}
  and plugging the expression for $\m \Sigma_{\m \Theta}^{-1}$ into \cref{eq:Sigmaetastart} and multiplying out we
  get the desired expressions for
  $\m \Sigma_{\eta, 1, 1}, \m \Sigma_{\eta, 1, 2}, \m \Sigma_{\eta, 2, 1}$ and
  $\m \Sigma_{\eta, 2, 2}$. The expressions for
  $\m C^{\eta^a}, \m C^{\eta^b}, \m C^{\eta^b,\eta^a}$
  follow from the covariances
  \begin{equation*}
    \begin{split}
      \cov(\m \eta^a_i(\tilde{\m t}), \m \eta^a_i(\tilde{\m t}) \mid \m \Theta)
      &=
        \m K^\eta_{\m \theta_\eta}(\tilde{\m t}, \tilde{\m t})\\
      \cov(\m \eta^a_i(\tilde{\m t}), \m \eta^a_j(\tilde{\m t}) \mid \m \Theta)
      &=
        -\frac{1}{n-1}\m K^\eta_{\m \theta_\eta}(\tilde{\m t}, \tilde{\m t})\\
      \cov(\m \eta^b_i(\tilde{\m t}), \m \eta^b_i(\tilde{\m t}) \mid \m \Theta)
      &=
        \m K^\eta_{\m \theta_\eta}(\tilde{\m t}, \tilde{\m t})\\
      \cov(\m \eta^b_i(\tilde{\m t}), \m \eta^b_j(\tilde{\m t}) \mid \m \Theta)
      &=
        -\frac{1}{n-1}\m K^\eta_{\m \theta_\eta}(\tilde{\m t}, \tilde{\m t})
    \end{split}
  \end{equation*}
  for all $i \ne j$ and
  \begin{equation*}
    \cov(\m \eta^b_i(\tilde{\m t}), \m \eta^a_j(\tilde{\m t}) \mid \m \Theta)
    =
    -\frac{1}{n-1}\m K^\eta_{\m \theta_\eta}(\tilde{\m t}, \tilde{\m t})
  \end{equation*}
  for all $i, j$.
  
  We now derive the long list of simplifications. We get
  \begin{equation*}
    \begin{split}
      \m C^{\eta^a,y^b}
      & \m S^{-1} (\m C^{\eta^a,y^b})^T\\
      &=
        \left(
        -\frac{1}{n-1} \m 1_{n_a} \otimes \m K^\eta_{\m \theta_\eta}(\tilde{\m t}, \m t^b)
        \right)
        \m S^{-1}
        \left(
        -\frac{1}{n-1} \m 1_{n_a}^T \otimes (\m K^\eta_{\m \theta_\eta}(\tilde{\m t}, \m t^b))^T
        \right)\\
      &= \frac{1}{(n-1)^2}(\m 1_{n_a} \otimes \m K^\eta_{\m \theta_\eta}(\tilde{\m t}, \m t^b))\m
        S^{-1}(\m 1_{n_a}^T \otimes \m K^\eta_{\m \theta_\eta}(\m t^b, \tilde{\m t})) \\
      &= \frac{1}{(n-1)^2}(\m 1_{n_a} \otimes \m K^\eta_{\m \theta_\eta}(\tilde{\m t}, \m t^b))(1
        \otimes \m S^{-1})(\m 1_{n_a}^T \otimes \m K^\eta_{\m \theta_\eta}(\m t^b, \tilde{\m t}))
      \\
      &= \frac{1}{(n-1)^2}\m 1_{n_a} 1 \m 1_{n_a}^T \otimes
        \m K^\eta_{\m \theta_\eta}(\tilde{\m t}, \m t^b)\m S^{-1}\m K^\eta_{\m \theta_\eta}(\m t^b, \tilde{\m t}) \\
      &= \frac{1}{(n-1)^2}\m 1_{n_a,n_a} \otimes
        \m K^\eta_{\m \theta_\eta}(\tilde{\m t}, \m t^b)\m S^{-1}\m K^\eta_{\m \theta_\eta}(\m t^b, \tilde{\m t}).
    \end{split}
  \end{equation*}
  Using \cref{eq:AiCt}, we get that
  \begin{equation}
    \label{eq:CaaAiCt}
    \begin{split}
      \m C^{\eta^a,y^a}
      & \m A^{-1} \m C ^T\\
      &=
        \left(
        \frac{n}{n-1}\m I_{n_a} \otimes \m K^\eta_{\m \theta_\eta}(\tilde{\m t}, \m t^a) -
        \frac{1}{n-1} \m 1_{n_a,n_a} \otimes \m K^\eta_{\m \theta_\eta}(\tilde{\m t}, \m t^a)
        \right)
        \left(
        \m 1_{n_a} \otimes \m A^{-1}_1 \m C^b
        \right)\\
      &=
        \frac{n}{n-1}\m 1_{n_a} \otimes \m K^\eta_{\m \theta_\eta}(\tilde{\m t}, \m t^a)
        \m A^{-1}_1 \m C^b
        -
        \frac{n_a}{n-1} \m 1_{n_a} \otimes \m K^\eta_{\m \theta_\eta}(\tilde{\m t}, \m t^a)\m
        A^{-1}_1 \m C^b\\
      &=
        \frac{n-n_a}{n-1}\m 1_{n_a} \otimes \m K^\eta_{\m \theta_\eta}(\tilde{\m t}, \m t^a)\m
        A^{-1}_1 \m C^b.
    \end{split}
  \end{equation}
  Using \cref{eq:CaaAiCt}, we get that
  \begin{equation*}
    \begin{split}
      \m C^{\eta^a,y^a}& \m A^{-1} \m C ^T \m S^{-1} \m C\m A^{-1} (\m
                         C^{\eta^a,y^a})^T\\
                       &=
                         \left(
                         \frac{n-n_a}{n-1}\m 1_{n_a} \otimes \m K^\eta_{\m \theta_\eta}(\tilde{\m
                         t}, \m t^a) \m A^{-1}_1 \m C^b
                         \right)
                         (1 \otimes \m S^{-1})\\
                         &\qquad\left(
                         \frac{n-n_a}{n-1}\m 1_{n_a}^T \otimes
                         (\m C^b)^T \m A^{-1}_1 \m K^\eta_{\m \theta_\eta}(\m t^a, \tilde{\m t})
                         \right)
      \\
                       &=
                         \left(
                         \frac{n-n_a}{n-1}
                         \right)^2
                         \m 1_{n_a,n_a} \otimes \m K^\eta_{\m \theta_\eta}(\tilde{\m t}, \m t^a)
                         \m A^{-1}_1 \m C^b \m S^{-1} (\m C^b)^T
                         \m A^{-1}_1 \m K^\eta_{\m \theta_\eta}(\m t^a, \tilde{\m t})
    \end{split}
  \end{equation*}
  and
  \begin{equation*}
    \begin{split}
      \m C^{\eta^a,y^a}
      & \m A^{-1}\m C^T \m S^{-1} (\m C^{\eta^a,y^b})^T\\
      &=
        \left(
        \frac{n-n_a}{n-1} \m 1_{n_a}
        \otimes
        \m K^\eta_{\m \theta_\eta}(\tilde{\m t}, \m t^a)\m A^{-1}_1 \m C^b
        \right)
        (1 \otimes \m S^{-1})
        \left(
        -\frac{1}{n-1}\m 1_{n_a}^T \otimes \m K^\eta_{\m \theta_\eta}(\m t^b, \tilde{\m t})
        \right)\\
      &=
        - \frac{n-n_a}{(n-1)^2}\m 1_{n_a}1 \m 1_{n_a}^T \otimes \m
        K^\eta_{\m \theta_\eta}(\tilde{\m t}, \m t^a) \m A^{-1}_1 \m C^b \m S^{-1} \m K^\eta_{\m \theta_\eta}(\m
        t^b, \tilde{\m t})\\
      &=
        - \frac{n-n_a}{(n-1)^2}\m 1_{n_a,n_a} \otimes \m
        K^\eta_{\m \theta_\eta}(\tilde{\m t}, \m t^a) \m A^{-1}_1 \m C^b \m S^{-1} \m K^\eta_{\m \theta_\eta}(\m
        t^b, \tilde{\m t})
    \end{split}
  \end{equation*}
  as well as its transpose
  \begin{equation*}
    \m C^{\eta^a,y^b} \m S^{-1} \m C \m A^{-1} (\m C^{\eta^a,y^a})^T
    =
    - \frac{n-n_a}{(n-1)^2} \m 1_{n_a, n_a}
    \otimes
    \m K^\eta_{\m \theta_\eta}(\tilde{\m t}, \m t^b) \m S^{-1}
    (\m C^b)^T \m A^{-1}_1 \m K^\eta_{\m \theta_\eta}(\m t^a, \tilde{\m t}).
  \end{equation*}
  We find
  \begin{align*}
      \m C^{\eta^a,y^a}
      & \m A^{-1} (\m C^{\eta^a,y^a})^T\\
      &=
        \left(\frac{n}{n-1} \m I_{n_a} \otimes \m K^\eta_{\m \theta_\eta}(\tilde{\m t}, \m t^a)
        -
        \frac{1}{n-1} \m 1_{n_a,n_a} \otimes \m K^\eta_{\m \theta_\eta}(\tilde{\m t}, \m
        t^a)\right)\\
        &\quad\left(
        \m I_{n_a} \otimes \m A^{-1}_0 +
        \frac{1}{n_a}\m 1_{n_a,n_a}
        \otimes
        (\m A^{-1}_1 - \m A^{-1}_0)
        \right)\\
        &\quad\left(
        \frac{n}{n-1} \m I_{n_a} \otimes \m K^\eta_{\m \theta_\eta}(\m t^a, \tilde{\m t})
        -
        \frac{1}{n-1} \m 1_{n_a,n_a}
        \otimes
        \m K^\eta_{\m \theta_\eta}(\m t^a, \tilde{\m t})
          \right)\\
      &= 
        \left(
        \frac{n}{n-1}
        \right)^2
        \m I_{n_a}
        \otimes
        \m K^\eta_{\m \theta_\eta}(\tilde{\m t}, \m t^a)
        \m A^{-1}_0
        \m K^\eta_{\m \theta_\eta}(\m t^a, \tilde{\m t})\\
        &\quad+
        \frac{n^2}{n_a(n-1)^2}
        \m 1_{n_a,n_a} \otimes
        \m K^\eta_{\m \theta_\eta}(\tilde{\m t}, \m t^a)(\m A^{-1}_1 - \m A^{-1}_0)\m K^\eta_{\m \theta_\eta}(\m
          t^a, \tilde{\m t})\\
      &\quad- \frac{n}{(n-1)^2} \m 1_{n_a,n_a} \otimes \m K^\eta_{\m \theta_\eta}(\tilde{\m t}, \m
        t^a)
        \m A^{-1}_0 \m K^\eta_{\m \theta_\eta}(\m t^a, \tilde{\m t})\\
      &\quad- \frac{n}{n_a(n-1)^2}n_a \m 1_{n_a, n_a}
        \otimes
        \m K^\eta_{\m \theta_\eta}(\tilde{\m t}, \m t^a)
        (\m A^{-1}_1 - \m A^{-1}_0)
        \m K^\eta_{\m \theta_\eta}(\m t^a, \tilde{\m t})\\
      &\quad- \frac{n}{(n-1)^2}\m 1_{n_a,n_a} \otimes
        \m K^\eta_{\m \theta_\eta}(\tilde{\m t}, \m t^a)\m A^{-1}_0 \m K^\eta_{\m \theta_\eta}(\m t^a, \tilde{\m
        t})\\
      &\quad- \frac{n}{n_a(n-1)^2}n_a \m 1_{n_a,n_a} \otimes
        \m K^\eta_{\m \theta_\eta}(\tilde{\m t}, \m t^a)(\m A^{-1}_1 - \m A^{-1}_0) \m K^\eta_{\m \theta_\eta}(\m
        t^a, \tilde{\m t})\\
      &\quad+ \frac{1}{(n-1)^2} n_a \m 1_{n_a,n_a} \otimes
        \m K^\eta_{\m \theta_\eta}(\tilde{\m t}, \m t^a) \m A^{-1}_0 \m K^\eta_{\m \theta_\eta}(\m t^a, \tilde{\m
        t})\\
      &\quad+ \frac{1}{n_a(n-1)^2}n_a^2 \m 1_{n_a,n_a}
        \otimes
        \m K^\eta_{\m \theta_\eta}(\tilde{\m t}, \m t^a)(\m A^{-1}_1 - \m A^{-1}_0) \m K^\eta_{\m \theta_\eta}(\m
        t^a, \tilde{\m t})\\
      &=
        \left(
        \frac{n}{n-1}
        \right)^2
        \m I_{n_a}
        \otimes
        \m K^\eta_{\m \theta_\eta}(\tilde{\m t}, \m t^a)\m A^{-1}_0 \m K^\eta_{\m \theta_\eta}(\m t^a, \tilde{\m
        t})\\
      &\quad+ 
        \left[
        \m 1_{n_a,n_a} \otimes \m K^\eta_{\m \theta_\eta}(\tilde{\m t}, \m t^a) \m A^{-1}_1 \m
        K^\eta_{\m \theta_\eta}(\m t^a, \tilde{\m t})
        \right]\\
        &\qquad \cdot \left(
        \frac{n^2}{n_a(n-1)^2} -
        \frac{n}{(n-1)^2} -
        \frac{n}{(n-1)^2} +
        \frac{n_a}{(n-1)^2}
          \right)\\
      &\quad+
        \left[
        \m 1_{n_a,n_a} \otimes
        \m K^\eta_{\m \theta_\eta}(\tilde{\m t}, \m t^a) \m A^{-1}_0
        \m K^\eta_{\m \theta_\eta}(\m t^a, \tilde{\m t})
        \right]\\
      &\qquad\cdot 
        \bigg(
        -\frac{n^2}{n_a(n-1)^2}-
        \frac{n}{(n-1)^2}+
        \frac{n}{(n-1)^2}-
        \frac{n}{(n-1)^2}+
      \\&\quad\quad\quad
        \frac{n}{(n-1)^2}+
        \frac{n_a}{(n-1)^2}-
        \frac{n_a}{(n-1)^2}
      \bigg)\\
      &=
        \left(\frac{n}{n-1}\right)^2 \m I_{n_a} \otimes
        \m K^\eta_{\m \theta_\eta}(\tilde{\m t}, \m t^a) \m A^{-1}_0 \m K^\eta_{\m \theta_\eta}(\m t^a, \tilde{\m
        t})
        \\&\quad+
        \m 1_{n_a,n_a}\otimes \bigg(
        \frac{(n-n_a)^2}{n_a(n-1)^2}
        \m K^\eta_{\m \theta_\eta}(\tilde{\m t}, \m t^a)\m A^{-1}_1 \m K^\eta_{\m \theta_\eta}(\m t^a, \tilde{\m
        t})\\
        &\quad\quad\quad\qquad\qquad-
        \frac{n^2}{n_a(n-1)^2} \m K^\eta_{\m \theta_\eta}(\tilde{\m t}, \m t^a)\m A^{-1}_0
        \m K^\eta_{\m \theta_\eta}(\m t^a, \tilde{\m t})
        \bigg).
  \end{align*}
  We find
  \begin{equation}
    \label{eq:AiCaat}
    \begin{split}
      \m A^{-1}
      &(\m C^{\eta^a,y^a})^T\\
      &=
        \left(
        \m I_{n_a} \otimes \m A^{-1}_0 + \frac{1}{n_a} \m 1_{n_a,n_a} \otimes
        (\m A^{-1}_1 - \m A^{-1}_0)
        \right)\\
        &\quad\left(
        \frac{n}{n-1} \m I_{n_a} \otimes \m K^\eta_{\m \theta_\eta}(\m t^a, \tilde{\m t})
        -
        \frac{1}{n-1}\m 1_{n_a,n_a} \otimes \m K^\eta_{\m \theta_\eta}(\m t^a, \tilde{\m t})
          \right)\\
      &= \frac{n}{n-1}\m I_{n_a} \otimes \m A^{-1}_0 \m K^\eta_{\m \theta_\eta}(\m t^a, \tilde{\m
        t})
      \\&\quad-
      \frac{1}{n-1} \m 1_{n_a,n_a} \otimes \m A^{-1}_0 \m K^\eta_{\m \theta_\eta}(\m t^a,
      \tilde{\m t})
      \\&\quad+
      \frac{n}{n_a(n-1)} \m 1_{n_a,n_a} \otimes
      (\m A^{-1}_1 - \m A^{-1}_0) \m K^\eta_{\m \theta_\eta}(\m t^a, \tilde{\m t})
      \\&\quad- \frac{1}{n_a(n-1)}n_a \m 1_{n_a,n_a} \otimes
      (\m A^{-1}_1 - \m A^{-1}_0) \m K^\eta_{\m \theta_\eta}(\m t^a, \tilde{\m t})
      \\&=
      \frac{n}{n-1}\m I_{n_a} \otimes \m A^{-1}_0 \m K^\eta_{\m \theta_\eta}(\m t^a, \tilde{\m
      t})
      \\&\quad+
      \m 1_{n_a,n_a} \otimes
      \left(
      -
      \frac{n}{n_a(n-1)} \m A^{-1}_0 \m K^\eta_{\m \theta_\eta}(\m t^a, \tilde{\m t})
      +
      \frac{n-n_a}{n_a(n-1)}\m A^{-1}_1 \m K^\eta_{\m \theta_\eta}(\m t^a, \tilde{\m t})
      \right)
    \end{split}
  \end{equation}
  and
  \begin{equation}
    \label{eq:AiCtSi}
    \begin{split}
      \m A^{-1} \m C^T \m S^{-1}
      =
      (\m 1_{n_a} \otimes \m A^{-1}_1 \m C^b)(1 \otimes \m S^{-1})
      =
      \m 1_{n_a} \otimes \m A^{-1}_1 \m C^b \m S^{-1}.
    \end{split}
  \end{equation}
  Using \cref{eq:AiCaat}, we find
  \begin{equation*}
    \begin{split}
      \m C^{\eta^b,y^a}
      & \m A^{-1}
        (\m C^{\eta^a,y^a})^T\\
      &=
        \left(
        - \frac{1}{n-1} \m 1_{n_b,n_a} \otimes
        \m K^\eta_{\m \theta_\eta}(\tilde{\m t}, \m t^a)
        \right)
        \m A^{-1} (\m C^{\eta^a,y^a})^T\\
      &=
        -\frac{n}{(n-1)^2} \m 1_{n_b,n_a} \otimes
        \m K^\eta_{\m \theta_\eta}(\tilde{\m t}, \m t^a) \m A^{-1}_0 \m K^\eta_{\m \theta_\eta}(\m t^a, \tilde{\m
        t})\\
      &\quad-
        \frac{1}{n-1}n_a \m 1_{n_b,n_a}
        \otimes
        \bigg(
        -\frac{n}{n_a(n-1)}
        \m K^\eta_{\m \theta_\eta}(\tilde{\m t}, \m t^a)
        \m A^{-1}_0 \m K^\eta_{\m \theta_\eta}(\m t^a, \tilde{\m t})\\
      &\qquad+
        \frac{n-n_a}{n_a(n-1)}\m K^\eta_{\m \theta_\eta}(\tilde{\m t}, \m t^a) \m A^{-1}_1
        \m K^\eta_{\m \theta_\eta}(\m t^a, \tilde{\m t})
        \bigg)\\
      &= - \frac{n-n_a}{(n-1)^2} \m 1_{n_b, n_a} \otimes
        \m K^\eta_{\m \theta_\eta}(\tilde{\m t}, \m t^a) \m A^{-1}_1 \m K^\eta_{\m \theta_\eta}(\m t^a, \tilde{\m t}).
    \end{split}
  \end{equation*}
    Using \cref{eq:AiCaat}, we find
  \begin{equation}
    \label{eq:CAiCaat}
    \begin{split}
      \m C
      &\m A^{-1} (\m C^{\eta^a,y^a})^T\\
      &= (\m 1_{n_a}^T \otimes (\m C^b)^T) \m A^{-1} (\m C^{\eta^a,y^a})^T\\
      &= \frac{n}{n-1} \m 1_{n_a}^T \otimes (\m C^b)^T \m A^{-1}_0 \m K^\eta_{\m \theta_\eta}(\m
        t^a, \tilde{\m t})\\
      &\quad+ n_a \m 1_{n_a}^T \otimes
        \left(
        - \frac{n}{n_a(n-1)}(\m C^b)^T \m A^{-1}_0 \m K^\eta_{\m \theta_\eta}(\m t^a, \tilde{\m
        t})
        +
        \frac{n-n_a}{n_a(n-1)}(\m C^b)^T \m A^{-1}_1 \m K^\eta_{\m \theta_\eta}(\m t^a, \tilde{\m t})
        \right)\\
      &=
        \frac{n-n_a}{n-1} \m 1_{n_a}^T
        \otimes
        (\m C^b)^T \m A^{-1}_1 \m K^\eta_{\m \theta_\eta}(\m t^a, \tilde{\m t}).
    \end{split}
  \end{equation}
  Using \cref{eq:AiCtSi} and \cref{eq:CAiCaat} we find
  \begin{equation*}
    \begin{split}
      \m A^{-1}
      &\m C^T \m S^{-1} \m C \m A^{-1} (\m C^{\eta^a,y^a})^T\\
      &=
        (\m 1_{n_a} \otimes \m A^{-1}_1 \m C^b\m S^{-1})
        \left(
        \frac{n-n_a}{n-1} \m 1_{n_a}^T \otimes
        (\m C^b)^T \m A^{-1}_1 \m K^\eta_{\m \theta_\eta}(\m t^a, \tilde{\m t})
        \right)\\
      &= \frac{n-n_a}{n-1} \m 1_{n_a,n_a} \otimes
        \m A^{-1}_1 \m C^b \m S^{-1} (\m C^b)^T
        \m A^{-1}_1 \m K^\eta_{\m \theta_\eta}(\m t^a, \tilde{\m t})
    \end{split}
  \end{equation*}
  and thus
  \begin{equation*}
    \begin{split}
      \m C^{\eta^b,y^a}
      & \m A^{-1} \m C^T \m S^{-1} \m C \m A^{-1} (\m C^{\eta^a,y^a})^T\\
      &=
        \left(
        -\frac{1}{n-1} \m 1_{n_b,n_a} \otimes
        \m K^\eta_{\m \theta_\eta}(\tilde{\m t}, \m t^a)
        \right)\\
      &\quad
        \left(
        \frac{n-n_a}{n-1}\m 1_{n_a,n_a}
        \otimes
        \m A^{-1}_1 \m C^b \m S^{-1} (\m C^b)^T \m A^{-1}_1 \m K^\eta_{\m \theta_\eta}(\m t^a, \tilde{\m t})
        \right)\\
      &=
        -\frac{n-n_a}{(n-1)^2}n_a \m 1_{n_b,n_a} \otimes
        \m K^\eta_{\m \theta_\eta}(\tilde{\m t}, \m t^a)\m A^{-1}_1 \m C^b \m S^{-1} (\m C^b)^T
        \m A^{-1}_1 \m K^\eta_{\m \theta_\eta}(\m t^a, \tilde{\m t}).
    \end{split}
  \end{equation*}
  Using \cref{eq:AiCtSi}, we find
  \begin{equation*}
    \begin{split}
      \m A^{-1}
      & \m C^T \m S^{-1} (\m C^{\eta^a,y^b})^T\\
      &= (\m 1_{n_a} \otimes \m A^{-1}_1 \m C^b\m S^{-1})
        \left(
        - \frac{1}{n-1} \m 1_{n_a}^T \otimes
        \m K^\eta_{\m \theta_\eta}(\m t^b, \tilde{\m t})
        \right)\\
      &=
        - \frac{1}{n-1} \m 1_{n_a,n_a}
        \otimes
        \m A^{-1}_1 \m C^b \m S^{-1} \m K^\eta_{\m \theta_\eta}(\m t^b, \tilde{\m t})
    \end{split}
  \end{equation*}
  and thus
  \begin{equation*}
    \begin{split}
      - \m C^{\eta^b,y^a}
      & \m A^{-1} \m C^T \m S^{-1} (\m C^{\eta^a,y^b})^T\\
      &= -
        \left(
        -\frac{1}{n-1}\m 1_{n_b,n_a} \otimes
        \m K^\eta_{\m \theta_\eta}(\tilde{\m t}, \m t^a)
        \right)
        \left(
        -\frac{1}{n-1}\m 1_{n_a,n_a} \otimes
        \m A^{-1}_1 \m C^b \m S^{-1} \m K^\eta_{\m \theta_\eta}(\m t^b, \tilde{\m t})
        \right)\\
      &= - \frac{1}{(n-1)^2} n_a \m 1_{n_b, n_a}
        \otimes
        \m K^\eta_{\m \theta_\eta}(\tilde{\m t}, \m t^a)\m A^{-1}_1 \m C^b \m S^{-1} \m K^\eta_{\m \theta_\eta}(\m
        t^b, \tilde{\m t}).
    \end{split}
  \end{equation*}
  Using \cref{eq:CAiCaat}, we find
  \begin{equation*}
    \begin{split}
      \m S^{-1}
      &\m C \m A^{-1} (\m C^{\eta^a,y^a})^T\\
      &= (1 \otimes \m S^{-1})
        \left(
        \frac{n-n_a}{n-1}\m 1_{n_a}^T \otimes (\m C^b)^T \m A^{-1}_1 \m
        K^\eta_{\m \theta_\eta}(\m t^a, \tilde{\m t})
        \right)\\
      &=
        \frac{n-n_a}{n-1}\m 1_{n_a}^T \otimes
        \m S^{-1} (\m C^b)^T \m A^{-1}_1 \m K^\eta_{\m \theta_\eta}(\m t^a, \tilde{\m t}).
    \end{split}
  \end{equation*}
  We find
  \begin{equation*}
    \begin{split}
      \m S^{-1}
      & (\m C^{\eta^a,y^b})^T\\
      &=
        (1 \otimes \m S^{-1})
        \left(
        -\frac{1}{n-1}\m 1_{n_a}^T\otimes \m K^\eta_{\m \theta_\eta}(\m t^b, \tilde{\m t})
        \right)\\
      &=
        -\frac{1}{n-1} \m 1_{n_a}^T \otimes \m S^{-1} \m K^\eta_{\m \theta_\eta}(\m t^b, \tilde{\m t})
    \end{split}
  \end{equation*}
  and
  \begin{equation*}
    \begin{split}
      \m C^{\eta^b,y^a}
      & \m A^{-1} (\m C^{\eta^b,y^a})^T\\
      &=
        \left(
        - \frac{1}{n-1} \m 1_{n_b,n_a} \otimes
        \m K^\eta_{\m \theta_\eta}(\tilde{\m t}, \m t^a)
        \right)\\
      &\quad\left(\m I_{n_a} \otimes \m A^{-1}_0 +
        \frac{1}{n_a}\m 1_{n_a,n_a}
        \otimes
        (\m A^{-1}_1 - \m A^{-1}_0)\right)\\
      &\quad
        \left(
        -\frac{1}{n-1}
        \m 1_{n_a, n_b} \otimes
        \m K^\eta_{\m \theta_\eta}(\m t^a, \tilde{\m t})
        \right)\\
      &=
        \frac{1}{(n-1)^2} n_a \m 1_{n_b,n_b} \otimes \m K^\eta_{\m \theta_\eta}(\tilde{\m t}, \m
        t^a)
        \m A^{-1}_0 \m K^\eta_{\m \theta_\eta}(\m t^a, \tilde{\m t})\\
      &\quad+
        \frac{1}{n_a(n-1)^2}n_a^2 \m 1_{n_b,n_b}
        \m K^\eta_{\m \theta_\eta}(\tilde{\m t}, \m t^a)
        (\m A^{-1}_1 - \m A^{-1}_0) \m K^\eta_{\m \theta_\eta}(\m t^a, \tilde{\m t})\\
      &=
        \frac{n_a}{(n-1)^2} \m 1_{n_b,n_b} \m K^\eta_{\m \theta_\eta}(\tilde{\m t}, \m t^a)
        \m A^{-1}_1 \m K^\eta_{\m \theta_\eta}(\m t^a, \tilde{\m t}).
    \end{split}
  \end{equation*}
  Using \cref{eq:AiCt}, we find
  \begin{equation*}
    \begin{split}
      \m C^{\eta^b,y^a}
      & \m A^{-1} \m C^T \m S^{-1}
        \m C \m A^{-1} (\m C^{\eta^b,y^a})^T\\
      &=
        \left(
        - \frac{1}{n-1} \m 1_{n_b, n_a} \otimes
        \m K^\eta_{\m \theta_\eta}(\tilde{\m t}, \m t^a)
        \right)\\
      &\quad 
        \left(
        \m 1_{n_a} \otimes
        \m A^{-1}_1 \m C^b
        \right)\\
      &\quad
        \left(
        1 \otimes \m S^{-1}
        \right)\\
      &\quad 
        \left(
        \m 1_{n_a}^T \otimes (\m C^b)^T \m A^{-1}_1
        \right)\\
      &\quad
        \left(
        - \frac{1}{n-1} \m 1_{n_a,n_b} \otimes \m K^\eta_{\m \theta_\eta}(\m t^a, \tilde{\m t})
        \right)\\
      &=
        \frac{1}{(n-1)^2} \m 1_{n_b,n_a}\m 1_{n_a} 1 \m 1_{n_a}^T \m 1_{n_a,n_b}
        \otimes
        \m K^\eta_{\m \theta_\eta}(\tilde{\m t}, \m t^a)
        \m A^{-1}_1 \m C^b \m S^{-1}
        (\m C^b)^T \m A^{-1}_1 \m K^\eta_{\m \theta_\eta}(\m t^a, \tilde{\m t})\\
      &=
        \left(
        \frac{n_a}{n-1}
        \right)^2
        \m 1_{n_b,n_b}
        \otimes
        \m K^\eta_{\m \theta_\eta}(\tilde{\m t}, \m t^a)\m A^{-1}_1 \m C^b
        \m S^{-1} (\m C^b)^T \m A^{-1}_1 \m K^\eta_{\m \theta_\eta}(\m t^a, \tilde{\m t}).
    \end{split}
  \end{equation*}
  Using \cref{eq:AiCtSi}, we find
  \begin{equation*}
    \begin{split}
      - \m C^{\eta^b,y^a}
      & \m A^{-1} \m C^T \m S^{-1}\\
      &=
        -
        \left(
        - \frac{1}{n-1} \m 1_{n_b,n_a} \otimes
        \m K^\eta_{\m \theta_\eta}(\tilde{\m t}, \m t^a)
        \right)
        \left(
        \m 1_{n_a} \otimes \m A^{-1}_1 \m C^b \m S^{-1}
        \right)\\
      &=
        \frac{1}{n-1} n_a \m 1_{n_b} \otimes
        \m K^\eta_{\m \theta_\eta}(\tilde{\m t}, \m t^a)
        \m A^{-1}_1
        \m C^b \m S^{-1},
    \end{split}
  \end{equation*}
  which completes the proof of the expressions for the posterior covariance matrix of $\m \eta(\tilde{\m t})$.

\subsubsection*{Posterior cross-covariance of $\m \mu(\tilde{\m t})$ and $\m \eta(\tilde{\m t})$}

Exactly as in the proof of \cref{prop:postmureg}, we get from the model specification in \cref{mod:eta} and then conditioning in the multivariate normal distribution that the joint posterior of $(\m \eta(\tilde{\m t}), \m \mu(\tilde{\m t}))$ is multivariate normal with cross-covariance (the lower left block of the joint posterior covariance matrix of $(\m \eta(\tilde{\m t}), \m \mu(\tilde{\m t}))$) given by $\m \Sigma^T_{\m \eta(\tilde{\m t}), \m \mu(\tilde{\m t}) \mid \m y(\m t), \m \Theta} = -\m C^{\mu, y}\m \Sigma_{\m \Theta}^{-1}\m C^{y,\eta}$.

We will simplify $-\m C^{\mu, y}\m \Sigma_{\m \Theta}^{-1}$. This can be
written
\begin{equation*}
  \begin{split}
    -\m C^{\mu, y}\m \Sigma_{\m \Theta}^{-1}
    &=
      -
      \begin{pmatrix}
        \m C^{\mu, a} & \m C^{\mu, b}
      \end{pmatrix}
      \begin{pmatrix}
        \m P_1\\
        \m P_2
      \end{pmatrix}\\
    &=
      - \m C^{\mu, a}\m P_1 - \m C^{\mu, b}\m P_2.
  \end{split}
\end{equation*}
By writing out $\m C^{\mu, a}$ and $\m P_1$ we get
\begin{equation*}
  \begin{split}
    \m C^{\mu, a}\m P_1
    &= \left(
      (\m 1_{n_a}^T \otimes \m K^\mu_{\m \theta_\mu}(\tilde{\m t}, \m t^a)
      \right)
      \m P_1\\
    &=
      \begin{pmatrix}
        \m H_1 + \m H_2 & \m H_3
      \end{pmatrix}
  \end{split}
\end{equation*}
where
\begin{equation*}
  \begin{split}
    \m H_1
    &=
      (\m 1_{n_a}^T \otimes \m K^\mu_{\m \theta_\mu}(\tilde{\m t}, \m t^a))\m A^{-1}\\
    &=
      (\m 1_{n_a}^T \otimes \m K^\mu_{\m \theta_\mu}(\tilde{\m t}, \m t^a))
      \left(
      \m I_{n_a} \otimes \m A_0 ^{-1} + \frac{1}{n_a} \m 1_{n_a,n_a} \otimes (\m
      A_1 ^{-1} - \m A_0 ^{-1})\right)\\
    &=
      \m 1_{n_a}^T \otimes \m K^\mu_{\m \theta_\mu}(\tilde{\m t}, \m t^a) \m A_0 ^{-1} +
      \m 1_{n_a}^T \otimes \m K^\mu_{\m \theta_\mu}(\tilde{\m t}, \m t^a)(\m A_1 ^{-1} - \m A_0
      ^{-1})\\
    &=
      \m 1_{n_a}^T \otimes \m K^\mu_{\m \theta_\mu}(\tilde{\m t}, \m t^a)\m A_1 ^{-1}
  \end{split}
\end{equation*}
and
\begin{equation*}
  \begin{split}
    \m H_2
    &=
      (\m 1_{n_a}^T \otimes \m K^\mu_{\m \theta_\mu}(\tilde{\m t}, \m t^a))
      (\m 1_{n_a} \otimes \m A_1 ^{-1} \m C^b)
      (\m S^{-1} \m C \m A^{-1})\\
    &=
      n_a \m K^\mu_{\m \theta_\mu}(\tilde{\m t}, \m t^a) \m A^{-1}_1 \m C^b \m S^{-1} \m C \m
      A^{-1}\\
    &=
      - \m H_3 \m C \m A^{-1}
  \end{split}
\end{equation*}
and
\begin{equation*}
  \begin{split}
    \m H_3
    &=
      - (\m 1_{n_a}^T \otimes \m K^\mu_{\m \theta_\mu}(\tilde{\m t}, \m t^a))
      (\m 1_{n_a} \otimes \m A^{-1}_1 \m C^b)
      \m S^{-1}\\
    &=
      - n_a \m K^\mu_{\m \theta_\mu}(\tilde{\m t}, \m t^a) \m A^{-1}_1 \m C^b \m S^{-1}.
  \end{split}
\end{equation*}
By writing out $\m P_2$ we get
\begin{equation*}
  \begin{split}
    \m C^{\mu, b} \m P_2 =
    \begin{pmatrix}
      - \m C^{\mu, b} \m S^{-1} \m C \m A^{-1} & \m C^{\mu, b} \m S^{-1}
    \end{pmatrix}
  \end{split}
\end{equation*}
which completes the proof.

\subsection*{Iterative Block Cholesky Factorization}

The block Cholesky algorithm \citep[Section 4.2.9]{golub2013matrix} is based on iteratively applying \cref{eq:blockchol} as we now describe.

\subsubsection*{Completely regular sampling design}

From the results in \cref{prop:postmureg} we have that
\begin{equation*}
    \m \Sigma_{\m \eta'(\tilde{\m t}) \mid \m y(\m t), \m \Theta}
    =
      \m I_{n-1} \otimes (\m V - \m W) + \m 1_{n-1,n-1} \otimes \m W
    =
      \begin{pmatrix}
        \m V & \m W & \cdots & \m W\\
        \m W & \m V & \ddots & \vdots\\
        \vdots & \ddots & \ddots & \m W\\
        \m W & \cdots & \m W & \m V
      \end{pmatrix}
\end{equation*}
with
\begin{equation*}
  \begin{split}
    \m W &= \frac{1}{n-1}
           \left(
           \frac{n}{n-1}
           \m K^\eta_{\m \theta_\eta}(\tilde{\m t}, \m t)
           \m \Sigma_0^{-1}
           \m K^\eta_{\m \theta_\eta}(\m t, \tilde{\m t})
           -
           \m K^\eta_{\m \theta_\eta}(\tilde{\m t}, \tilde{\m t})
           \right)\\
    \m V &= \m W + \frac{n}{n-1}
                 \left(
                 \m K^\eta_{\m \theta_\eta}(\tilde{\m t}, \tilde{\m t}) - \frac{n}{n-1}
                 \m K^\eta_{\m \theta_\eta}(\tilde{\m t}, \m t) \m \Sigma_0^{-1}
                 \m K^\eta_{\m \theta_\eta}(\m t, \tilde{\m t})
                 \right).
  \end{split}
\end{equation*}
To perform the iterative block Cholesky factorization we first find the Cholesky factor of the upper left $1\times 1$ block matrix, which is just $\ch \m V$. We then use
\cref{eq:blockchol} with $\m A = \m B = \m V$, $\m C = \m W$ to find the
Cholesky factor of the upper $2\times 2$ block matrix
\begin{equation*}
  \ch
  \begin{pmatrix}
    \m V & \m W\\
    \m W & \m V
  \end{pmatrix}=
  \begin{pmatrix}
    \ch \m V & \m 0\\
    \m W (\ch \m V)^{-T} & \ch \m S
  \end{pmatrix}
\end{equation*}
where $\m S = \m V - \m W\m V ^{-1}\m W$. This now enables us to use
\cref{eq:blockchol} with $\m B = \m V$, $\m C = (\m W, \m W)$ and
\begin{equation*}
  \m A = \begin{pmatrix}
    \m V & \m W\\
    \m W & \m V
  \end{pmatrix}
\end{equation*}
to find
\begin{equation*}
  \ch
  \begin{pmatrix}
    \m V & \m W & \m W\\
    \m W & \m V & \m W\\
    \m W & \m W & \m V
  \end{pmatrix}
  =
  \begin{pmatrix}
    \ch
    \begin{pmatrix}
      \m V & \m W\\
      \m W & \m V      
    \end{pmatrix}
    & \m 0\\
    \begin{pmatrix}
      \m W & \m W
    \end{pmatrix}
    \left[
    \ch
    \begin{pmatrix}
      \m V & \m W\\
      \m W & \m V
    \end{pmatrix}
    \right]^{-T}
    & \ch \m S
  \end{pmatrix}
\end{equation*}
where
\begin{equation*}
  \m S = \m V -
  \begin{pmatrix}
    \m W & \m W
  \end{pmatrix}
  \begin{pmatrix}
    \m V & \m W\\
    \m W & \m V
  \end{pmatrix}^{-1}
  \begin{pmatrix}
    \m W \\ \m W
  \end{pmatrix}.
\end{equation*}
We continue in the same way, always reusing the result from the previous step,
until we end with the Cholesky factor of the entire matrix.

We will now give a fast way of calculating the matrix division $\m C(\ch \m A)^{-1}$ in each step, after first motivating it by showing why the iterative block Cholesky factorization algorithm is not in general faster than the standard Cholesky factorization algorithm. Let $\m M_i = \m I_i \otimes (\m V - \m W) + \m 1_{i,i} \otimes \m W$, so it is
the upper left $i \times i$ block submatrix of $\m \Sigma_{\m \eta'(\tilde{\m t}) \mid \m y(\m t), \m \Theta}$, and let
$\m S_i = \m V - (\m 1_{i-1}^T \otimes \m W)\m M_{i-1}^{-1}(\m 1_{i-1} \otimes
\m W)$ be the Schur complement used in the $i$'th step of the iterative block
Cholesky algorithm. In the $i$'th step, we already know $\ch \m M_{i-1}$ from
the previous step and now calculate $\ch \m M_i$ by doing as follows. We need the Cholesky decomposition of $\m S_i$,
which is of order $O(J_p^3)$ as $\m S_i \in \R^{J_p \times J_p}$, and we have to
calculate the matrix left division
\begin{equation*}
  (\m 1_{i-1}^T \otimes \m W)(\ch \m M_{i-1})^{-T} =
  \left[(\ch \m M_{i-1}) \backslash (\m 1_{i-1} \otimes \m W)\right]^T
\end{equation*}
which is $O(((i-1)J_p)^2 J_p) = O((i-1)^2J_p^3)$, since we do forward
substitution for each of the $J_p$ columns of $\m 1_{i-1} \otimes \m W$. Combining all the steps for $i = 1, \ldots, n-1$, the entire block Cholesky
factorization is at least
$\sum_{i = 1}^{n-1} \left[O((i-1)^2J_p^3) + O(J_p^3)\right] = O(n^3 J_p^3)$, which
is the same time complexity as just directly calculating the Cholesky factor of
$\m \Sigma_{\m \eta'(\tilde{\m t}) \mid \m y(\m t), \m \Theta} \in \R^{(n-1)J_p \times (n-1)J_p}$. Therefore, something more than the general iterative
block Cholesky algorithm is needed to obtain a speedup. Using the structure in our problem (namely, that all diagonal blocks and all off-diagonal blocks are equal), we will now show how to obtain a speedup.

Using the structure described above, we get that 
\begin{equation*}
  (\ch \m M_{i-1}) \backslash (\m 1_{i-1} \otimes \m W)
  =
  \begin{pmatrix}
    \ch \m M_{i-2} & \m 0_{(i-2)J_p, J_p}\\
    (\m 1_{i-2}^T \otimes \m W)(\ch \m M_{i-2})^{-T} & \ch \m S_{i-1}
  \end{pmatrix}
  \backslash
  \begin{pmatrix}
    \m 1_{i-2} \otimes \m W\\
    \m W
  \end{pmatrix}
\end{equation*}
so the left-division corresponds
to finding $\m x_1 \in \R^{(i-2)J_p \times J_p}$ and $\m x_2 \in \R^{J_p \times J_p}$ such that
\begin{equation*}
  \begin{split}
    \begin{pmatrix}
      \m 1_{i-2} \otimes \m W\\
      \m W
    \end{pmatrix}
    &=
    \begin{pmatrix}
      \ch \m M_{i-2} & \m 0_{(i-2)J_p, J_p}\\
      (\m 1_{i-2}^T \otimes \m W)(\ch \m M_{i-2})^{-T} & \ch \m S_{i-1}
    \end{pmatrix}
    \begin{pmatrix}
      \m x_1\\
      \m x_2
    \end{pmatrix}\\
    &=
      \begin{pmatrix}
        \ch (\m M_{i-2})\m x_1\\
        (\m 1_{i-2}^T \otimes \m W)(\ch \m M_{i-2})^{-T}\m x_1 + \ch (\m S_{i-1})\m x_2
      \end{pmatrix}
  \end{split}
\end{equation*}
since these will satisfy
\begin{equation*}
  \left[(\ch \m M_{i-1}) \backslash (\m 1_{i-1} \otimes \m W)\right]^T
  =
  (\m x_1^T, \m x_2^T).
\end{equation*}
From the top row we directly get
$\m x_1 = \ch (\m M_{i-2}) \backslash (\m 1_{i-2} \otimes \m W)$. Plugging this
into row two gives that
\begin{equation*}
  \m x_2 = \ch (\m S_{i-1}) \backslash
  \left[
    \m W -
    (\m 1_{i-2}^T \otimes \m W)
    \m M_{i-2}^{-1}
    (\m 1_{i-2} \otimes \m W)
  \right]
\end{equation*}
Since $\m x_1^T = [\ch (\m M_{i-2}) \backslash (\m 1_{i-2} \otimes \m W)]^T$,
which we already calculate in step $i-1$, we only have to calculate $\m x_2 ^T$ in step $i$. We already
calculate $\ch \m S_{i-1}$ in step $i-1$ by using $\m x_1$ to do the computation
\begin{equation*}
  \begin{split}
    (\m 1_{i-2}^T \otimes \m W) &\m M_{i-2}^{-1} (\m 1_{i-2} \otimes \m W)\\
                                &= \left[(\ch \m M_{i-2}) \backslash (\m 1_{i-2}
                                  \otimes \m W)\right]^T\left[(\ch \m M_{i-2})
                                  \backslash (\m 1_{i-2} \otimes \m W)\right]\\
                                &= \m x_1^T \m x_1
  \end{split}
\end{equation*}
which is $O((i-2) J_p^3)$ as we multiply the $J_p \times (i-2)J_p$ matrix
$\m x^T_1$ and the $(i-2)J_p \times J_p$ matrix $\m x_1$. To get $\m x_2$, it thus only remains to use the already computed terms to perform the final left division
$\ch (\m S_{i-1}) \backslash \left[ \m W - (\m 1_{i-2}^T \otimes \m W) \m
  M_{i-2}^{-1} (\m 1_{i-2} \otimes \m W) \right]$, which is
$O(J_p^3)$ since we do forward substitution for each of the $J_p$ columns. This
gives us a total time complexity of
$\sum_{i=1}^{n-1}[O((i-2)J_p^3) + O(J_p^3)] = O(n^2J_p^3)$. We have thus obtained a reduction from time complexity $O(n^3J_p^3)$ to $O(n^2 J_p^3)$. As shown in the benchmark results in \cref{sec:benchmarks}, this has significant practical importance concerning the runtime. 

\subsubsection*{Partially regular sampling design}

In the irregular case, we have from \cref{prop:postirreg} that
\begin{equation*}
  \m \Sigma_{\m \eta(\tilde{\m t}) \mid \m y(\m t), \m \Theta} =
  \begin{pmatrix}
    \m \Sigma_{\eta,1,1} & \m \Sigma_{\eta,1,2}\\
    \m \Sigma_{\eta,2,1} & \m \Sigma_{\eta,2,2}
  \end{pmatrix}
\end{equation*}
where
$\m \Sigma_{\eta,1,1} = \m I_{n_a} \otimes (\m V - \m W) + \m 1_{n_a, n_a}
\otimes \m W$ with $\m V$ and $\m W$ given in
\cref{prop:postirreg}. We can therefore reuse the iterative block Cholesky
procedure from above (with all the simplifications) to get
$\ch (\m \Sigma_{\eta,1,1})$. Afterwards, we simply use \cref{eq:blockchol} a
single time to get $\ch (\m \Sigma_{\m \eta'(\tilde{\m t}) \mid \m y(\m t), \m \Theta})$, while discarding the part that has
to do with $\eta_n$. To do this, we need the Cholesky decomposition of
$\m S = \m \Sigma_{\eta,2,2} - \m \Sigma_{\eta,2,1}\m \Sigma_{\eta,1,1}^{-1}\m
\Sigma_{\eta,1,2}$, which is of time complexity
$O\left[\left(\sum_{i = 1}^{n_b} J^b_i\right)^3\right]$, so it will scale poorly
with the number of irregularly sampled functions. However, the iterative block Cholesky algorithm
gives us the same scaling as before for the  part of the covariance matrix that corresponds to the regularly sampled functions. Once we have $\ch (\m \Sigma_{\m \eta'(\tilde{\m t}) \mid \m y(\m t), \m \Theta})$, we can use \cref{eq:blockchol} one final time to get the Cholesky decomposition of the entire joint posterior covariance matrix for $\m \eta'(\tilde{\m t})$ and $\m \mu(\tilde{\m t})$, which is
\begin{equation*}
        \ch   \begin{pmatrix}
      \m \Sigma_{\m \eta'(\tilde{\m t}) \mid \m y(\m t), \m \Theta} & \m \Sigma_{\m \eta'(\tilde{\m t}), \m \mu(\tilde{\m t}) \mid \m y(\m t), \m \Theta}\\
      \m \Sigma^T_{\m \eta'(\tilde{\m t}), \m \mu(\tilde{\m t}) \mid \m y(\m t), \m \Theta} & \m \Sigma_{\m \mu(\tilde{\m t}) \mid \m y(\m t), \m \Theta}
  \end{pmatrix} =
    \begin{pmatrix}
      \ch (\m \Sigma_{\m \eta'(\tilde{\m t}) \mid \m y(\m t), \m \Theta}) & \m 0\\
      \m \Sigma^T_{\m \eta'(\tilde{\m t}), \m \mu(\tilde{\m t}) \mid \m y(\m t), \m \Theta}(\ch (\m \Sigma_{\m \eta'(\tilde{\m t}) \mid \m y(\m t), \m \Theta}))^{-T} & \ch \m S_\mu
    \end{pmatrix}
\end{equation*}
where
\begin{equation*}
  \m S_\mu := \m \Sigma_{\m \mu(\tilde{\m t}) \mid \m y(\m t), \m \Theta} - \m \Sigma^T_{\m \eta'(\tilde{\m t}), \m \mu(\tilde{\m t}) \mid \m y(\m t), \m \Theta}\m \Sigma^{-1}_{\m \eta'(\tilde{\m t}) \mid \m y(\m t), \m \Theta}\m \Sigma_{\m \eta'(\tilde{\m t}), \m \mu(\tilde{\m t}) \mid \m y(\m t), \m \Theta}.
\end{equation*}

\end{document}